\documentclass[aps,prb,preprint,superscriptaddress]{revtex4-1}

\usepackage{color}
\usepackage{graphicx}  
\usepackage{dcolumn}   
\usepackage{bm}        
\usepackage{amssymb}   
\newcommand{\degr}{$^{\circ}$}

\newcommand{\Co}{Sc$_3$CoC$_4$}
\newcommand{\ixrd}{$I_\mathrm{XRD}(T,p)$}
\newcommand{\rhoT}{$\rho(T)$}
\newcommand{\MT}{$M(T)$}
\newcommand{\Tconset}{$T_\mathrm{c}^{onset}$}

\begin{document}


\title{The structure of the superconducting high-pressure
  phase of \Co}


\author{Jan Langmann}
\author{Marcel Vöst}
\author{Dominik Schmitz}
\author{Christof Haas}
\affiliation{CPM,
  Institut f\"ur Physik, Universit\"at Augsburg, 
  D-86159 Augsburg, Germany}
\author{Georg Eickerling}
\email{georg.eickerling@uni-a.de}
\affiliation{CPM,
  Institut f\"ur Physik, Universit\"at Augsburg, 
  D-86159 Augsburg, Germany}
\author{Anton Jesche}
\affiliation{Experimentalphysik VI,
  Zentrum f\"ur Elektronische Korrelation und Magnetismus,
  Institut f\"ur Physik, Universit\"at Augsburg,
  D-86159 Augsburg, Germany}
\author{Michael Nicklas}
\affiliation{
  Max Planck Institute for Chemical Physics of Solids,
  N\"othnitzer Straße 40, D-01087 Dresden, Germany}
\author{Arianna Lanza}
\affiliation{
  Center for Nanotechnology Innovation@NEST, Istituto Italiano
  di Tecnologia, I-56127 Pisa, Italy}
\author{Nicola Casati}
\affiliation{Swiss Light Source,
  Paul Scherrer Institut, CH-5232 Villigen, Switzerland}
\author{Piero Macchi}
\affiliation{Dipartimento di Chimica,
  Materiali ed Ingegneria Chimica ``G. Natta'', Politecnico
  di Milano, I-20133 Milano, Italy}

\author{Wolfgang Scherer}
\email{wolfgang.scherer@uni-a.de}
\affiliation{CPM,
  Institut f\"ur Physik, Universit\"at Augsburg, 
  D-86159 Augsburg, Germany}


\date{\today}

\begin{abstract}
  We investigate pressure-induced structural changes to the
  Peierls-type distorted low-temperature phase of the low-dimensional
  \Co\ as a possible origin of its pressure-enhanced superconduc-
  tivity. By means of cryogenic high-pressure x-ray diffraction
  experiments we could reveal subtle, but significant structural
  differences between the low-temperature phase at ambient and
  elevated pressures. We could thus establish the structure of the
  superconducting phase of the title compound which interestingly
  still shows the main features of the Peierls-type distorted
  low-temperature phase.  This indicates that in contrast to other
  low-dimensional materials a suppression of periodic structural
  distortions is no prerequisite for superconducitivity in the
  transition metal carbide.
\end{abstract}


\maketitle


\section{Introduction}\label{sec:intro}

Structurally low-dimensional materials and dimensionality-driven
physical effects are making their way into technical
applications. Quantum dots (0D) are actively deployed in display
technology\cite{Kim11,Patel12} or under intensive research for future
uses in quantum computing.\cite{Li18} Nano-wires (1D) enable great
improvements in photo- and chemo-electric detectors and thermoelectric
devices.\cite{Dasgupta14} Transistors fabricated of atomically thin
graphene layers (2D) might become an integral part of post-silicon
microprocessors.\cite{Schwierz10,Hills19} Futhermore, 3D
superstructures of thin metal layers -- arranged in the right way to
break inversion symmetry -- might provide promising candidates for
diodes in superconducting electronics.\cite{Ando20}

\begin{figure}[htb]
  \centering
  \includegraphics[width=0.4\textwidth]{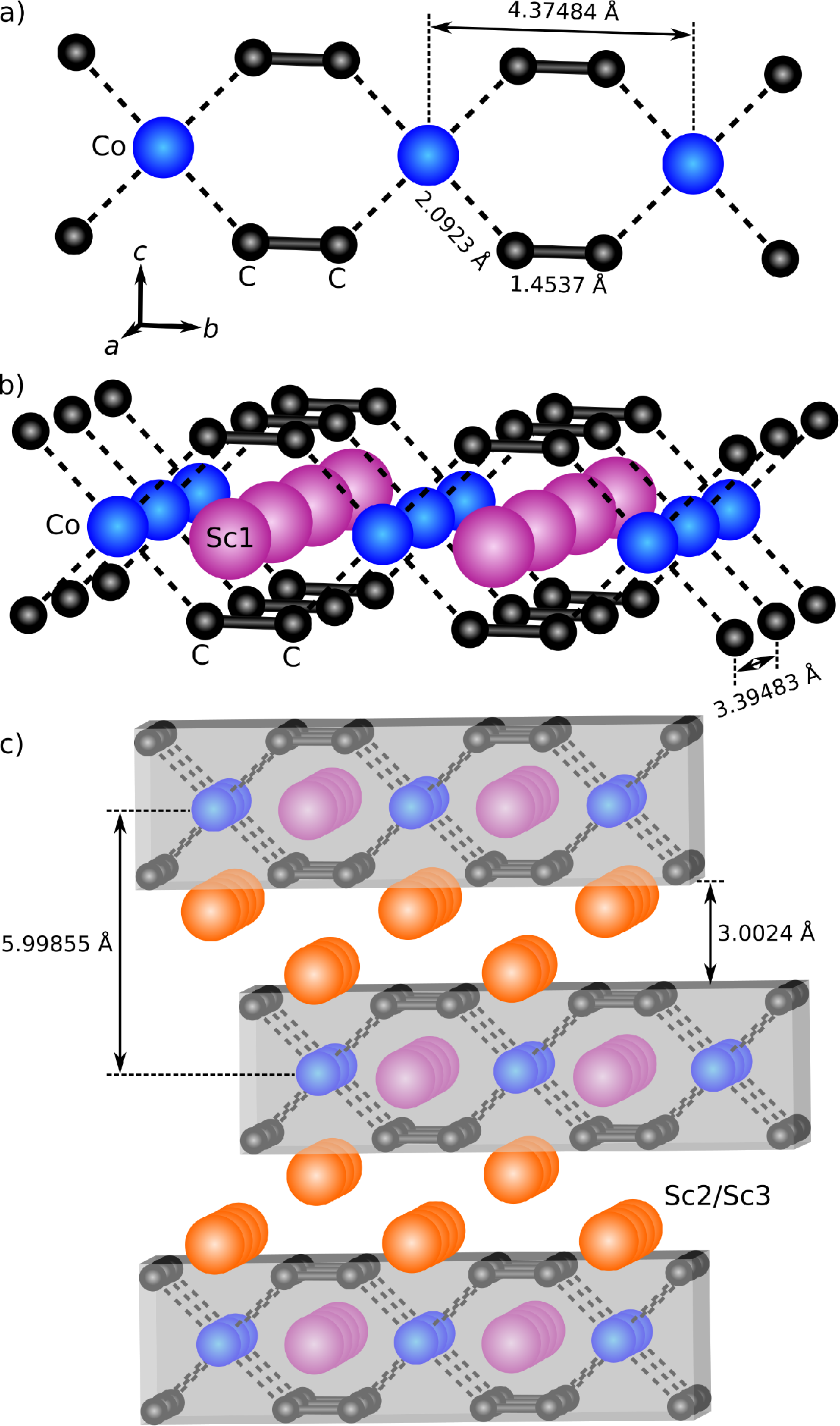}
  \caption{(a)~Infinite [Co(C$_2$)$_2$]$_\infty$ ribbon as basic
    quasi-1D building unit of the \Co\ structure; (b)~composition of
    a quasi-2D layer by stacking [Co(C$_2$)$_2$]$_\infty$ ribbons and
    Sc1 atoms along the $a$ axis of the orthorhombic high-temperature
    phase; (c)~decoupling of the quasi-2D layers along the $c$ axis by
    interleaved Sc2 and Sc3 atoms. Salient inter-atomic distances are
    specified (structural data from Ref.~\citenum{Eickerling13}).}
  \label{fig:sc3coc4-structure}
\end{figure}

As indicated by the last example, the combination of structural low
dimensionality with superconductivity can stimulate intriguing
effects, even though W. A. Little's prediction of room-temperature
superconductivity in quasi-one-dimensional
materials\cite{Little64,Little65} still remains an unobserved
phenomenon at ambient pressure. Instead, a rich playing field of
different ordering phenomena interacting with superconductivity has
unfolded, \textit{e.g.}  structural transitions, charge- and
spin-density waves, and antiferromagnetism.\cite{Pyon12, Yang12,
  Kiswandhi13, Hodeau78, Regueiro92, Cho18, Chang12, Chikina20,
  Knafo16, Chubukov08, Dai12} Superconducting compounds with
intrinsically low-dimensional character have a special appeal to
solid-state sciences, although the synthesis of large and defect-free
single crystals is often challenging. This, for example, becomes
evident from the large number of publications devoted to the
well-known quasi-one-dimensional NbSe$_3$\cite{Meerschaut75, Hodeau78,
  Regueiro92, Yang19} or quasi-two-dimensional
graphite/graphene\cite{Novoselov04, Cao18} and transition-metal
dichalcogenides.\cite{Williams74, Lee19, Hughes77, Shu19, Pyon12,
  Oike18}

The transition-metal carbide \Co\ crystallizes in a structure type
combining quasi-1D and quasi-2D features.  Quasi-1D
[Co(C$_2$)$_2$]$_\infty$ ribbons extending along the crystallographic
$b$ axis of the orthorhombic unit cell
(Fig.~\ref{fig:sc3coc4-structure}a) are formed by covalent bonds
between the cobalt atoms and C$_2$ moieties.\cite{Rohrmoser07}
Alternating stacking of the [Co(C$_2$)$_2$]$_\infty$ ribbons and
scandium atoms (Sc1) along the $a$ axis leads to quasi-2D layers
(Fig.~\ref{fig:sc3coc4-structure}b). Therein, neighboring
[Co(C$_2$)$_2$]$_\infty$ ribbons with a separation of 3.39483(3)~\AA\
are held together by subtle Sc-C$_2$ interactions. Additional scandium
atom layers (Sc2 and Sc3; Fig.~\ref{fig:sc3coc4-structure}c) are
interleaved along the $c$ axis resulting in a large interlayer
distance of 5.99855(5)~\AA\ between adjacent Sc1-Co-C
layers.\cite{jeitschko_carbon_1989, Tsokol86, Rohrmoser07, Scherer10,
  Scheidt11, Scherer12, Eickerling13, He15}

Superconductivity in \Co\ emerges below
$T_{\mathtt{c}} \approx$~4.5~K\cite{Scheidt11, Scherer10,
  Eickerling13} and is anticipated by a Peierls-type structural
transition below 72~K.\cite{Eickerling13} Therein, the orthorhombic
high-temperature (HT) phase structure (space group $Immm$) is
transformed into the monoclinic low-temperature (LT) phase structure
(space group $C2/m$) by a doubling of the translational period along
the [Co(C$_2$)$_2$]$_\infty$ ribbons.\cite{Scherer10, Scherer12,
  Eickerling13, Langmann20} The exact degree and mode of interaction
between this structural HT$\rightarrow$LT phase transition and the onset of
superconductivity at even lower temperatures is, however, not fully
established yet. Furthermore, high-pressure studies of the electrical
resistivity and magnetization in polycrystalline samples by Wang
\textit{et al.}\cite{Wang16} revealed a drastic increase of the
superconducting volume at virtually constant $T_\mathrm{c}$
values. The authors rationalized this behavior by a pressure- and
temperature-controlled coexistence of the HT and LT phase in the
compound, whereby only the HT phase was supposed to become
superconducting.\cite{Wang16} But no structural information to verify
this hypothesis has been provided up to now. Therefore, we performed
high-pressure and low-temperature single-crystal x-ray diffraction
studies in combination with physical property measurements to explore
the pressure- and temperature-dependent structure-property
relationship in \Co.

\section{Methods} \label{sec:experimental}

Single- and polycrystalline samples of \Co\ were synthesized by
arc-melting according to the method described in the
literature\cite{Vogt09, Rohrmoser07, He15} and in addition from a
lithium metal flux.\cite{Jesche14,Haas19} Needle-like samples were
obtained from arc-melting and platelet-like samples from
crystallization in a lithium flux (full details of the synthesis and
characterization methods employed can be found in the Supplemental
Material\footnote{see Supplemental Material at [URL will be inserted
  by publisher] for information on synthesis and properties of the
  investigated samples, details of the magnetization, resistivity and
  x-ray diffraction experiments under ambient and high-pressure
  conditions, phonon dispersion relations under uniaxial strain, and
  remarks on the analysis of the experimental data.}).

Magnetization measurements on a single-crystalline \Co\ sample were
performed at various pressures up to 1.48~GPa using a miniature
Ceramic Anvil Cell (mCAC)\cite{tateiwa_high_2014,
  tateiwa_note_2013,tateiwa_magnetic_2012,tateiwa_miniature_2011,
  Kobayashi07} assembled with a Cu:Be gasket. The respective pressures
were determined at low temperatures by reference to the pressure
dependence of $T_\mathrm{c}$ for an additional lead piece inside the
pressure chamber.\cite{eiling_pressure_1981,bireckoven_diamond_1988}
Both, single-crystal and lead pressure gauge, were surrounded by
Daphne 7373\cite{Yokogawa07,Murata08} serving as a pressure
transmitting medium. Supplemental ambient-pressure measurements before
and after the high-pressure study were performed by gluing the sample
to a glass rod with GE Varnish. For all magnetization measurements a
QUANTUM DESIGN MPMS3 SQUID magnetometer was employed. The
superconducting properties of the \Co\ sample were investigated by
cooling the pressure cell or glass rod to 1.8~K under
zero-field-cooling conditions and recording the temperature-dependent
magnetization while heating from 1.8~K to 9~K in a magnetic field of
5~Oe.

High-pressure electrical resistivity measurements up to 1.26~GPa were
performed employing a piston-cylinder-type pressure cell and silicon
oil as pressure-transmitting medium. The single-crystalline \Co\
whisker was contacted by a four-point configuration using silver
conductive paint and gold filaments. The pressure inside the pressure
chamber was determined at low temperatures by measuring the shift of
the superconducting transition temperature of a piece of
lead.\cite{eiling_pressure_1981,bireckoven_diamond_1988} For details
of the setup see Ref.~\citenum{Nicklas15}. The temperature-dependent
resistivity measurements were carried out for various applied
pressures upon cooling and heating cycles between 1.8~K and 300~K in a
QUANTUM DESIGN PPMS using a LINEAR RESEARCH LR700 resistance
bridge.\cite{Schmitz18} Additional ambient-pressure measurements of
single-crystalline \Co\ whiskers four-point contacted with
silver-epoxy resin were taken without surrounding pressure cell and
using the standard DC-resistivity option of a QUANTUM DESIGN PPMS.
Uniaxial strain was created by gluing both ends of a whisker to a
sapphire substrate using large droplets of silver-epoxy resin.

Pressure-dependent lattice parameters at room temperature were
obtained from Le Bail fits\cite{LeBail88,Berar91} of synchrotron powder
x-ray diffraction data with the software JANA2006.\cite{Petricek14}
The respective diffraction experiments were carried out at the X04SA
Materials Science (MS) beamline at the Swiss Light Source
(SLS)\cite{Willmott13,Fisch15} using a PSI Mythen~II one-dimensional
detector\cite{Bergamaschi10} and a membrane-driven diamond anvil cell
(DAC).  The pressure chamber was filled with finely ground and sieved
\Co\ powder (nominal sieve opening 32~$\mu$m), and a 4:1 volume
mixture of methanol and ethanol\cite{piermarini_hydrostatic_1973} was
used as pressure-transmitting medium. $\alpha$-quartz powder was added for
pressure calibration by reference to its well-known equation of
state.\cite{Angel97}

The single-crystal x-ray diffraction data in this work was collected
on a HUBER four-circle Eulerian cradle goniometer equipped with a
DECTRIS Pilatus CdTe 300K pixel detector and an INCOATEC AgK$_\alpha$
microfocus sealed-tube x-ray source ($\lambda =$~0.56087~\AA).

High-pressure low-temperature x-ray diffraction studies of \Co\ single
crystals up to a pressure of 5.5~GPa were carried out using a Diacell
Tozer-type DAC\cite{graf_nonmetallic_2011,boehler_new_2004} and Daphne
7575 as pressure-transmitting medium.\cite{murata_development_2016}
Ruby spheres inside the pressure chamber allowed a pressure
determination at room temperature \textit{via} the ruby fluorescence
method.\cite{piermarini_calibration_1975,Dewaele08,Kantor} Sample
cooling to temperatures above 20~K was achieved utilizing an ARS
closed-cycle helium cryocooler with exchangeable vacuum and radiation
shields surrounding the Tozer-type DAC. The temperature-dependence of
selected reflection intensities at various applied pressures was
tracked with a stainless steel vacuum chamber featuring kapton
windows. To collect x-ray diffraction data for structure
determinations at pressures of 0~GPa and 4~GPa and temperatures of
approx. 40~K and 110~K the stainless steel vacuum chamber was replaced
by a beryllium vacuum dome providing a larger accessible reciprocal
space fraction.

A similar experimental setup featuring the closed-cycle helium
cryocooler and an outer and inner beryllium vacuum and radiation
shield was used to obtain single-crystal x-ray diffraction data at
ambient pressure and sample temperatures of 11~K, 70~K and 100~K.

For high-pressure x-ray diffraction measurements on a \Co\
single-crystal up to 10.1~GPa at room temperature a
Boehler-plate-type DAC\cite{boehler_new_2006,boehler_new_2004}
was employed. The filling procedure and pressure determination method
were analogous to the experiments with the Tozer-type DAC described
above, but with a 4:1 volume mixture of methanol and ethanol
\cite{piermarini_hydrostatic_1973} as pressure-transmitting medium.

Obtained x-ray diffraction intensities were evaluated using the EVAL14
suite of programs\cite{Duisenberg92,Duisenberg03} and subjected to
scaling and absorption correction using the programs
SADABS/TWINABS.\cite{Krause15} More information on the handling of
parasitic scattering and shadowing of the x-ray beam by high-pressure
or low-temperature equipment is available in the Supplemental
Material.\cite{Note1} Structural refinements were performed with the
program JANA2006.\cite{Petricek14}

Density Functional Theory (DFT) calculations on the HT phase of \Co\
were performed employing the VASP
code.\cite{kresse_efficiency_1996,kresse_efficient_1996,kresse_ab_1994,
  kresse_ab_1993} The PBE density functional,\cite{perdew96, perdew97}
an energy cutoff for the plane wave basis set of 500~eV and a
Brillouin grid sampling of 4$\times$4$\times$2 were used
throughout. The starting geometry for the ambient pressure HT
structure was adopted from the optimizations performed in
Ref.\citenum{Langmann20}, which are based on the same set of
parameters.

Pressure-dependent geometry relaxations were performed at pressures of
2, 4, 6, 8 and 10~GPa. Optimizations were stopped when forces were
smaller than 0.001~eV/\AA. Single-point SCF calculations enforcing
uniaxial strain were performed by reducing the $a$, $b$ and $c$
lattice parameters of the relaxed HT ambient pressure structure by
$\pm0.02$~\AA\ and $\pm0.04$~\AA.

All phonon dispersion calculations employing the finite displacement
approach in a 2$\times$2$\times$2 supercell were performed with the
PHONOPY code\cite{Togo15} and VASP as force calculator using the same
parameters as specified above.

\section{Results and Discussion}\label{sec:results}

\begin{figure*}[htb]
  \centering
  \includegraphics[width=1.0\textwidth]{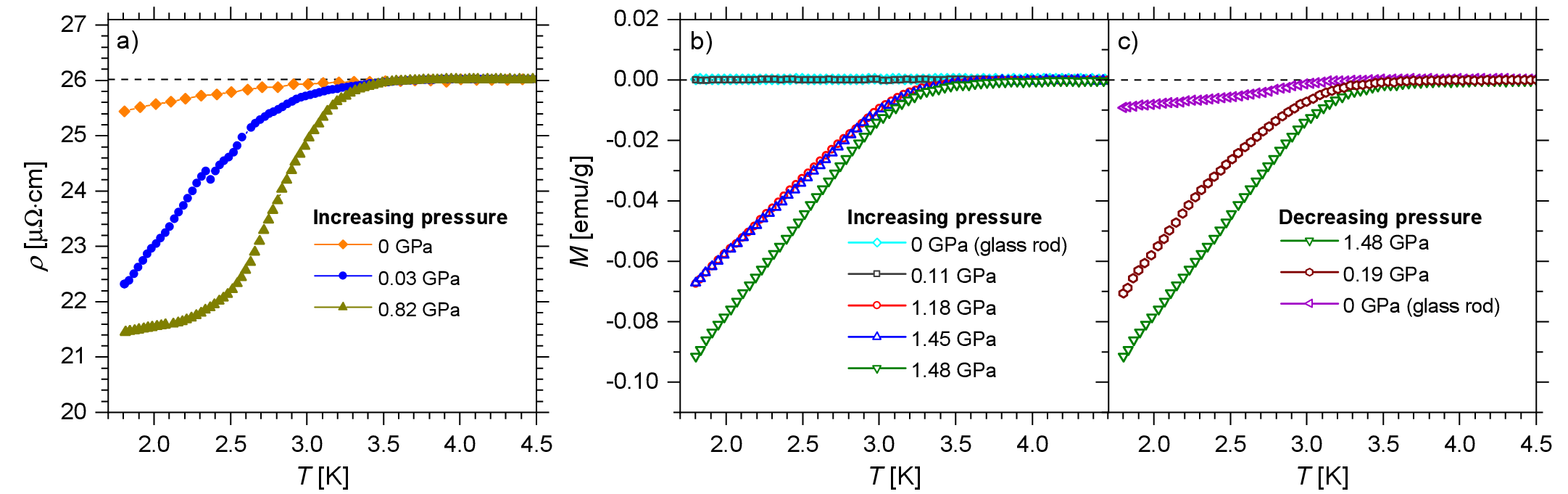}
  \caption{Temperature- and pressure-dependent development of (a)~the
    electrical resistivity \rhoT\ and (b, c)~the magnetization \MT\
    after zero-field-cooling in a magnetic field of 5~Oe for the
    superconducting transition of \Co. Data points in (a) and (b) were
    recorded while increasing the pressure and data points in (c)
    while decreasing the pressure. Note that the ambient-pressure
    measurements of \rhoT\ (a) and \MT\ (b, c) were performed without
    a pressure cell.  For better comparability, data points were
    brought to overlap at 4.5~K by applying shifts along the
    $\rho$-/$M$-axis.}
  \label{fig:mag-rho-p}
\end{figure*}

Starting point of our study are the results published earlier by Wang
\textit{et al.}\cite{Wang16} These authors found a significant
increase in the superconducting volume fraction of polycrystalline
\Co\ samples under the application of modest hydrostatic pressures.
In the present study, we performed physical property measurements and
x-ray diffraction experiments on single-crystalline samples.  This
allows us to explore potential structure-property relationships in
\Co\ and gain deeper insight into the origins of pressure-enhanced
superconductivity in the low-dimensional material. Also for
single-crystalline samples a clear superconducting signature is only
observed in the electrical resistivity \rhoT\ (see
Fig.~\ref{fig:mag-rho-p}a) and the magnetization \MT\
(Fig.~\ref{fig:mag-rho-p}b and Fig.~\ref{fig:mag-rho-p}c) after
application of pressure. It is noteworthy that the enhanced
superconducting signal persists for several hours after decreasing the
pressure from 1.48~GPa to 0.19~GPa. It remains remanently present even
after removing the sample from the pressure cell (see
Fig.~\ref{fig:mag-rho-p}c). This hints to a potential hysteretical
behavior of the inherent structural changes induced by the
application of pressure. Degradation of the sample quality as a
possible origin of this behavior could be excluded by means of x-ray
diffraction before and after performing a high-pressure experiment at
4.5~GPa and 27~K (see Supplemental Material\cite{Note1}).

In other low-dimensional compounds like the transition-metal
dichalcogenides $1T$-TiSe$_2$, $2H$-TaSe$_2$ and $2H$-NbSe$_2$ the
pressure-induced emergence of superconductivity is intimately linked
to the suppression of a periodic structural distortion at low
temperatures, \textit{i.e.}  a commensurate or incommensurate
charge-density wave.\cite{Brown, Joe14, Wakabayashi78, Moncton75,
  Moncton77, Kusmartseva09, Freitas16, Berthier76, Feng12} We
therefore tried to clarify, whether the Peierls-type distortion
leading to the low-temperature (LT) phase\cite{Eickerling13,
  Langmann20} might be suppressed upon application of pressure to
enhance the superconductivity in \Co.\cite{Wang16} The structural
properties of the ambient-pressure low-temperature phase have been
studied earlier\cite{Eickerling13} and provide the starting point of
this pressure- and temperature-dependent study. Atom displacements and
bond lengths mentioned hereafter were determined in an
ambient-pressure x-ray diffraction experiment on a high-quality
single-crystalline needle of \Co\ at 11~K (see experimental section
and Supporting Material\cite{Note1} for further details). All bond
lengths and displacements in this work are given with their threefold
standard deviation, while crystallographic directions are always
specified with respect to the axes of the orthorhombic HT phase (space
group $Immm$).

\begin{figure*}[htb]
  \centering
  \includegraphics[width=0.95\textwidth]{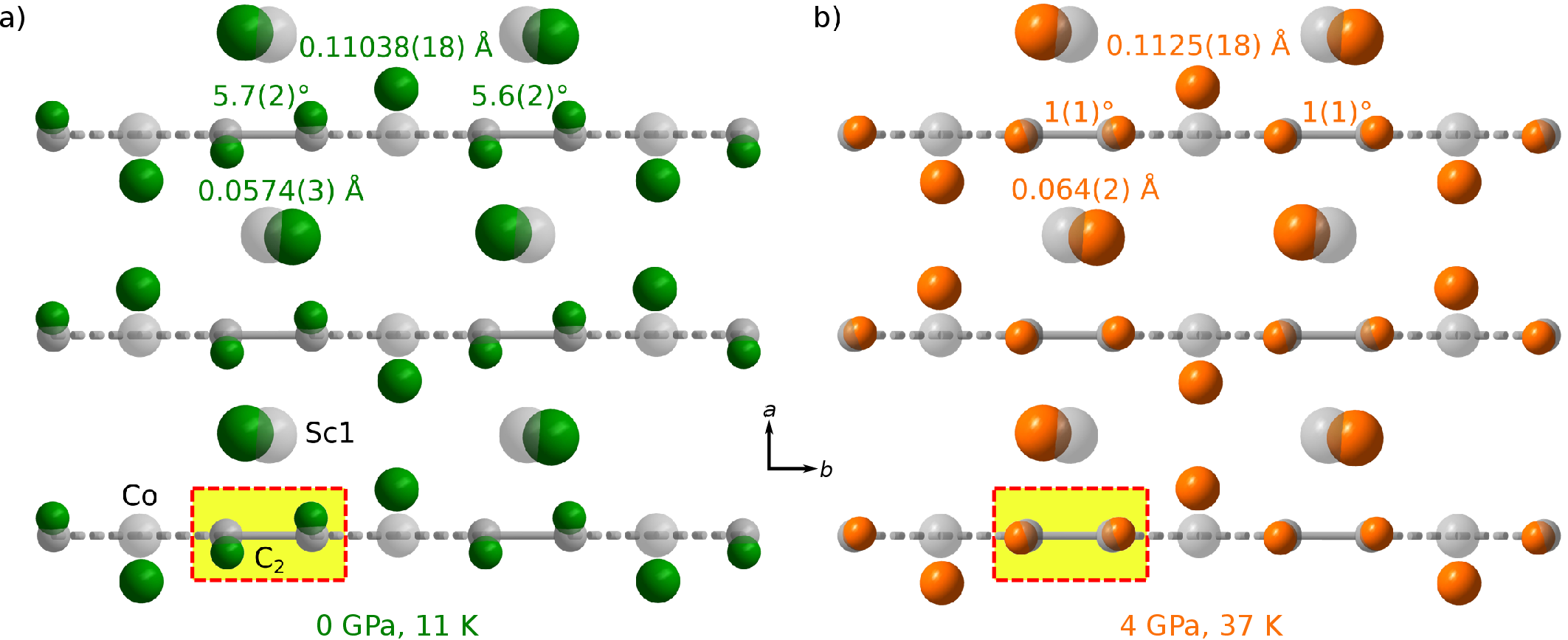}
  \caption{Overlays of the refined atomic positions within a layered
    building unit of \Co\ at room temperature (gray, semi-transparent;
    atomic positions from Ref.~\citenum{Eickerling13}) and after
    cooling to low temperatures (a)~without or (b)~with applied
    pressure (colored, non-transparent). All atom displacements are
    exaggerated seven-fold, Sc2 and Sc3 atoms have been omitted for
    clarity.  Specified values of atom displacements and rotation
    angles are given with their threefold standard deviation.}
  \label{fig:atom-shifts}
\end{figure*}

The LT phase of \Co\ (space group $C2/m$) is characterized by
modulated displacements of Co, Sc1 and C atoms from their HT phase
positions in the quasi-2D layers of the \Co\ structure (see
Fig.~\ref{fig:atom-shifts}a).  Precisely, the cobalt atoms along a
[Co(C$_2$)$_2$]$_\infty$ ribbon experience shifts of
$\pm$0.11038(18)~\AA\ relative to their crystallographic $2d$ site in
the HT phase (information on the calculation of the atom displacements
is provided in the Supplemental Material\cite{Note1}). Hence, Co--Co
distances within chains of cobalt atoms along the $a$ axis display
alternating larger (3.5985(9)~\AA) and smaller (3.1569(6)~\AA) values
compared to the constant separation of 3.3948(12)~\AA\ in the HT
phase.\cite{Eickerling13} This modulation of the Co atomic positions
is complemented by a modulation of the Sc1 atomic positions.  Their
displacements of $\pm$0.0574(3)~\AA\ with regard to the $2b$ HT
positions point along the $b$ axis and alternate along the $a$ axis,
\textit{i.e.}  their displacement direction is perpendicular to the
modulation of the Co atomic positions. As can be seen in
Fig.~\ref{fig:atom-shifts}a, the modulation of the Co and Sc1 atoms is
correlated in such a way that the Sc1 atoms are shifted towards long
Co--Co contacts and evade short Co--Co contacts.  In analogy to the
arrangement of the cobalt atoms, this displacement pattern turns
chains of equispaced scandium atoms along the $b$ axis
(4.3748(12)~\AA) above and below the [Co(C$_2$)$_2$]$_\infty$ ribbons
into chains with alternating longer (4.5015(12)~\AA) and shorter
(4.2718(12)~\AA) Sc1--Sc1 distances.

As a consequence of the HT$\rightarrow$LT transition the
[Co(C$_2$)$_2$]$_\infty$ ribbons are no longer planar, which is also
reflected by rotations of the C$_2$ units about rotation axes parallel
to $c$. Due to the lack of a crystallographic $m$ plane perpendicular
to $a$ rotations of adjacent C$_2$ units about the $b$ axis in the
same direction (conrotatory) or opposite directions (disrotatory) are
both allowed by symmetry. The potential importance of the carbon atoms
for superconductivity in \Co\ can be derived from isotopic
substitution experiments: replacement of $^{12}$C by $^{13}$C leads to
a systematic suppression of the superconducting onset temperature
\Tconset\ with an isotope coefficient $\alpha$ of 0.58.\cite{Haas17}
This observation is in line with the predictions of a Density
Functional Theory (DFT) study by Zhang \textit{et al.}\cite{Zhang12}
The authors proposed that rotations of the C$_2$ units and cobalt and
scandium atom displacements are integral parts of key phonon modes
coupling conduction electrons into superconducting Cooper pairs.

\begin{figure*}[htb]
  \centering
  \includegraphics[width=0.95\textwidth]{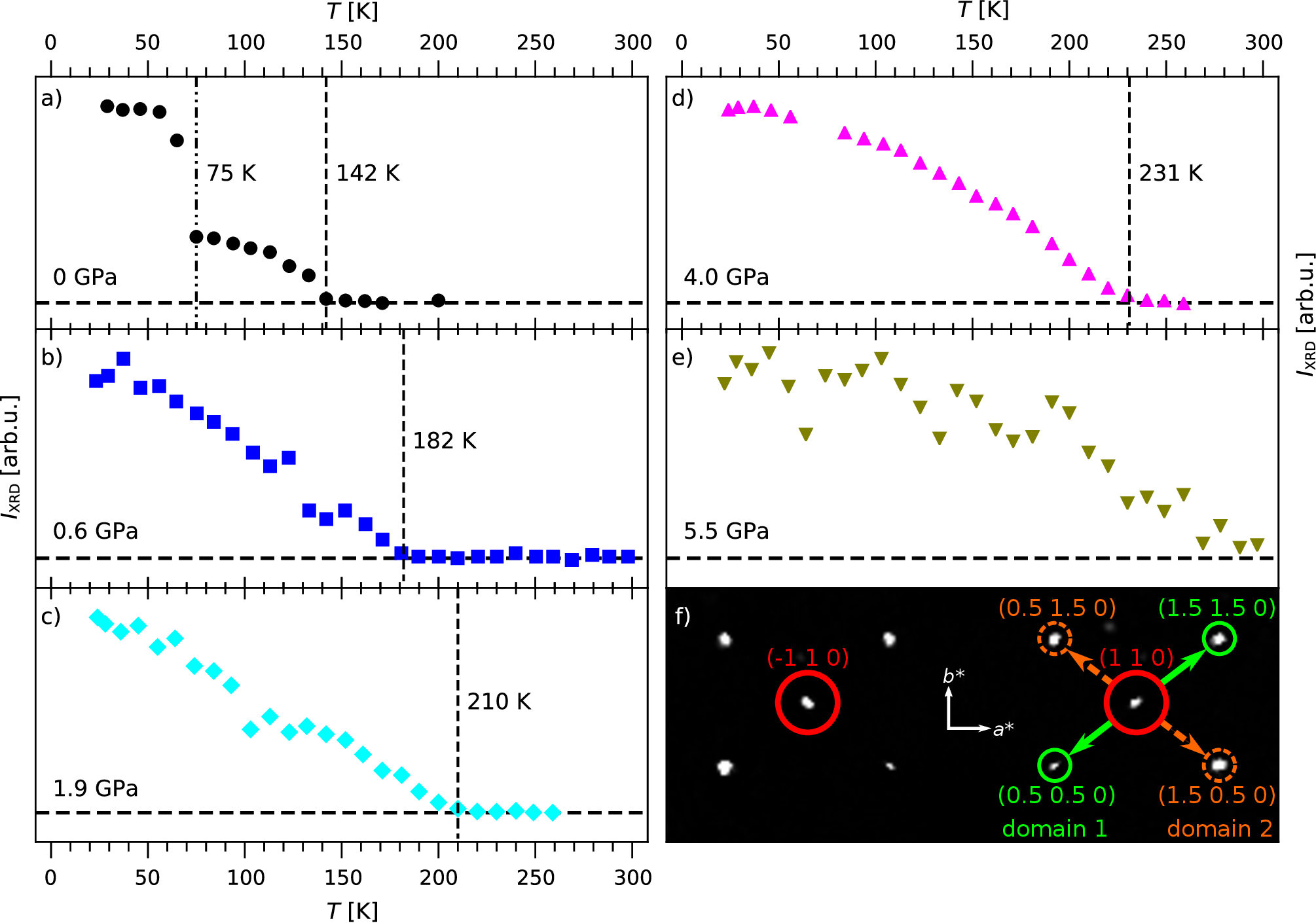}
  \caption{Temperature-dependence of the scattered intensity \ixrd\ at
    the positions of representative superstructure reflections for
    pressures between 0~GPa (a) and 5.5~GPa (e). A section of the
    $(hk0)$ plane (f) illustrates the location of the superstructure
    reflections for twin domains~1 and 2 with respect to the main
    reflections.}
  \label{fig:t-p-dep-I-superstruc}
\end{figure*}

Yet, rotations of the C$_2$ units are experimentally more difficult to
assess by x-ray diffraction than shifts of the heavy atoms cobalt and
scandium. The latter displacements invariably lead to the appearance
of prominent superstructure reflections with
$k = (+\frac{1}{2}, +\frac{1}{2}, 0)$ and also with
$k' = \left(+\frac{1}{2}, -\frac{1}{2}, 0\right)$ due to systematic
pseudo-merohedric twinning\footnote{This twinning process is due to a
  $t2$ step followed by an $i2$ step in the symmetry reduction
  (\textit{translationengleiche} and isomorphic groub-subgroup
  relationship, respectively) from the space-group $Immm$ of the HT
  phase structure to the space-group $C2/m$ of the LT phase structure,
  see Refs.~\citenum{Eickerling13} and \citenum{Vogt09}.} (see
Fig.~\ref{fig:t-p-dep-I-superstruc}f).\cite{Eickerling13,Vogt09,Langmann20}
By contrast, carbon atom displacements may only contribute to the
superstructure reflections in the case of disrotatory displacements of
neighboring C$_2$ units (a more detailed discussion can be found in
the Supplemental Material\cite{Note1}). Conrotatory displacements make
no contribution to the intensities of the superstructure reflections,
and only a minor contribution to the intensities of the main
reflections, \textit{i.e.} even a hypothetical rotation by 15$^\circ$
changes the average main reflection intensity by less than 1~$\%$. To
obtain precise intensity information for main and superstructure
reflections we therefore employed long exposure times, a
high-brilliance microfocus x-ray source and a noise-reduced pixel
detector with high dynamic range for the single-crystal x-ray
diffraction experiments discussed in the following (see the
experimental section and the Supporting Material\cite{Note1} for more
details).

Our structural model at 11~K and ambient pressure is characterized by
rotations of the two symmetry-independent C$_2$ units in the same
direction with rotation angles of 5.6(2)$^\circ$ and 5.7(2)$^\circ$
(highlighted in Fig.~\ref{fig:atom-shifts}a). Similar carbon atom
shifts resulting in somewhat smaller but still conrotatory rotation
angles of 2.8(4)$^\circ$ and 3.0(4)$^\circ$ have been found earlier at
9~K.\cite{Eickerling13} Notably, the observed conrotatory
displacements of subsequent C$_2$ units along the
[Co(C$_2$)$_2$]$_\infty$ ribbons result in the formation of shorter
(2.098(4)~\AA) and longer (2.113(3)~\AA) Co-C distances in contrast to
an alternative scenario with disrotatory displacements which would
minimize all Co--C distances. This rules out that strengthening of
Co--C bonds provides the only driving force of the carbon atom shifts
in the LT phase structure.

\begin{figure}[htb]
  \centering
  \includegraphics[width=0.49\textwidth]{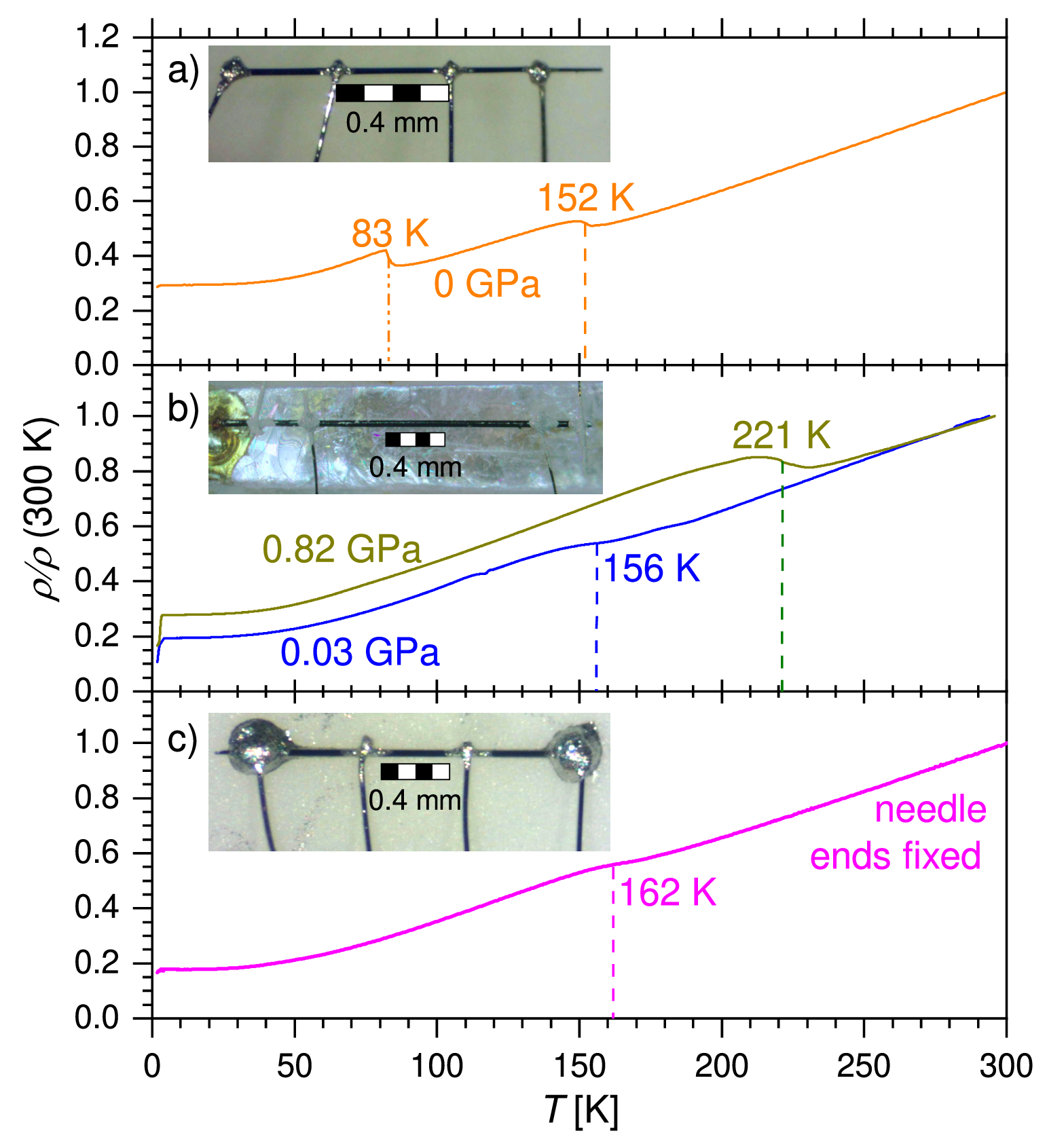}
  \caption{Temperature-dependent electrical resistivity \rhoT\ of \Co\
    single-crystals (a)~at ambient pressure and without fixation to a
    substrate, (b)~at hydrostatic pressures of 0.03~GPa and 0.82~GPa
    and (c)~glued at both ends on top of a sapphire chip.  Insets:
    Photographic images of the respective samples after their
    preparation for measurements (a) to (c).}
  \label{fig:res-p}
\end{figure}

In the next steps of our analysis we will aim at establishing possible
pressure-dependent structure-property relationships to gain further
insight into the potential origins of the characteristic
superconducting behavior of \Co. Measurements of the electrial
resistivity \rhoT\ already hint towards potential pressure-induced
modifications of our reference LT phase structure: As outlined
earlier,\cite{Langmann20} the extended phonon softening process
leading to a static Peierls-type structural distortion of \Co\ at
ambient pressure is delimited by two pronounced anomalies in \rhoT\ at
152~K and 83~K (Fig.~\ref{fig:res-p}a).  Our high-pressure experiments
on single-crystalline samples in accordance with Wang \textit{et
  al.}\cite{Wang16} show that only a single broad anomaly at 156~K
can be observed in \rhoT\ after application of 0.03~GPa (see blue
curve in Fig.~\ref{fig:res-p}b).  This anomaly further shifts towards
higher temperatures with increasing pressure (dark yellow curve in
Fig.~\ref{fig:res-p}b). A similar result, \textit{i.e.} a suppression
of the first anomaly at lower temperature in \rhoT\ and an upward
shift of the second one by approx. 10~K, can be obtained by fixing a
single-crystalline \Co\ needle at two points along its long axis
(\textit{i.e.} parallel to the crystallographic $a$ axis) on top of a
sapphire chip (see Fig.~\ref{fig:res-p}c).

In order to investigate the origin of these pressure-dependent changes
in \rhoT\ we performed x-ray diffraction experiments at variable
pressure and temperature. Inspection of Bragg intensities as well as
atomic positions from structural refinements should reveal, whether
\textit{(i)}~only the distortion pattern during the HT$\rightarrow$LT phase
transition changes under pressure, \textit{(ii)}~the structurally
distorted LT phase is suppressed in favor of the undistorted HT
phase,\cite{Wang16} or \textit{(iii)}~a designated and structurally
distinct high-pressure LT phase of \Co\ is formed.

\begin{figure}[htb]
  \centering
  \includegraphics[width=0.49\textwidth]{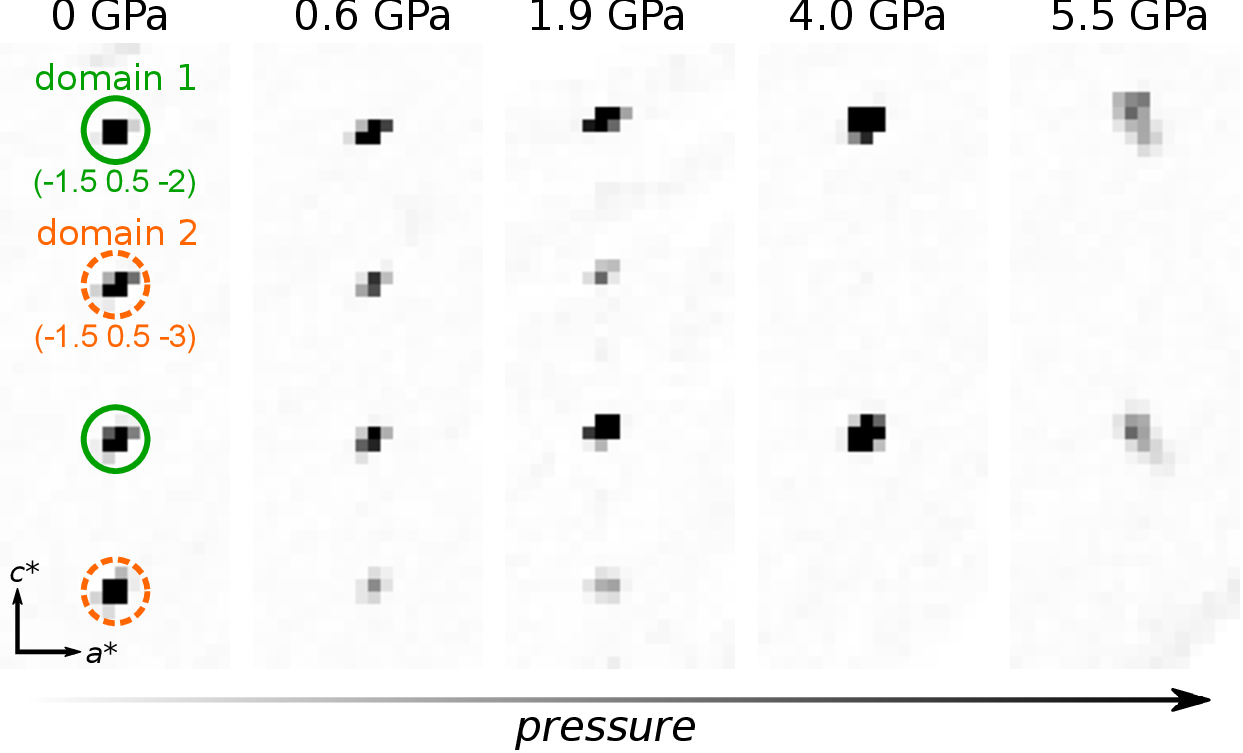}
  \caption{Pressure-induced detwinning of a \Co\ sample at $T =$~22~K,
    as observed by x-ray diffraction in case of characteristic Bragg
    reflections from two twin domains (twin law [[-1~0~0], [0~1~0],
    [0~0~1]], \textit{i.e.} an $m$ plane perpendicular to the $a$ axis
    of the orthorhombic HT unit cell). The displayed sections of the
    $(h, 0.5, l)$ plane contain only superstructure reflections and
    were recorded after applying pressures up to 5.5~GPa and cooling
    to 22~K. The decreased scattering intensity at 5.5~GPa is due to a
    beginning deterioration of the sample crystallinity.}
  \label{fig:detwinning}
\end{figure}

In support of hypothesis \textit{(i)}, our measurements indicate that
the low-temperature superstructure Bragg reflections with
$k = (+\frac{1}{2}, +\frac{1}{2}, 0)$ can be observed up to pressures
of 5.5~GPa. No indication for a pressure-induced structural phase
transition connected with a change of the space group can be found.
Instead, the collected diffraction data can still be described by a
monoclinic lattice and complies with space group $C2/m$.  Differences
between pressurized and unpressurized samples, however, may be
recognized by comparing reconstructions of the $(h,0.5,l)$ reciprocal
space plane at 22~K for pressures between 0~GPa and 5.5~GPa
(Fig.~\ref{fig:detwinning}). Note that at ambient pressure \Co\
crystals in the LT phase are systematically twinned with each of the
two differently oriented twin domains contributing half of the
superstructure reflections to the $(h,0.5,l)$ plane (green and orange
circles in Fig.~\ref{fig:detwinning}). Cooling crystals below the
HT$\rightarrow$LT phase transition temperature at elevated pressures
reveals that one half of the superstructure reflections shows
increasing intensities, whereas the other half shows decreasing
intensities ($p \leq$ 1.9~GPa). The latter superstructure reflections
finally vanish completely for $p >$ 1.9~GPa, thus indicating a
pressure-induced detwinning of the crystal.\footnote{We note that a
  preference for one of the twin domains emerges well below the
  quasi-hydrostatic limit of the employed pressure transmitting medium
  (Daphne 7575) at 3.9~GPa~- 4~GPa ($T =$~298~K, see
  Ref.~\citenum{murata_development_2016}) in contrast to the usually
  required uni-directional pressures in other reported examples of
  detwinning (see Refs.~\citenum{Grube99}, \citenum{Gugenberger94},
  and \citenum{Niesen13}). Solidification of the pressure medium upon
  cooling does not induce significant non-hydrostaticity, as
  demonstrated in Ref.~\citenum{Stasko20}}

Taking into account the otherwise unchanged characteristics of the LT
phase in reciprocal space we may connect the strongly modified
behavior of the anomalies in \rhoT\ under pressure (compare
Fig.~\ref{fig:res-p}a and Fig.~\ref{fig:res-p}b) to changes along the
pathway from the HT to the LT phase. Therefore, we determined the
temperature- and pressure-dependence of the scattered intensity \ixrd\
at the positions of representative superstructure reflections (see
Fig.~\ref{fig:t-p-dep-I-superstruc}). Consistent with \rhoT, the
increase of \ixrd\ with decreasing temperature, which occurs in two
steps at ambient pressure (Fig.~\ref{fig:t-p-dep-I-superstruc}a),
renders into a continuous increase under pressure
(Fig.~\ref{fig:t-p-dep-I-superstruc}b to
Fig.~\ref{fig:t-p-dep-I-superstruc}e). We note, however, that a
quantitative comparison of applied pressure values in \rhoT\ and \MT\
studies with those of the x-ray diffraction experiments is hampered by
the differing sample environments and pressure determination methods
employed (see experimental section and Supplemental
Material\cite{Note1}).

From our data we may conclude that the LT phase is stabilized
substantially, as is indicated by a shift of the onset temperature of
the superstructure reflection intensities from 142~K at 0~GPa
(Fig.~\ref{fig:t-p-dep-I-superstruc}a) to 231~K at 4~GPa
(Fig.~\ref{fig:t-p-dep-I-superstruc}d). A pressure of 5.5~GPa
preserves the superstructure reflections up to room temperature
(Fig.~\ref{fig:t-p-dep-I-superstruc}e, see also the Supplemental
Material\cite{Note1}), although at the cost of a degradation of the
sample crystallinity. Interestingly, the isoelectronic and
isostructural transition metal carbides Sc$_3$IrC$_4$ and
Sc$_3$RhC$_4$ also show a periodically distorted structure in analogy
to the LT phase of \Co\ at room temperature but without prior cooling
or pressure application.\cite{Zhang07,Vogt05} There are, however,
neither hints to superconductivity nor to the existence of an
undistorted high-temperature phase comparable to \Co\ for these highly
related compounds. In particular, systematic twinning as indicator of
a potential $t_2$ HT$\rightarrow$LT transition has not been observed
in the iridium and rhodium congeners of \Co.\cite{Vogt05} This
discrepancy may be due to the fact that the $3d$ metal cobalt is
characterized by a significantly smaller covalent radius and weaker
transition metal-carbon bonds in comparison with the $4d$ and $5d$
group members rhodium and iridium. The resulting higher structural
flexibility of \Co\ could thus be a prerequisite to allow the
existence of both, a HT and a LT phase structure.

\begin{figure*}[htb]
  \centering
  \includegraphics[width=0.89\textwidth]{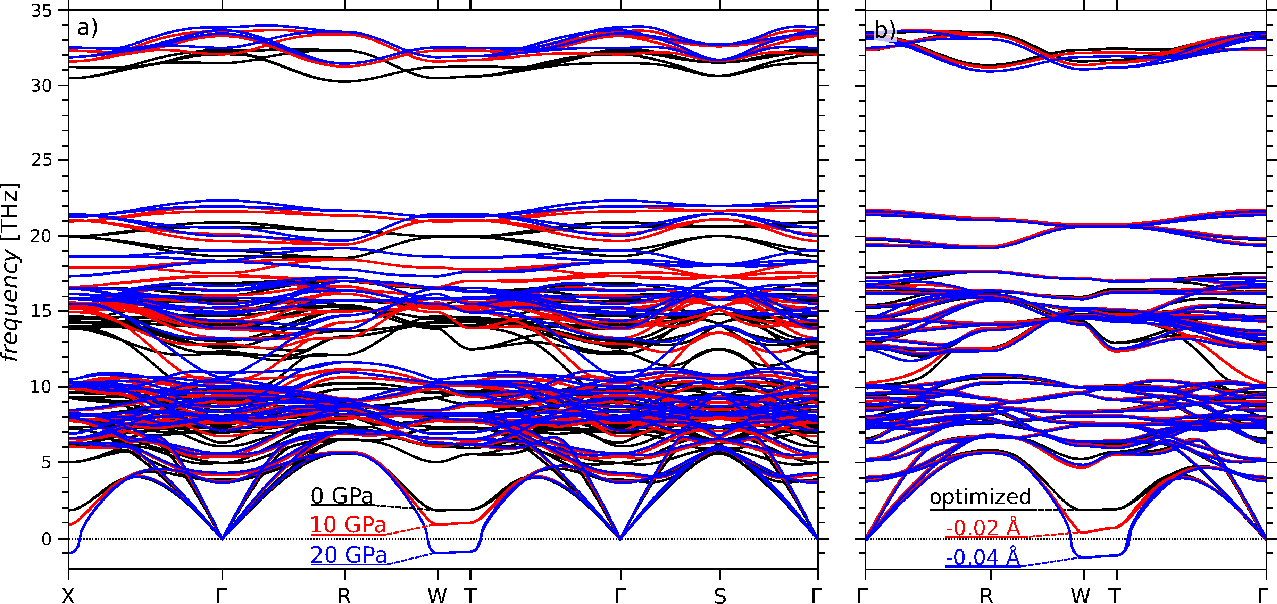}
  \caption{Response of the phonon dispersion of HT \Co\ (DFT study) to
    (a)~hydrostatic pressure, and (b)~to a reduction of the lattice
    parameter $a$.}
  \label{fig:phonon-disp-pressure}
\end{figure*}

\begin{figure*}[htb]
  \centering
  \includegraphics[width=0.95\textwidth]{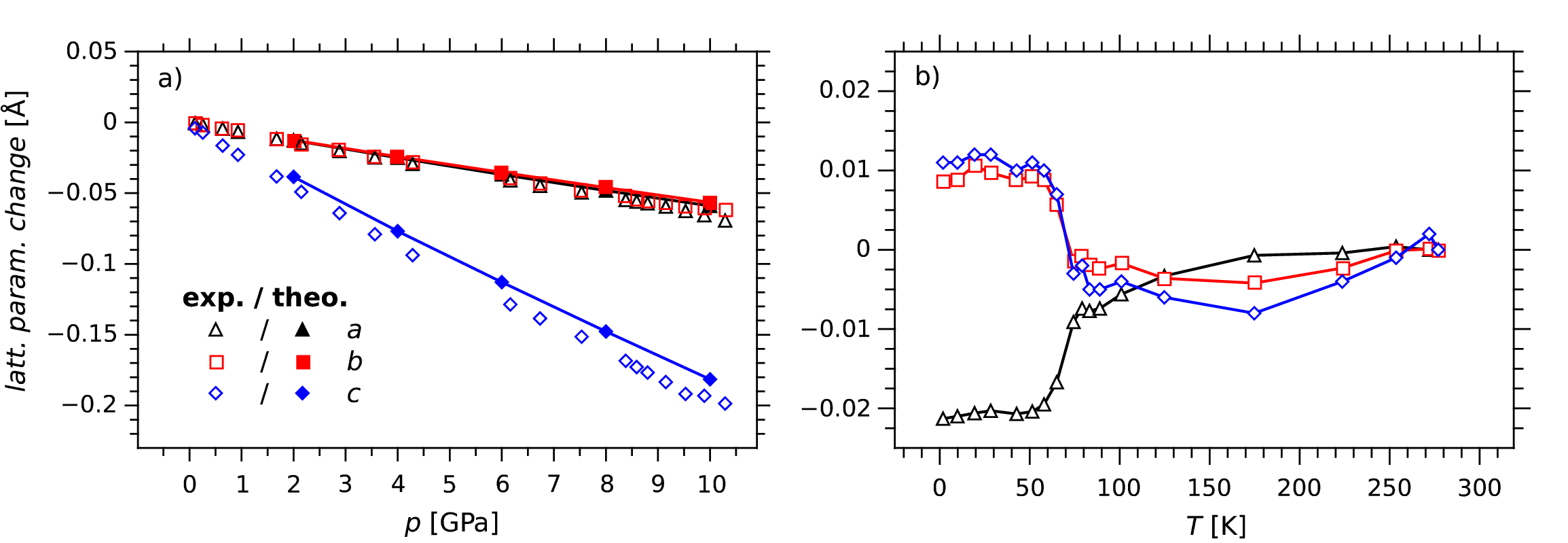}
  \caption{Changes of the experimental (empty symbols) and theoretical
    (filled symbols) lattice parameters of \Co\ (a)~with varying
    pressure at constant temperature and (b)~with varying temperature at
    ambient pressure (data from Ref.~\citenum{Eickerling13}).}
  \label{fig:lattice-params}
\end{figure*}

The occurrence of a subtle competition between the HT and LT phase in
\Co\ is reflected in the extended phonon softening regime preceding
the HT$\rightarrow$LT transition at ambient-pressure.\cite{Langmann20}
Also its signatures in the temperature dependencies of \rhoT\
(Fig.~\ref{fig:res-p}a) and \ixrd\
(Fig.~\ref{fig:t-p-dep-I-superstruc}a) react sensitively and in a
highly related way already to small changes in pressure. Application
of a pressure below 0.6~GPa is sufficient to induce a cross-over of
both physical properties from a course with two anomalies limiting the
phonon softening regime towards low and high
temperatures\cite{Langmann20} to a course with a single anomaly (see
Fig.~\ref{fig:res-p}b and Fig.~\ref{fig:t-p-dep-I-superstruc}b to
Fig.~\ref{fig:t-p-dep-I-superstruc}e). Unfortunately, x-ray scattering
and absorption by the employed pressure cell did not permit the
investigation of the very weak and diffuse x-ray scattering features
in analogy to Ref.~\citenum{Langmann20}. Pressure-dependent \textit{ab
  initio} phonon dispersion calculations for the HT phase structure,
however, can provide more insight into the underlying causes for the
modification of \rhoT\ and \ixrd\ under pressure.
Fig.~\ref{fig:phonon-disp-pressure}a illustrates the presence of a
soft branch between the high-symmetry points W and T of the phonon
dispersion already at ambient-pressure conditions. The phonon mode
along the branch is characterized by displacements of the cobalt and
scandium atoms (Sc1) in the $ab$ plane anticipating their
displacements in the LT phase of \Co\ (see
Fig.~\ref{fig:atom-shifts}a).\cite{Langmann20} On progressing along
the path from T to W the LT-phase-like pattern of atomic displacements
at T is modified by modulations of decreasing wave length along the
$c$ axis.  Yet, carbon atom contributions to the mode in analogy to
the displacements shown in Fig.~\ref{fig:atom-shifts}a are
absent. These can be found in a separate medium-frequency phonon mode
at $\Gamma$ with still unclear behavior upon cooling below the
HT$\rightarrow$LT phase transition temperature. So far, there is only
evidence for a profound temperature-dependence of the W--T phonon
branch.  Approaching the HT$\rightarrow$LT transition temperature from
above results in a reduction of the phonon frequency at T to
zero.\cite{Langmann20} The same W--T phonon branch also displays an
extraordinary sensitivity to hydrostatic pressure (see
Fig.~\ref{fig:phonon-disp-pressure}a and the Supplemental
Material\cite{Note1}).  Its frequency is subjected to a strong and
steady red shift with increasing pressure, while the frequencies of
all other phonon branches show the expected blue shift. These trends
indicate a gradual destabilization of the HT phase structure with
increasing pressure in line with the experimentally observed
pressure-induced enhancement of the transition temperature from the HT
to the LT phase.\footnote{The fact that the calculations do not
  predict a preferential phonon frequency reduction at T may be
  related to shortcomings of the standard GGA functional employed
  within this study to properly resolve the very flat energy surface
  region separating the HT and LT phase of \Co.}

Despite the red frequency shift of the W--T phonon branch in our
calculations (Fig.~\ref{fig:phonon-disp-pressure}a) and the
preservation of the low-temperature superstructure reflections upon
heating to room temperature at 5.5~GPa in our diffraction experiments
(Fig.~\ref{fig:t-p-dep-I-superstruc}e), application of pressure alone
does not suffice to induce a transition of \Co\ from the HT to the LT
phase structure. No superstructure reflections could be observed in
single-crystal XRD experiments up to the destruction of the sample at
10.1~GPa, when the pressure cell was kept constantly at room
temperature (see the Supplemental Material\cite{Note1}). An overlay of
structural models at 0.2~GPa and 4.2~GPa in the Supplemental
Material\cite{Note1} illustrates that the pressure-induced shifts of
the atomic positions remain negligible under these conditions.
Likewise, no phase transition could be inferred from the
pressure-dependence of the lattice parameters obtained from
room-temperature powder XRD experiments
(Fig.~\ref{fig:lattice-params}a). The experimentally observed linear
decrease of the lattice parameters upon application of up to 10.1~GPa
with a stronger absolute compression of $c$ ($\Delta c =$ 0.20~\AA)
compared to $a$ and $b$ ($\Delta a =$ 0.07~\AA, $\Delta b =$ 0.06~\AA)
is very well reproduced by DFT studies of the compressibility behavior
of the HT phase (Fig.~\ref{fig:lattice-params}a).

This behavior might be related to the strongly differing development
of the lattice parameters of \Co\ either upon cooling or upon
increasing hydrostatic pressure. Previous variable-temperature neutron
diffraction studies between 277~K and 1.8~K showed that a reduction of
temperature is accompanied by increases of the $b$ and $c$ parameters
by approx. 0.01~\AA, and a decrease in the $a$ parameter by
approx. 0.02~\AA\ (Fig.~\ref{fig:lattice-params}b).\cite{Eickerling13}
An increase of hydrostatic pressure, by contrast, results in the
compression of all lattice parameters. Thus, a strongly anisotropic
pressure response of \Co\ may be suspected in accordance with the
low-dimensional structure of the compound. The validity of this
hypothesis is underlined by the fact that the application of uniaxial
stress along the long axis of single-crystalline \Co\ needles leads to
pronounced changes in the temperature-dependent electrical resistivity
(see Fig.~\ref{fig:res-p}c).  We therefore performed a DFT study
probing the response of the phonon dispersion to a uniaxial
compression along each of the three unit cell axes by varying the HT
phase lattice parameters independently (refer to the Supplemental
Material\cite{Note1} for further details). The strongest frequency
reduction along the W--T phonon branch is obtained by a reduction of
the $a$ parameter correlating with the distance between adjacent
[Co(C$_2$)$_2$]$_\infty$ ribbons (see red and blue curves in
Fig.~\ref{fig:phonon-disp-pressure}b). Negative frequencies along the
path indicate the instability of the HT phase structure after a
compression of the $a$ lattice parameter by more than
approx. 0.02~\AA. A similar dispersion behavior is only achieved by
the application of hydrostatic pressure in the range of 20~GPa.

After pointing out the destabilization of the HT phase under pressure
we will now focus on the pressure-induced structural changes to the LT
phase. Although the space group $C2/m$ applies to the LT phase
structure under ambient-pressure and high-pressure conditions, some
degrees of freedom for the atom arrangement remain. We find
for example that the distance between adjacent
[Co(C$_2$)$_2$]$_\infty$ ribbons is reduced from 3.378(7)~\AA\ at
0~GPa to 3.34(2)~\AA\ at 4~GPa. An even smaller compression from
6.0105(3)~\AA\ to 5.9849(8)~\AA\ is found for the interlayer distance
between adjacent quasi-2D Sc1-Co-C layers. Further free parameters in
the HT$\rightarrow$LT phase transition involve the magnitude of the cobalt and
scandium atom shifts, and the extent and relative orientation of the
C$_2$ unit rotations. Fig.~\ref{fig:atom-shifts}a and
Fig.~\ref{fig:atom-shifts}b visualize the effect on the atom
positions, when \Co\ is cooled to $T <$ 40~K with and without an
applied pressure of 4~GPa, respectively. In
Fig.~\ref{fig:atom-shifts}a the atomic positions at 11~K and ambient
pressure are marked by green non-transparent spheres, while
Fig.~\ref{fig:atom-shifts}b shows the atomic positions after applying
a pressure of 4~GPa and cooling to 37~K (orange non-transparent
spheres).

It becomes evident that the general displacement pattern of the cobalt
and scandium atoms in the LT phase remains rather invariant upon
pressure application: At both 0~GPa and 4~GPa, the cobalt atoms along
the [Co(C$_2$)$_2$]$_\infty$ ribbons are shifted in positive and
negative $a$ direction by similar extents of 0.11038(18)~\AA\ and
0.1125(18)~\AA, respectively. Also the shifts of the Sc1 atoms along
the $b$ axis are remarkably similar with values of 0.0574(3)~\AA\ at
0~GPa and 0.064(2)~\AA\ at 4~GPa and 37~K. By contrast, the rotation
angles of the C$_2$ units obtained from the structural refinements
display a distinct pressure dependency (highlighted in
Fig.~\ref{fig:atom-shifts}a and Fig.~\ref{fig:atom-shifts}b): At 0~GPa
and 11~K, the C$_2$ units are subjected to significant rotations out
of their HT phase positions with rotation angles between 5.6(2)\degr\
and 5.7(2)\degr. Nearly vanishing rotation angles of 1(1)\degr\ are,
however, observed at 4~GPa and 37~K (see the Supplemental
Material\cite{Note1} for further details). This observation might be
linked to the pressure-induced detwinning of \Co\ samples: The two
possible twin domains in the ambient-pressure LT phase mainly differ
by a different rotation sense of the C$_2$ units in an otherwise
nearly unchanged arrangement of cobalt and scandium atoms. A vanishing
C$_2$ rotation angle makes these twin domains nearly identical leaving
only marginal differences in the cobalt and scandium atom positions
(further information in the Supplemental Material\cite{Note1}).  As a
result, the realization of a single-domain state extending over the
entire sample volume might be favored. The absence of further
anomalies in the electrical resistivity between 37~K and the
superconducting transition temperature at 4.5~K
(Fig.~\ref{fig:res-p}b) implies that this detwinned high-pressure
low-temperature phase is the one hosting superconductivity in \Co.

\begin{figure}[htb]
  \centering
  \includegraphics[width=0.49\textwidth]{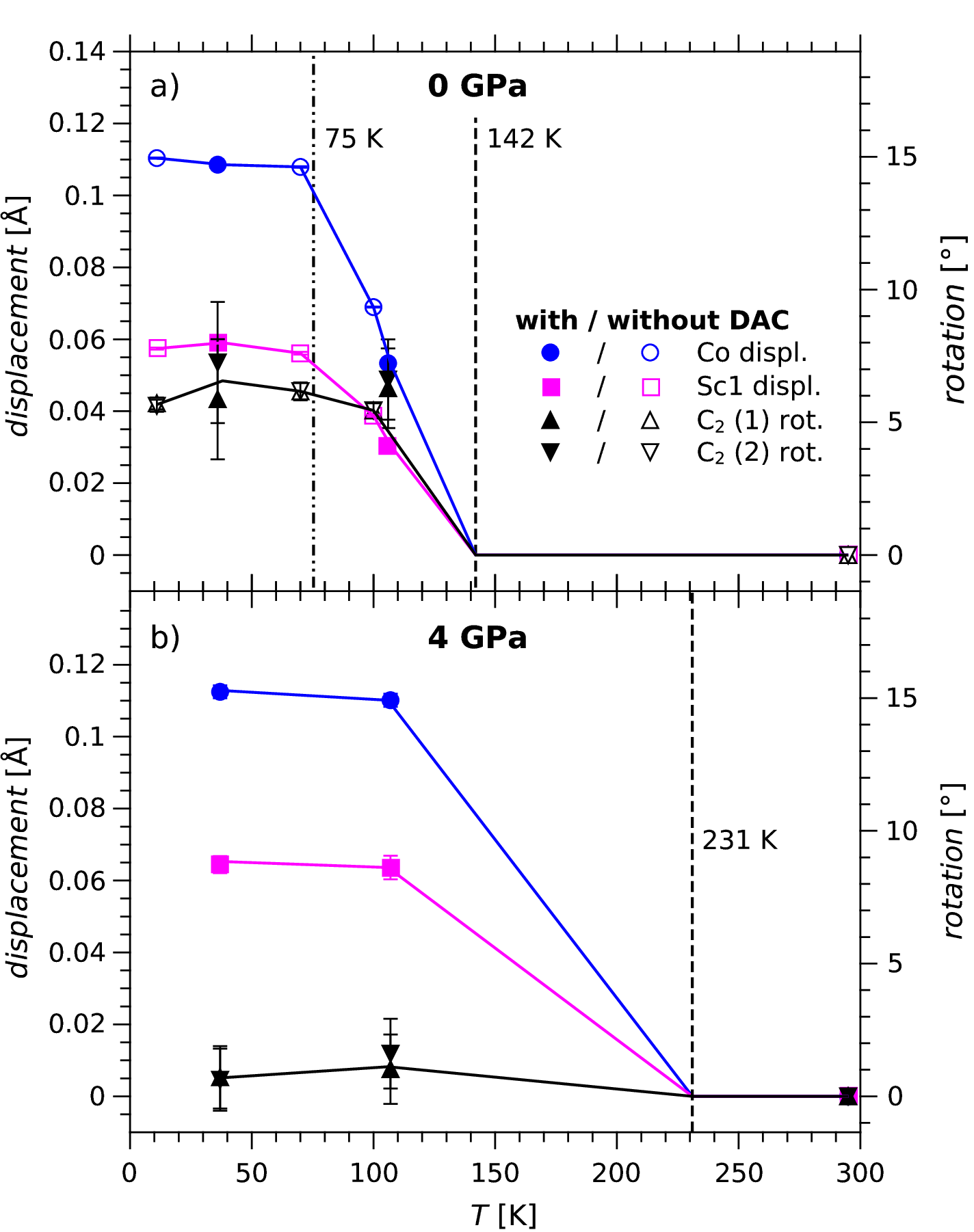}
  \caption{Temperature-dependence of the displacements of the Co (blue
    circles) and Sc1 atoms (magenta squares) and the rotation angles
    of the two symmetry-inequivalent C$_2$-units (C$_2$~(1) and
    C$_2$~(2), black triangles) for (a)~0~GPa and (b)~4~GPa. Note that
    error bars refer to the threefold standard deviation and that the
    carbon atom displacements corresponding to each C$_2$ rotation
    angle (right ordinate) are specified on the left
    ordinate. Critical temperatures which mark the onset of the
    formation of superstructure reflections were taken from
    Fig.~\ref{fig:t-p-dep-I-superstruc}a and
    Fig.~\ref{fig:t-p-dep-I-superstruc}d. In case of the 0~GPa study
    also the jump-like increase of the superstructure reflection
    intensity connected to the completion of the Peierls-type
    transition at 75~K was marked by an additional dotted line. Solid
    lines serve as guides to the eye.}
  \label{fig:atom-shifts-overview}
\end{figure}

In a last step we proceed to link the observed temperature- and
pressure-dependent changes in the superstructure reflection intensity
\ixrd\ (see Fig.~\ref{fig:t-p-dep-I-superstruc}) to changes observed
in the atomic positions. To do so we performed x-ray diffraction
experiments at 0~GPa and 4~GPa for selected temperatures above and
below the step-like increase of \ixrd\ observed at approx. 80~K and
0~GPa. Fig.~\ref{fig:atom-shifts-overview} gives an overview of the
observed temperature-dependence of the Co (blue circles) and Sc1 atom
displacements (magenta rectangles) and the C$_2$-unit rotations (black
triangles) at 0~GPa (a) and 4~GPa (b). Data points from x-ray
diffraction experiments with and without usage of a DAC are indicated
by filled and open symbols, respectively.

It turns out that the structural models obtained at ambient pressure
and above 75~K represent intermediate steps in the progression of \Co\
towards its final state below 75~K (see
Fig.~\ref{fig:atom-shifts-overview}a). Consistent with the non-zero
value of the superstructure reflection intensity in the same
temperature- and pressure-regime
(Fig.~\ref{fig:t-p-dep-I-superstruc}a) shifts of the atoms from their
positions in the HT phase structure are already present at this stage
(Fig.~\ref{fig:atom-shifts-overview}a).  Thereby, larger values of
\ixrd\ correspond to larger shifts of the atomic positions.  It should
be emphasized, however, that the phonon softening process during the
HT$\rightarrow$LT transition is not yet completed for temperatures above 75~K, so
that all atom displacements are likely to be of a correlated, but
still dynamical nature.\cite{Langmann20} When the atom displacements
become rather static on cooling below 75~K, a step-like increase and a
subsequent saturation of the superstructure reflection intensity is
observed.

As already stated above, direction and maximal displacements of the Co
and Sc1 atoms are not altered significantly by appyling a pressure of
4~GPa. Yet, the absence of a step in \ixrd\ and the higher onset
temperature of 231~K (0~GPa: 142~K) correlate with the attainment of
these maximum displacements at temperatures significantly above 75~K
(see Fig.~\ref{fig:atom-shifts-overview}b). The main structural
difference between 0 and 4~GPa can thus be attributed to the different
extent of the out-of-plane rotation of the C$_2$ units at all
investigated temperatures. Hence, the application of pressure
effectively prevents the displacement of the carbon atoms from their
HT phase positions, but does not suppress the formation of a periodic
structural distortion.

\section{Conclusion} \label{sec:conclusion}

To conlude, we suggest a connection between the occurence of volume
superconductivity and subtle structural modifications to the known
Peierls-type distorted low-temperature (LT) phase of
\Co\cite{Eickerling13,Langmann20}\cite{Wang16} under pressure.  We
demonstrated that the differences between the ambient- and
elevated-pressure LT phase structure are limited to a reduction of the
C$_2$ rotations out of the [Co(C$_2$)$_2$]$_\infty$ ribbon plane from
5.6$^{\circ}$ - 5.7$^{\circ}$ to nearly zero. This brings the C$_2$
moieties back to their high-temperature (HT) phase positions in an
otherwise still distorted arrangement of cobalt and scandium atoms.

On an atomistic level the changed equilibrium position of the C$_2$
units may affect phononic and electronic properties of the
electron-phonon coupling driven superconductor \Co.  The importance of
carbon atom vibrations for the emergence of superconductivity is
highlighted by $^{12}$C/$^{13}$C isotope substitution
experiments indicating a clearly non-zero isotope coefficient of
0.58.\cite{Haas17} A key role of C$_2$ librational modes in the
coupling of conduction electrons into Cooper pairs is also pointed out
by DFT studies employing the Eliashberg formalism.\cite{Zhang12} Thus,
establishing structure-property relationships in favor of
pressure-induced volume superconductivity presents an interesting,
but challenging task for future theoretical studies.

But the subtle changes in the C$_2$ rotation also affect the
properties of \Co\ on a macroscopic level: They render the two
possible twin domains in LT-\Co\ nearly indistinguishable and set the
stage to the realization of detwinned, single-domain crystals above
1.9~GPa. Thus, a continuous superconducting sample volume may only be
realized after the disappearance of twin domain walls from pressurized
\Co. Such a barrier function of twin domain walls for superconducting
currents has been investigated recently by Song \textit{et
  al.}\cite{Song12} for FeSe.

We finally note that our results point to the simultaneous existence
of volume superconductivity and a Peierls-type distorted phase at
elevated pressures. There seems to be no pressure-adjustable
competition between periodic structural distortion and
superconductivity like in many other structurally low-dimensional
materials.\cite{Berthier76, Cho18, Chang12, Chikina20, Freitas16,
  Oike18} Subtle pressure-induced modifications of the atom
arrangement in the distorted phase might already suffice to reconcile
both phenomena in \Co.

\begin{acknowledgments}
  We thank the group of Prof. Ch. Kuntscher for their continuous
  support in high-pressure studies.
\end{acknowledgments}

%

\end{document}



\title{Supplemental Material for: The structure of the superconducting
  high-pressure phase of \Co}


\author{Jan Langmann}
\author{Marcel Vöst}
\author{Dominik Schmitz}
\author{Christof Haas}
\affiliation{CPM,
  Institut f\"ur Physik, Universit\"at Augsburg, 
  D-86159 Augsburg, Germany}
\author{Georg Eickerling}
\email{georg.eickerling@uni-a.de}
\affiliation{CPM,
  Institut f\"ur Physik, Universit\"at Augsburg, 
  D-86159 Augsburg, Germany}
\author{Anton Jesche}
\affiliation{Experimentalphysik VI,
  Zentrum f\"ur Elektronische Korrelation und Magnetismus,
  Institut f\"ur Physik, Universit\"at Augsburg,
  D-86159 Augsburg, Germany}
\author{Michael Nicklas}
\affiliation{
  Max Planck Institute for Chemical Physics of Solids,
  N\"othnitzer Straße 40, D-01087 Dresden, Germany}
\author{Arianna Lanza}
\affiliation{
  Center for Nanotechnology Innovation@NEST, Istituto Italiano
  di Tecnologia, I-56127 Pisa, Italy}
\author{Nicola Casati}
\affiliation{Swiss Light Source,
  Paul Scherrer Institut, CH-5232 Villigen, Switzerland}
\author{Piero Macchi}
\affiliation{Dipartimento di Chimica,
  Materiali ed Ingegneria Chimica ``G. Natta'', Politecnico
  di Milano, I-20133 Milano, Italy}
\author{Wolfgang Scherer}
\email{wolfgang.scherer@uni-a.de}
\affiliation{CPM,
  Institut f\"ur Physik, Universit\"at Augsburg, 
  D-86159 Augsburg, Germany}


\date{\today}


\pacs{}

\maketitle


\tableofcontents

\newpage

\section{Synthesis of samples} \label{sec:synthesis}

All \Co\ single crystals in this work were grown from polycrystalline
samples. These were obtained by arc-melting pieces of the constituent
elements scandium (Smart Elements, 5N), cobalt (Cerametek Materials,
5N5) and carbon (Alfa Aesar, 5N5) in a purified argon atmosphere
(500~mbar). Homogeneity was ensured by re-melting the sample several
times (see also Refs.~\citenum{Rohrmoser07} and \citenum{Vogt09}).

\begin{figure}[htb]
  \centering
  \vspace{0.5cm}
  \includegraphics[height=5.4cm]{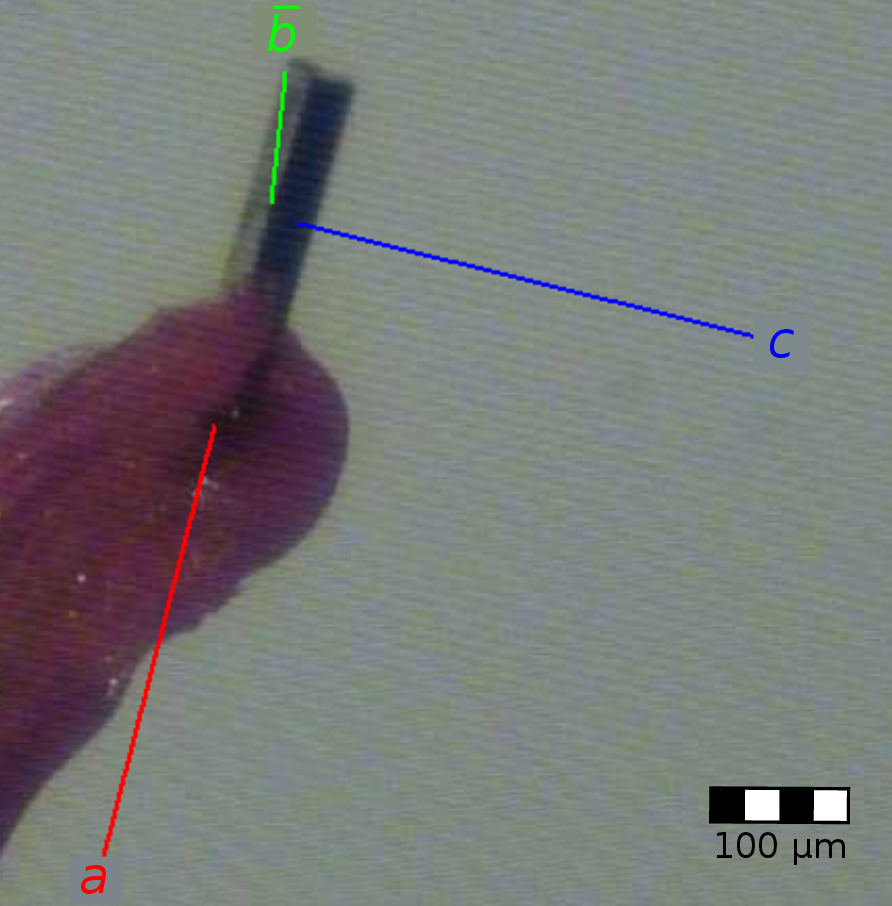}
  \caption{Photographic image of the \Co\ single crystal used in the
    ambient-pressure low-temperature x-ray diffraction experiments
    without pressure cell. Crystal axes $a$, $b$ and $c$ referring to
    the orthorhombic high-temperature phase unit cell are indicated by
    colored lines.}
  \label{fig:crystal-ambient-pressure}
\end{figure}

Single-crystalline needles of \Co\ (see
Fig.~\ref{fig:crystal-ambient-pressure}) develop spontaneously at the
surface of polycrystalline samples after rapid quenching from high
temperatures in the arc-melting furnace. Thereby, a large temperature
gradient from sample top to bottom is achieved using a water-cooled
copper crucible (see also Refs.~\citenum{He15} and \citenum{Haas19}).

\begin{figure}[htb]
  \centering
  \includegraphics[height=5.4cm]{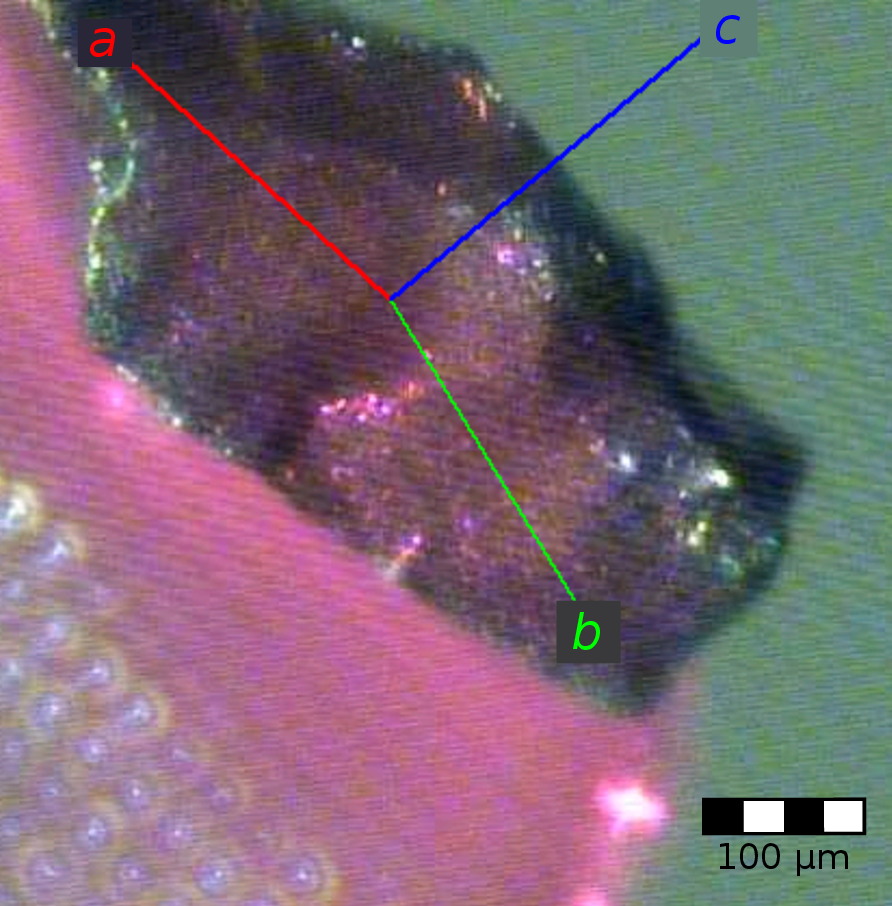}
  \caption{Photographic image of the single-crystalline \Co\ sample
    used in the high-pressure magnetization measurements. Crystal axes
    $a$, $b$ and $c$ referring to the orthorhombic high-tem\-pe\-ra\-ture
    phase unit cell are indicated by colored lines.}
  \label{fig:crystal-magnetization}
\end{figure}

\begin{figure}[htb]
  \centering
  \includegraphics[height=5.4cm]{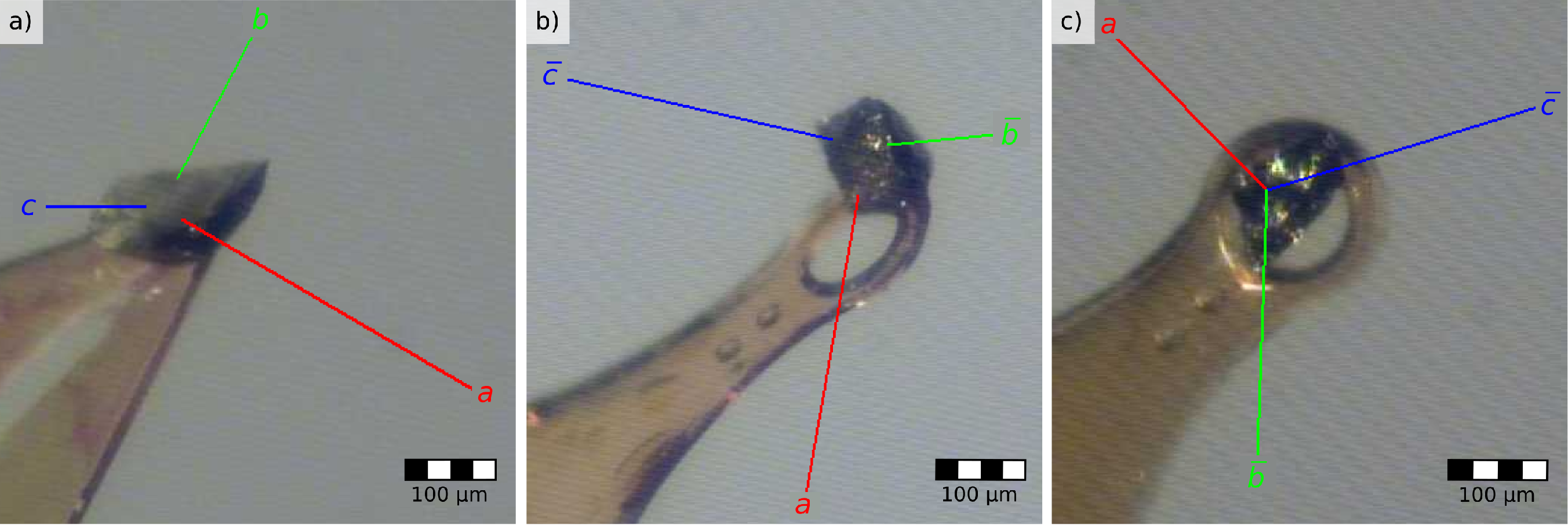}
  \caption{Photographic images of the \Co\ single crystals used in the
    high-pressure low-temperature x-ray diffraction experiments for
    (a)~tracking of superstructure reflection intensities,
    (b)~structure determinations, and (c)~tests for sample degradation
    under the high-pressure/low-temperature conditions. Crystal axes
    $a$, $b$ and $c$ referring to the orthorhombic high-temperature
    phase unit cell are indicated by colored lines.}
  \label{fig:crystals-tdac}
\end{figure}

\begin{figure}[htb]
  \centering
  \vspace{0.5cm}
  \includegraphics[height=5.4cm]{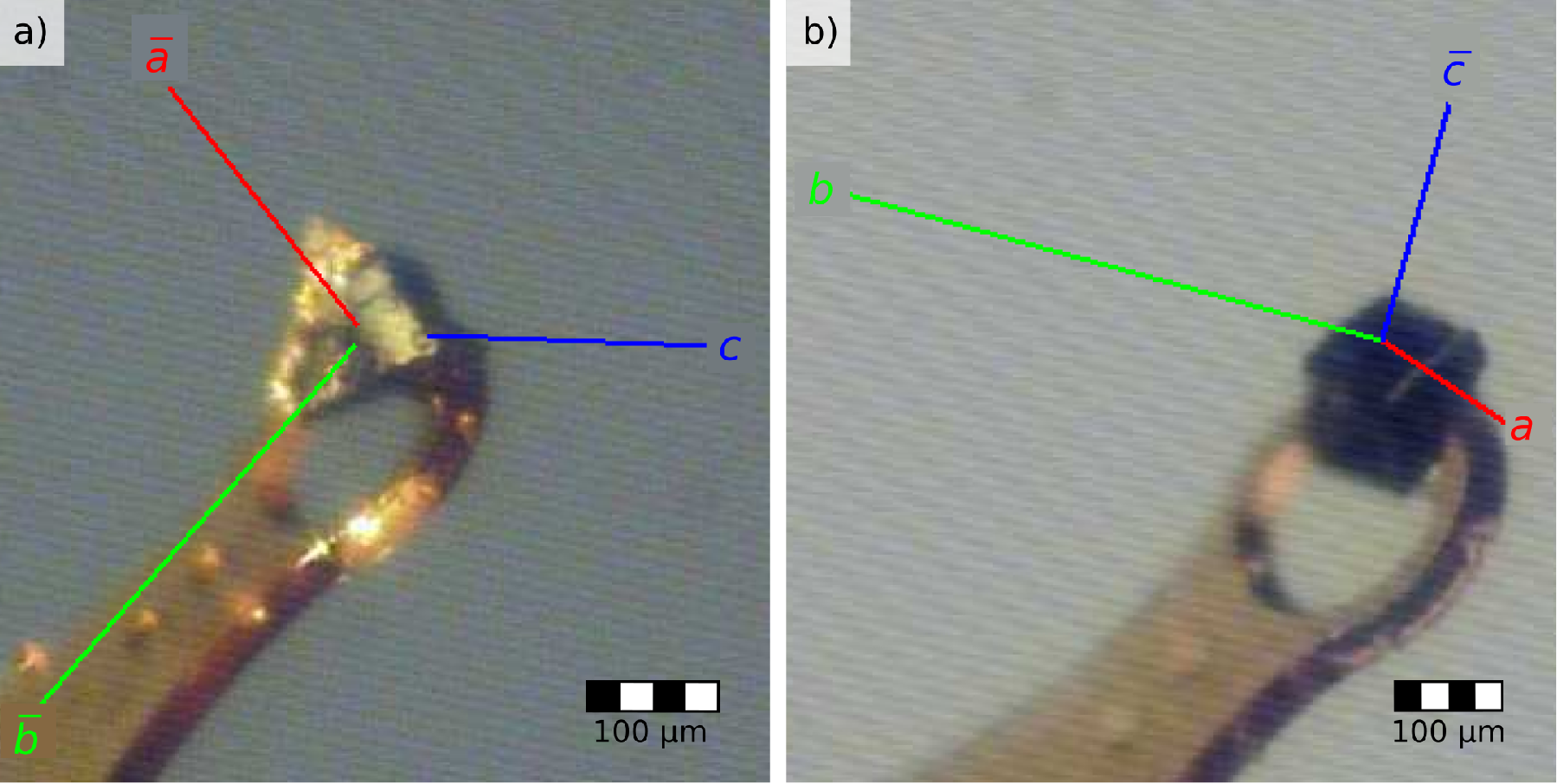}
  \caption{Photographic images of the \Co\ single crystals used in the
    high-pressure x-ray diffraction experiments at room temperature
    for (a)~structure determination, and (b)~reciprocal space
    mapping. Crystal axes $a$, $b$ and $c$ referring to the
    orthorhombic high-temperature phase unit cell are indicated by
    colored lines.}
  \label{fig:crystals-bdac}
\end{figure}

Larger plate-like single crystals (see
Fig.~\ref{fig:crystal-magnetization}, Fig.~\ref{fig:crystals-tdac} and
Fig.~\ref{fig:crystals-bdac}) were grown by crystallization from a
lithium flux.\cite{Jesche14} For this purpose, powder from a ground
polycrystalline sample of \Co\ and small lithium pieces (Alfa Aesar,
3N) were filled into a tantalum ampoule in a stoichiometric ratio of
1:20. The tantalum ampoule was closed by welding under argon
atmosphere (500 mbar), sealed into an evacuated quartz ampoule and
placed in a muffle furnace.  After heating to 600 $^\circ$C with a
heating rate of 20 $^\circ$C/h the temperature was held constant for 24
hours and then reduced to 500 $^\circ$C with a cooling rate of 100
$^\circ$C/h.  The heat treatment was terminated after two to four weeks
by quenching the samples in a water bath. Single crystals were
obtained by opening and turning the tantalum ampoule under inert
conditions.  Remaining flux medium was dissolved in ethanol (see also
Ref.~\citenum{Haas19}).

\newpage

\section{Investigated samples} \label{sec:samples}

\FloatBarrier
\hspace{1cm}

\newpage

\section{High-pressure magnetization measurements}
\label{sec:magnetization}

\begin{figure}[htb]
  \centering
  \includegraphics[width=0.64\textwidth]{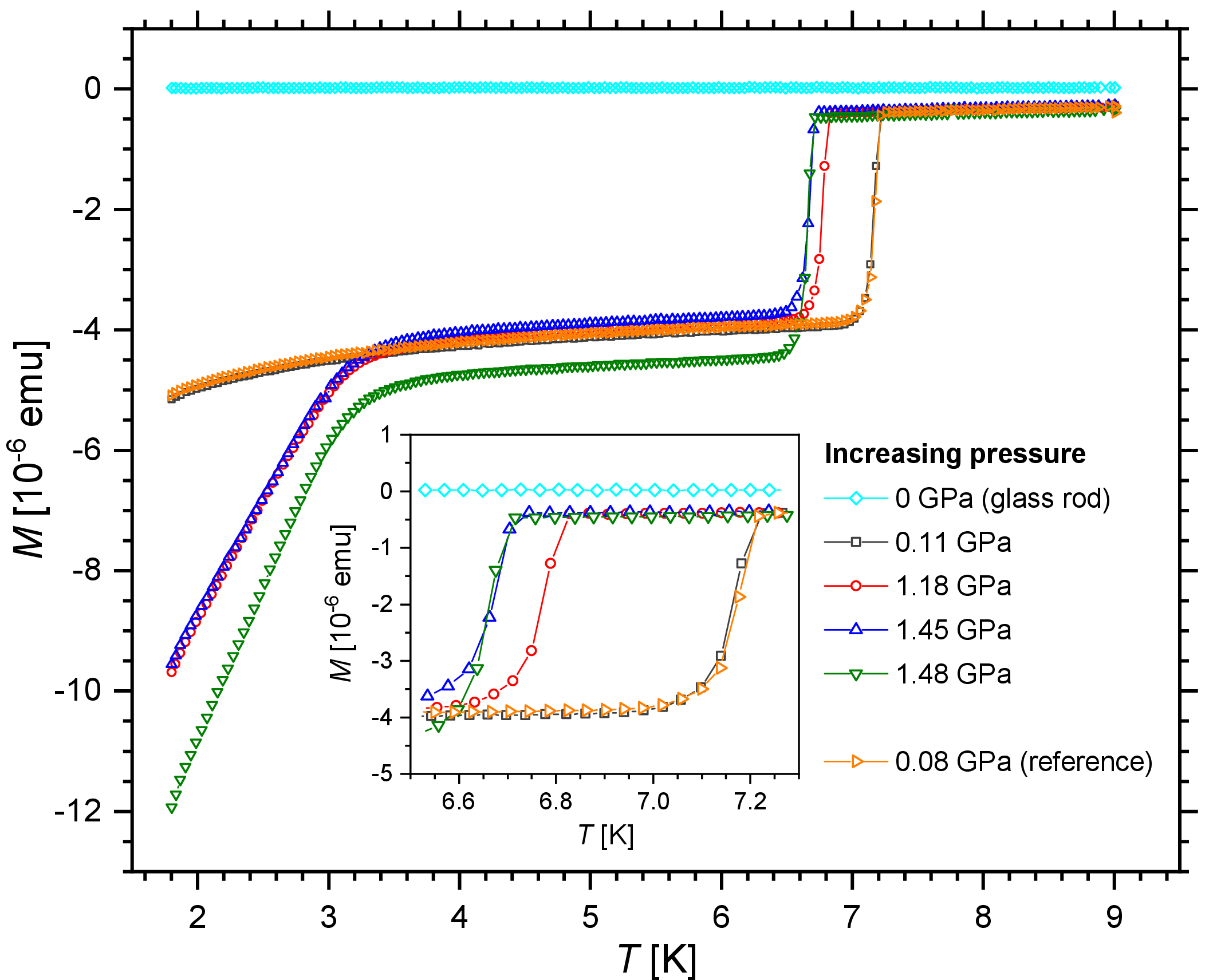}
  \caption{Unprocessed magnetization data \MT\ collected with
    increasing pressure. Data points of a reference measurement with
    only a lead piece and pressure-transmitting medium inside the
    pressure chamber are plotted with orange triangles.}
  \label{fig:magn-raw-data-incr}
\end{figure}

\begin{figure}[htb]
  \centering
  \includegraphics[width=0.64\textwidth]{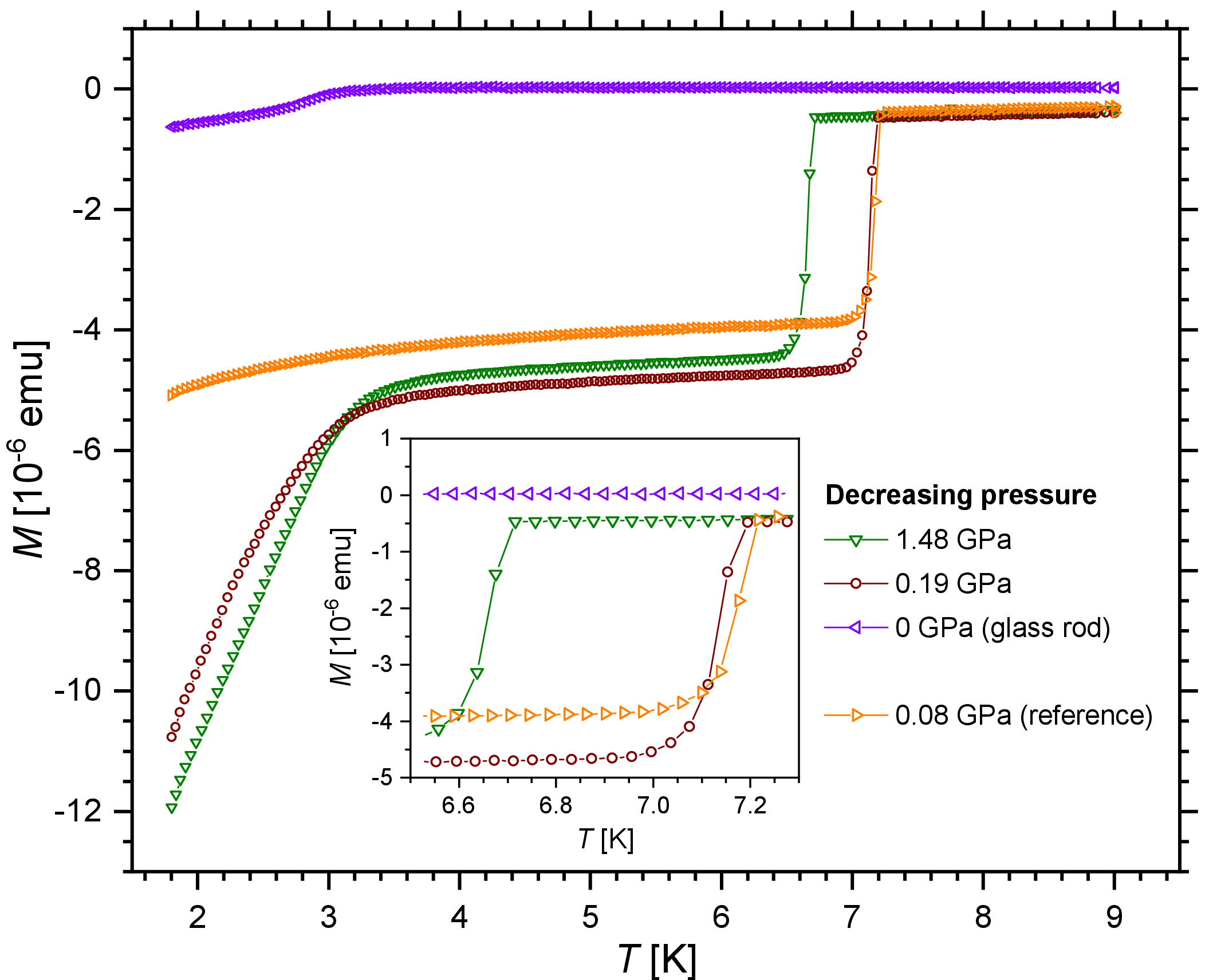}
  \caption{Unprocessed magnetization data \MT\ collected with
    decreasing pressure. Data points of a reference measurement with
    only a lead piece and pressure-transmitting medium inside the
    pressure chamber are plotted with orange triangles.}
  \label{fig:magn-raw-data-decr}
\end{figure}

\begin{figure}[htb]
  \centering
  \includegraphics[width=0.9\textwidth]{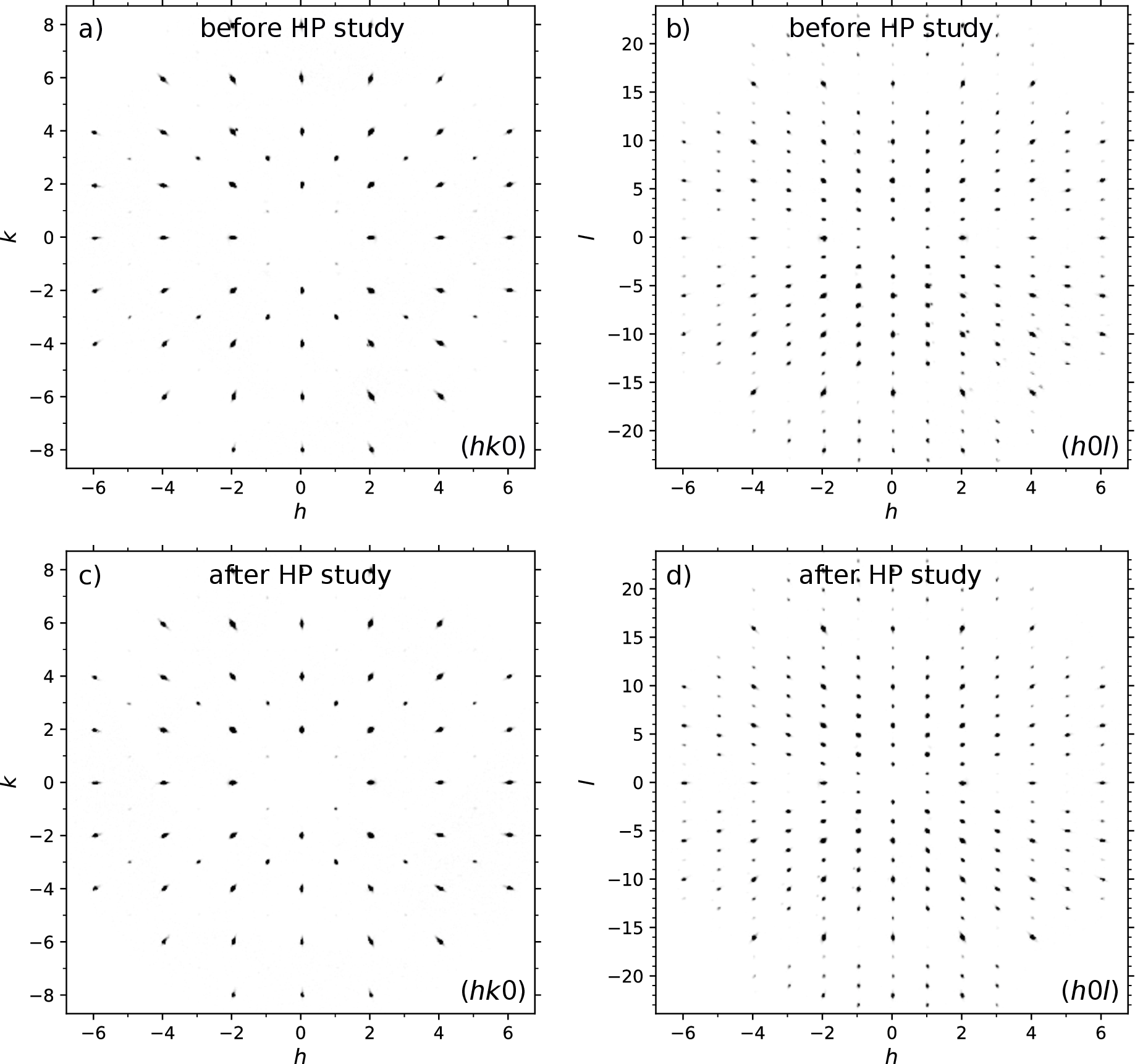}
  \caption{Reconstructions of reciprocal-space planes $(hk0)$
    and $(h0l)$ from room-temperature x-ray diffraction data
    collected for a \Co\ single crystal before (a, b) and after
    performing a high-pressure magnetization study on it
    (c, d).}
  \label{fig:rec-space-before-after-mag}
\end{figure}

For performing high-pressure magnetization measurements a miniature
Ceramic Anvil Cell
(mCAC)\cite{tateiwa_high_2014,tateiwa_note_2013,tateiwa_magnetic_2012,
  tateiwa_miniature_2011,Kobayashi07} was used. The ceramic anvils had
a culet diameter of 1.8~mm and the prefabricated Cu:Be gasket had a
pressure chamber diameter of 900~$\mu$m and a thickness of
900~$\mu$m. Placing a piece of lead inside the pressure chamber
allowed to infer the internal pressure from its superconducting
transition temperature
$T_\mathrm{c}$.\cite{eiling_pressure_1981,bireckoven_diamond_1988} As
pressure transmitting medium served Daphne 7373 with a
quasi-hydrostatic limit of 2.2~GPa -
2.3~GPa.\cite{Yokogawa07,Murata08} After a background measurement of
the closed mCAC with a piece of lead (dimensions: 105~$\mu$m $\times$
255~$\mu$m $\times$ 300~$\mu$m) and Daphne 7373 inside its pressure
chamber (orange triangles in Fig.~\ref{fig:magn-raw-data-incr} and
Fig.~\ref{fig:magn-raw-data-decr}) the actual sample measurement with
an additional \Co\ single crystal (dimensions: 225~$\mu$m $\times$
270~$\mu$m $\times$ 600~$\mu$m; mass: 69.2~$\mu$g; photographic image
in Fig.~\ref{fig:crystal-magnetization}) at pressures of 0.11~GPa,
1.18~GPa, 1.45~GPa, 1.48~GPa and 0.19~GPa (pressure release
measurement) took place (see Fig.~\ref{fig:magn-raw-data-incr} and
Fig.~\ref{fig:magn-raw-data-decr} for an overview of the raw data
collected upon increasing and decreasing the applied pressure,
respectively). Reciprocal space planes reconstructed from x-ray
diffraction data collected before and after the high-pressure study
can be found in Fig.~\ref{fig:rec-space-before-after-mag}. As
comparative measurements serve magnetization measurements, where the
same \Co\ single crystal was fixed on a glass rod with GE Varnish
before (cyan diamonds in Fig.~\ref{fig:magn-raw-data-incr}) and after
the high-pressure study (violet triangles in
Fig.~\ref{fig:magn-raw-data-decr}). All magnetization measurements
were performed under zero-field-cooling conditions employing a MPMS3
superconducting quantum interference device (SQUID) magnetometer
(QUANTUM DESIGN). The measured temperature range was between 1.8~K and
9~K with a heating rate of 0.2~K/min at 5~Oe. For analyzing the data,
the background measurement was in each case subtracted.

\FloatBarrier

\newpage

\section{High-pressure electrical resistivity measurements}
\label{sec:resistivity}

\begin{figure}[htb]
  \centering
  \includegraphics[width=1.0\textwidth]{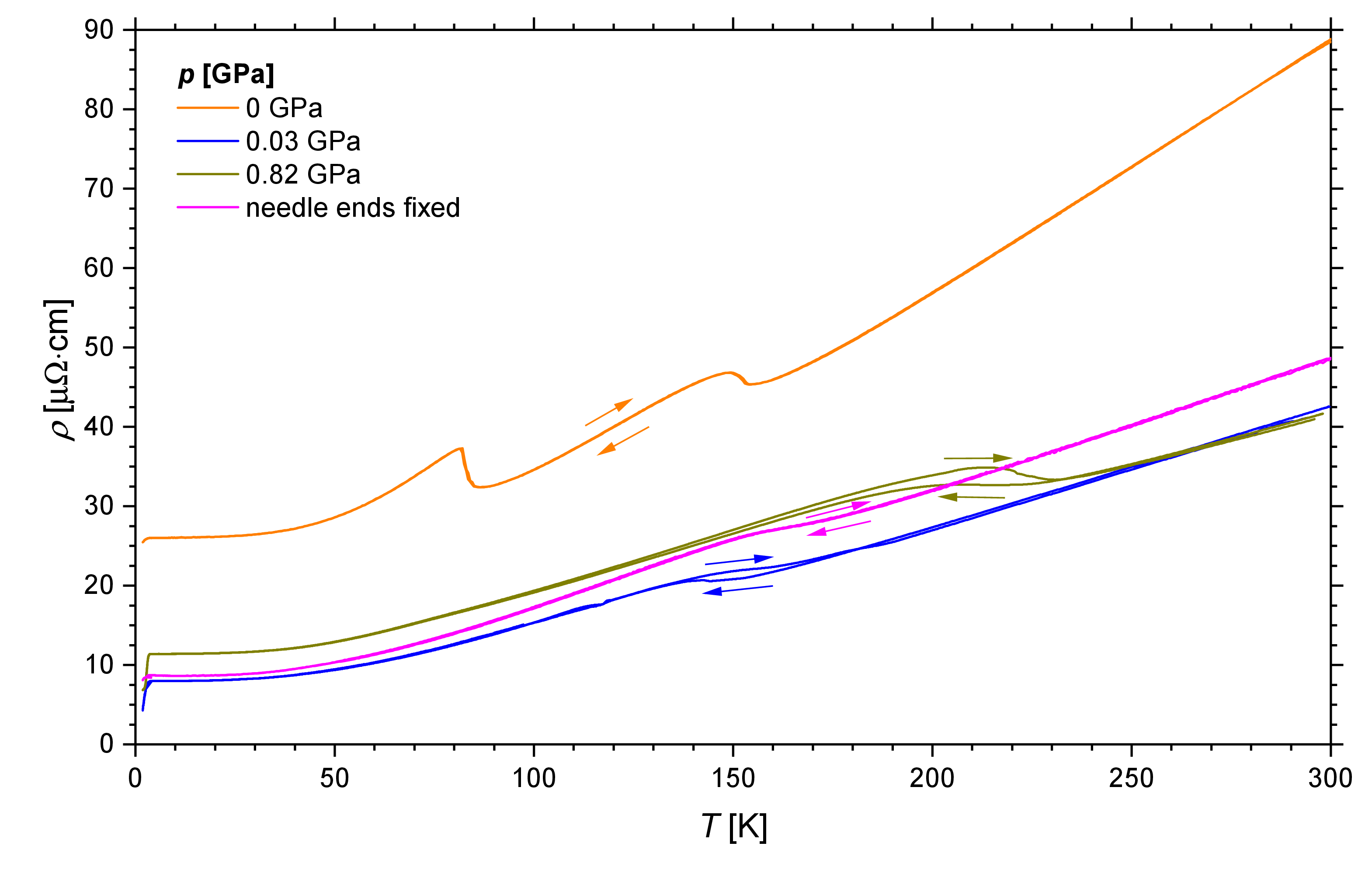}
  \caption{Unprocessed electrical resistivity data \rhoT\ for the
    measurements under hydrostatic pressures of 0.03~GPa (blue) and
    0.82~GPa (dark yellow). For comparison, \rhoT\ is also given for
    measurements under ambient pressure (orange) and uniaxial stress
    (magenta) that were performed on different samples.}
  \label{fig:resistivity-raw-data}
\end{figure}

For high-pressure electrical resistivity measurements a nonmagnetic
piston-cylinder-type pressure cell was used.\cite{Nicklas15} By using
silver conductive paint a crystalline \Co\ whisker (dimensions:
15~$\mu$m $\times$ 35~$\mu$m $\times$ 2740~$\mu$m) was four-point
contacted \textit{via} gold filaments which were attached to copper
wires.  Beforehand and in order to stabilize this setup, the whisker
was at one side slightly fixed to a mica plate employing GE-7031
varnish. On the backside of this mica plate a lead platelet was
attached, which itself was connected \textit{via} the four-point
method and served as pressure manometer. A silicon oil was utilized as
pressure-transmitting medium inside the Teflon cap of the
high-pressure cell. For determining the resistivity a LINEAR RESEARCH
LR700 resistance bridge was used. In order to cool the high-pressure
cell setup from 300~K to 1.8~K the cryostat function of a physical
property measurement system (PPMS; QUANTUM DESIGN) was employed.
Monitoring the temperature of the high-pressure cell was possible by a
thermometer (LAKE SHORE) fixed inside the shell of the pressure
cell. For each pressure point the resistivity was measured during the
cooling process to 1.8~K.  After that the pressure was determined by
fine-sliced resistivity measurements in a temperature range of 0.3~K
around the superconducting transition temperature of
Pb.\cite{eiling_pressure_1981,bireckoven_diamond_1988} Lastly followed
the resistivity measurement while heating the setup from 1.8~K to
300~K.\cite{Schmitz18} The unprocessed resistivity data for pressures
of 0.03~GPa and 0.82~GPa is given in
Fig.~\ref{fig:resistivity-raw-data}.

\FloatBarrier

\hspace{1cm}

\newpage

\section{High-pressure powder x-ray diffraction studies}
\label{sec:PXRD}

High-pressure powder x-ray diffraction experiments were performed at
the X04SA Materials Science (MS) beamline at Swiss Light Source
(SLS)\cite{Willmott13,Fisch15} employing a primary beam energy of
25~keV and a PSI Mythen~II one-dimensional
detector.\cite{Bergamaschi10} A radiation wave length of 0.49573~\AA\
was determined from the LaB$_6$ cell parameters refined for an initial
calibration measurement on a NIST SRM 660a powder standard.

Pressures up to 10.29(4)~GPa were generated by a diamond anvil cell
(DAC) equipped with 0.5~mm wide culets and stainless steel gaskets
providing a pressure chamber with 200~$\mu$m diameter and 45~$\mu$m~-
60~$\mu$m thickness. The pressure cell was loaded with finely ground
and sieved \Co\ powder (nominal sieve opening 32~$\mu$m) and a 4:1
MeOH:EtOH volume mixture serving as a pressure-transmitting
medium.\cite{piermarini_hydrostatic_1973} Powdered $\alpha$-quartz was
added as an internal pressure gauge exploiting the well-established
compression behavior of its lattice parameters.\cite{Angel97} With the
exception of the DAC filling process all sample preparation steps were
performed under an inert argon atmosphere to prevent a potential
sample degradation.

During data collection the pressure cell was oscillated continuously
by $\pm$3\degr\ around the $\omega$ axis, \textit{i.e.} perpendicular
to the primary beam,\cite{Fisch15} to bring a large number of
differently oriented crystallites into diffracting position. Remote
operation of the DAC by a helium-gas-driven membrane system allowed to
change the pressure while maintaining a consistent sample
orientation.

\begin{figure}[htb]
  \centering
  \includegraphics[width=1.0\textwidth]{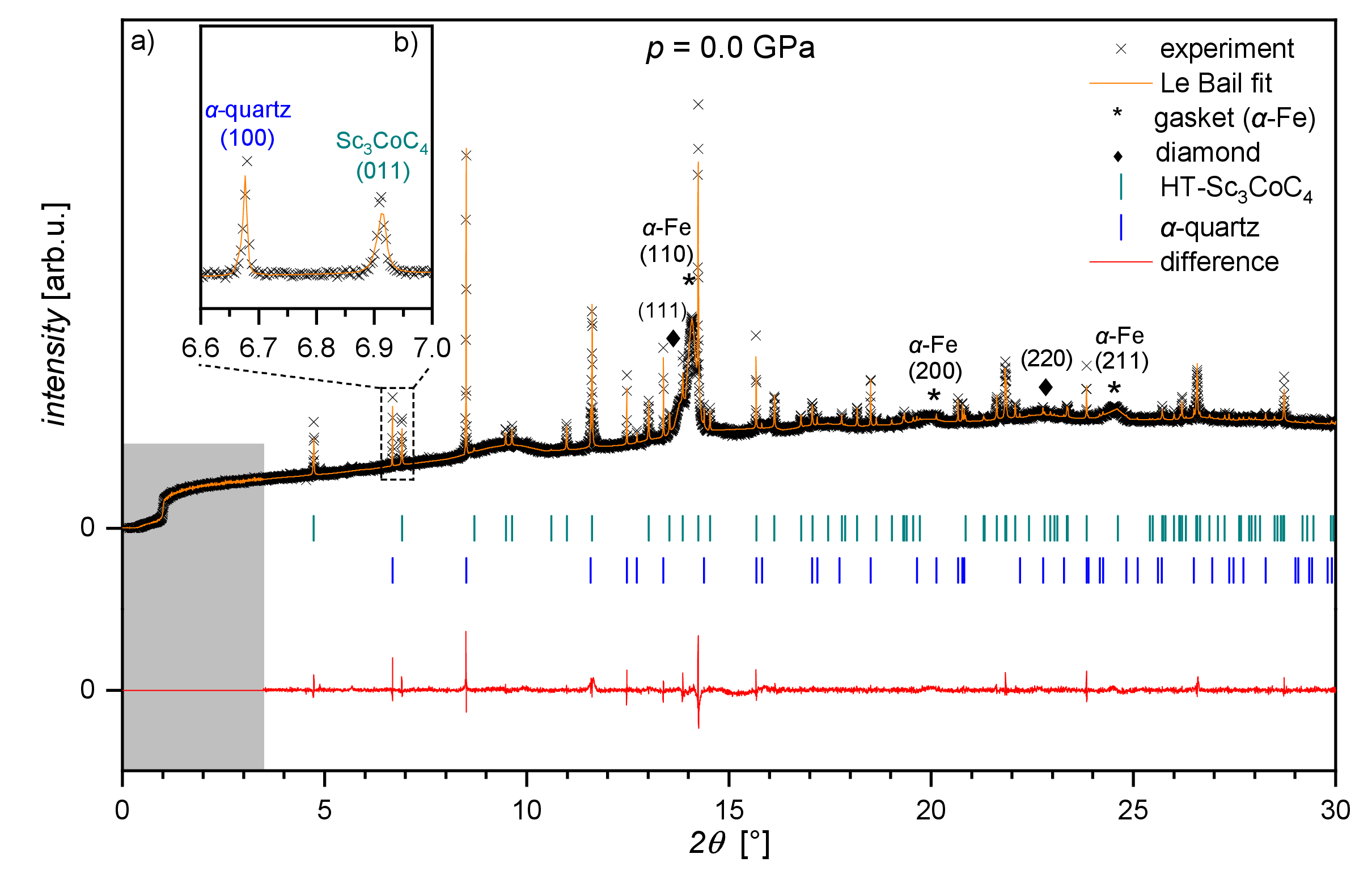}
  \caption{(a)~Room-temperature powder x-ray diffraction pattern
    (black crosses), Le Bail fit (orange line) and according
    difference plot (red line) for a \Co\ sample at a pressure of
    0.0~GPa ($\lambda =$~0.49573~\AA).  Expected reflection positions
    for \Co\ in its high-temperature phase and $\alpha$-quartz are
    indicated by green and blue bars, respectively. Asterisks and
    diamonds mark the positions of parasitic reflections from the
    gasket and the pressure-cell diamonds, while regions excluded from
    the Le Bail fit are shaded in gray. (b)~Enlarged view of the
    $(011)$ reflection for \Co\ and the $(100)$ reflection for
    $\alpha$-quartz.}
  \label{fig:pxrd-0GPa}
\end{figure}

\begin{figure}[htb]
  \centering
  \includegraphics[width=1.0\textwidth]{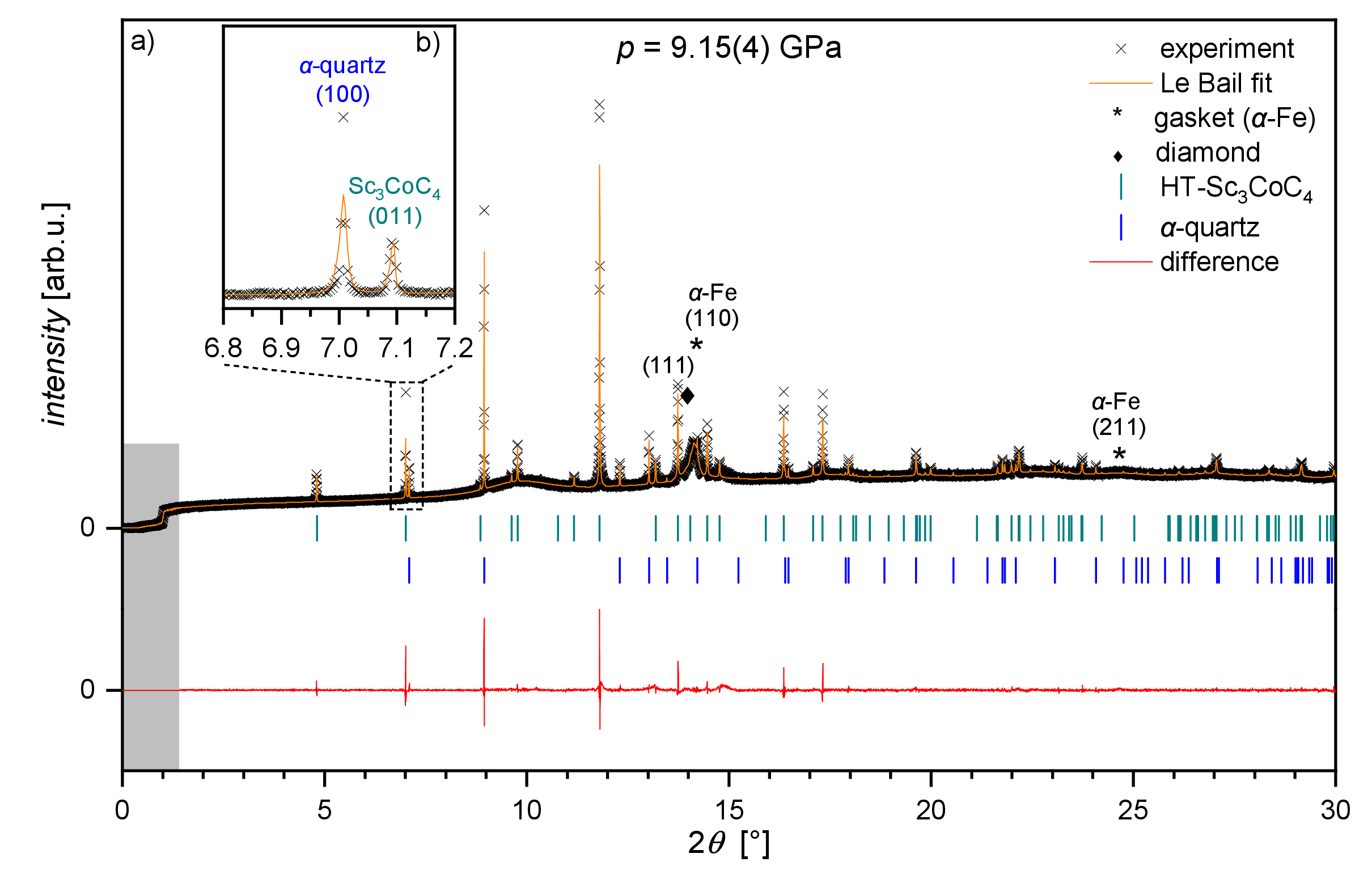}
  \caption{(a)~Room-temperature powder x-ray diffraction pattern
    (black crosses), Le Bail fit (orange line) and according
    difference plot (red line) for a \Co\ sample at a pressure of
    9.15(4)~GPa ($\lambda =$~0.49573~\AA).  Expected reflection
    positions for \Co\ in its high-temperature phase and
    $\alpha$-quartz are indicated by green and blue bars,
    respectively. Asterisks and diamonds mark the positions of
    parasitic reflections from the gasket and the pressure-cell
    diamonds, while regions excluded from the Le Bail fit are shaded
    in gray. (b)~Enlarged view of the $(011)$ reflection for \Co\ and
    the $(100)$ reflection for $\alpha$-quartz.}
  \label{fig:pxrd-9p15GPa}
\end{figure}

\begin{figure}[htb]
  \centering
  \includegraphics[width=1.0\textwidth]{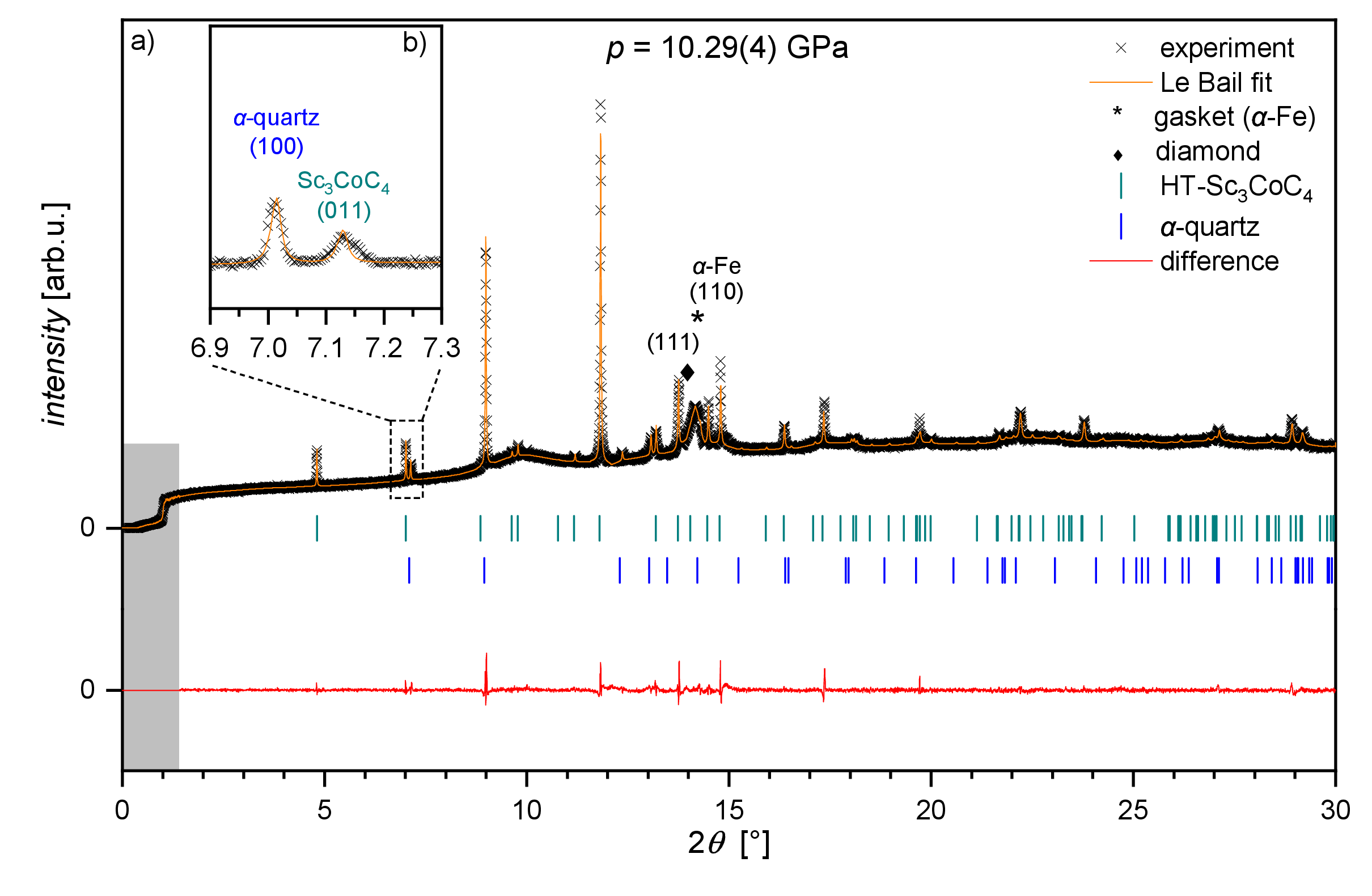}
  \caption{(a)~Room-temperature powder x-ray diffraction pattern
    (black crosses), Le Bail fit (orange line) and according
    difference plot (red line) for a \Co\ sample at a pressure of
    10.29(4)~GPa ($\lambda =$~0.49573~\AA).  Expected reflection
    positions for \Co\ in its high-temperature phase and
    $\alpha$-quartz are indicated by green and blue bars,
    respectively. Asterisks and diamonds mark the positions of
    parasitic reflections from the gasket and the pressure-cell
    diamonds, while regions excluded from the Le Bail fit are shaded
    in gray. (b)~Enlarged view of the $(011)$ reflection for \Co\ and
    the $(100)$ reflection for $\alpha$-quartz.}
  \label{fig:pxrd-10p29GPa}
\end{figure}

Exemplary powder x-ray diffraction patterns obtained at pressures of
0~GPa, 9.15(4)~GPa, and 10.29(4)~GPa (see Fig.~\ref{fig:pxrd-0GPa},
Fig.~\ref{fig:pxrd-9p15GPa}, and Fig.~\ref{fig:pxrd-10p29GPa})
illustrate the absence of additional reflections within the
investigated pressure range between 0~GPa and 10.29(4)~GPa. Only a
simultaneous broadening of the reflection profiles for \Co\ and
$\alpha$-quartz is observed for pressures above 9.15(4)~GPa
(Fig.~\ref{fig:pxrd-10p29GPa}) pointing to an incipient
non-hydrostaticity of the employed pressure medium (quasi-hydrostatic
limit $\approx$10~GPa\cite{piermarini_hydrostatic_1973}). Thus, all
observed reflections can be attributed to \Co\ in its high-temperature
phase (space group $Immm$), the pressure calibrant $\alpha$-quartz or
parasitic scattering from the surroundings of the pressure chamber,
\textit{i.e.} from the DAC body including the gasket and the
diamonds. In Le Bail fits\cite{LeBail88} of the diffraction data with
the program JANA2006\cite{Petricek14} the latter reflections were
added to the background description (specified by 100 points), whereas
\Co\ and $\alpha$-quartz reflections in the 2$\theta$ range between
3\degr\ and 30\degr\ were modelled by purely Lorentzian profiles (fit
lines and difference plots at selected pressure can be found in
Fig.~\ref{fig:pxrd-0GPa} to Fig.~\ref{fig:pxrd-10p29GPa}, an overview
of profile $Rp$ factors and lattice parameters is available in
Tab.~\ref{tab:pxrd-lebail}). Standard deviations of the refined
lattice parameters are specified with applied Berar's
correction\cite{Berar91} as implemented in JANA2006.\cite{Petricek14}

\begin{table}[htbp]
  \centering
  \begin{tabular}{cccccccc}
    \hline
    \textbf{\textit{p}~[GPa]} & \textbf{\textit{GoF}} & \textbf{\textit{Rp}} &
    \textbf{\textit{wRp}} & \textbf{\textit{a}~[\AA]} & \textbf{\textit{b}~[\AA]} &
    \textbf{\textit{c}~[\AA]} & \textbf{\textit{V}~[\AA\textsuperscript{3}]}\\
    \hline
    0 & 1.21 & 1.44 & 1.91 & 3.39685(15) & 4.37375(17) & 11.9971(6) & 178.240(13)\\
    0.106(1) & 1.22 & 1.56 & 1.97 & 3.39617(12) & 4.37311(14) & 11.9929(4) & 178.116(11)\\
    0.255(1) & 1.24 & 1.60 & 1.99 & 3.39496(9) & 4.37225(11) & 11.9899(3) & 177.974(8)\\
    0.637(3) & 1.12 & 1.46 & 1.78 & 3.39235(10) & 4.36986(10) & 11.9807(3) & 177.603(8)\\
    0.932(6) & 1.27 & 1.56 & 2.02 & 3.39015(11) & 4.36772(16) & 11.9742(4) & 177.304(10)\\
    1.674(14) & 1.22 & 1.49 & 1.94 & 3.38515(13) & 4.36153(16) & 11.9589(5) & 176.566(12)\\
    2.148(6) & 1.23 & 1.55 & 1.95 & 3.38207(13) & 4.35809(13) & 11.9480(5) & 176.106(11)\\
    2.882(4) & 1.20 & 1.51 & 1.90 & 3.37662(13) & 4.35417(16) & 11.9330(5) & 175.443(12)\\
    3.562(18) & 1.40 & 1.71 & 2.22 & 3.37207(7) & 4.34944(18) & 11.9181(5) & 174.799(11)\\
    4.286(9) & 1.20 & 1.47 & 1.91 & 3.36744(18) & 4.3451(2) & 11.9033(6) & 174.167(16)\\
    6.164(5) & 1.30 & 1.63 & 2.05 & 3.35590(16) & 4.3339(2) & 11.8685(6) & 172.618(15)\\
    6.734(8) & 1.20 & 1.54 & 1.89 & 3.35206(11) & 4.33076(15) & 11.8586(4) & 172.151(10)\\
    7.533(4) & 1.31 & 1.67 & 2.07 & 3.34729(12) & 4.32606(13) & 11.8457(4) & 171.533(10)\\
    8.380(12) & 1.30 & 1.57 & 2.03 & 3.34208(15) & 4.32134(18) & 11.8287(5) & 170.833(13)\\
    8.59(2) & 1.28 & 1.58 & 2.00 & 3.34090(14) & 4.31987(16) & 11.8243(5) & 170.652(12)\\
    8.80(3) & 1.25 & 1.50 & 1.96 & 3.33961(14) & 4.31831(16) & 11.8204(4) & 170.467(11)\\
    9.15(4) & 1.30 & 1.59 & 2.05 & 3.33732(17) & 4.3164(2) & 11.8137(6) & 170.180(14)\\
    9.53(6) & 1.18 & 1.45 & 1.86 & 3.3343(2) & 4.3149(3) & 11.8053(6) & 169.844(19)\\
    9.89(4) & 1.27 & 1.60 & 1.99 & 3.3313(3) & 4.3132(3) & 11.8040(8) & 169.61(2)\\
    10.29(4) & 1.18 & 1.50 & 1.88 & 3.3275(3) & 4.3121(4) & 11.7985(11) & 169.29(3)\\
    \hline                            
  \end{tabular}
  \caption{Pressure-dependent variation of the profile $Rp$ factors and
    refined lattice parameters for the Le Bail fits of all
    high-pressure powder x-ray diffraction patterns.}
  \label{tab:pxrd-lebail}
\end{table}
\FloatBarrier

\newpage

\section{Single-crystal x-ray diffraction experiments}
\label{sec:SCXRD}
\subsection{Data collection and reduction}

All single-crystal x-ray diffraction studies were performed using a
four-circle Eulerian cradle goniometer (HUBER) and a
sample-to-detector distance of 7~cm. Reflection intensities were
collected by a Pilatus 300K pixel detector with a CdTe detection layer
(DECTRIS). A micro-focus AgK$_\alpha$ tube ($\lambda =$~0.56087~\AA)
with a montel multilayer optic (INCOATEC) served as x-ray source.

\begin{figure}[h]
  \centering
  \includegraphics[width=0.9\textwidth]{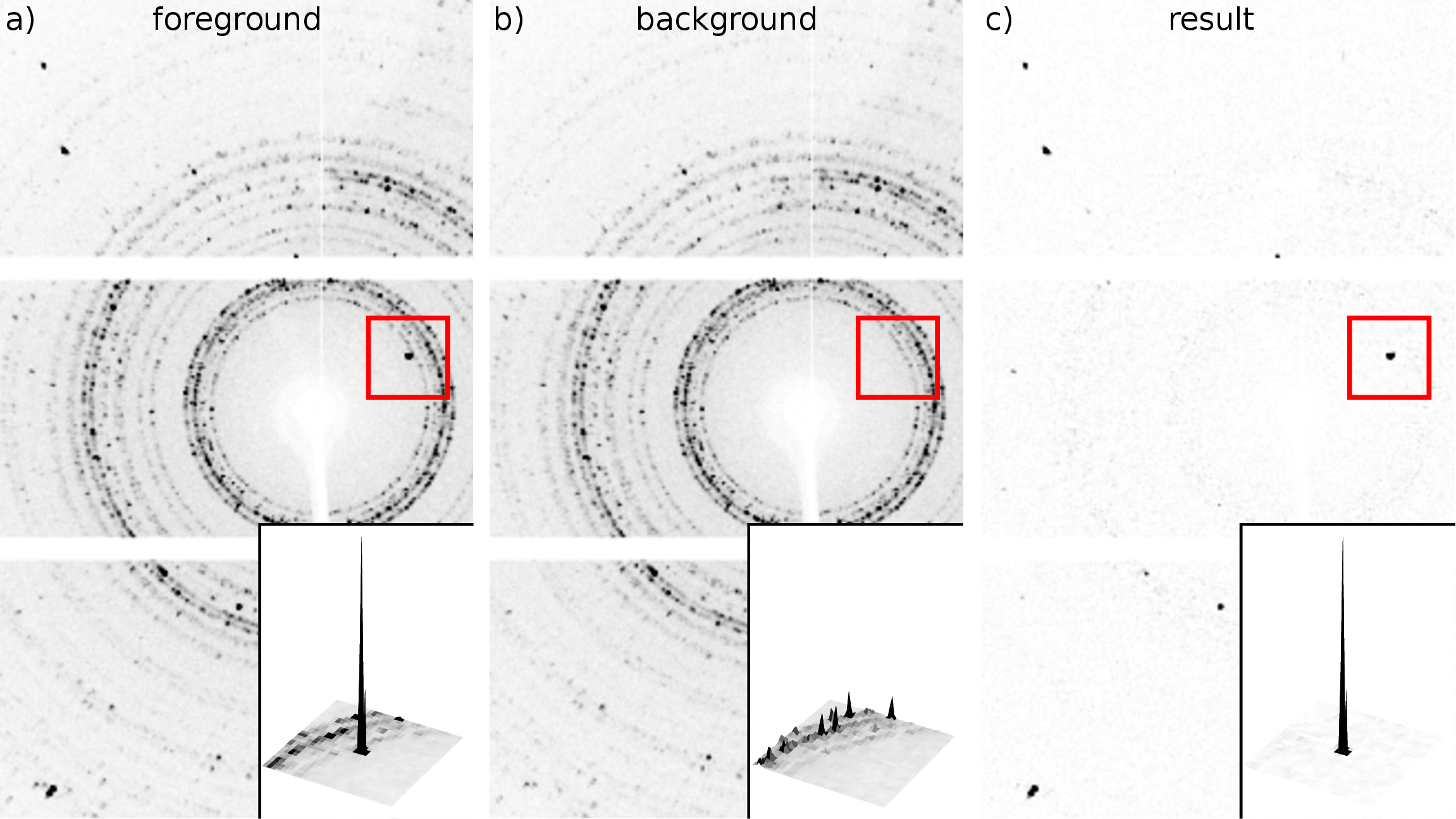}
  \caption{Background subtraction procedure for ambient-pressure
    low-temperature measurements without pressure cell: For each x-ray
    diffraction image (a) an individual background (b) is determined
    by translating the sample out of the beam.  Subtraction of the
    background leads to the result displayed in (c). Insets show 3D
    profiles of the regions marked by red rectangles. Note that the
    inset in (b) is plotted with an eight times enlarged $z$ scale.}
  \label{fig:bg-subtraction}
\end{figure}

Sample cooling to a minimum temperature of 11~K was done with a
closed-cycle helium cryocooler (ARS). Two alternative types of heat
and radiation shields (outer and optional inner shield) were used:
Stainless steel shields equipped with Kapton windows that add only a
weak and continuous background to the collected x-ray diffraction
images but strongly restrict the accessible portion of reciprocal
space, and beryllium shields (domes) that allow the measurement of a
larger number of reflection intensities but create stronger parasitic
scattering in the form of speckled rings (see
Fig.~\ref{fig:bg-subtraction}b). The type of heat and radiation
shields employed in each experiment is specified in
Tab.~\ref{tab:single-crystal-xrd-experiments} and
Tab.~\ref{tab:single-crystal-xrd-experiments1}.

High pressure was generated by commercially available diamond anvil
cells (DACs; see Tab.~\ref{tab:single-crystal-xrd-experiments} and
Tab.~\ref{tab:single-crystal-xrd-experiments1}) of Diacell Tozer-type
(ALMAX EASYLAB) and of Boehler-plate-type (ALMAX EASYLAB) equipped
with conical Boehler-Almax anvils (culet diameter
600~$\mu$m).\cite{graf_nonmetallic_2011,boehler_new_2006,boehler_new_2004}
Due to its smaller size the first DAC type was mounted inside the
sample chamber of the closed-cycle cryostat, whereas the latter DAC
type was only operated at room temperature (employed
pressure-transmitting media are listed in
Tab.~\ref{tab:single-crystal-xrd-experiments} and
Tab.~\ref{tab:single-crystal-xrd-experiments1}). Specified pressure
values were always determined ahead of the x-ray diffraction
experiment at room temperature using the fluorescence signal of ruby
spheres inside the pressure
chamber.\cite{piermarini_calibration_1975,Dewaele08,Kantor}

\begin{figure}[h]
  \centering
  \includegraphics[width=0.9\textwidth]{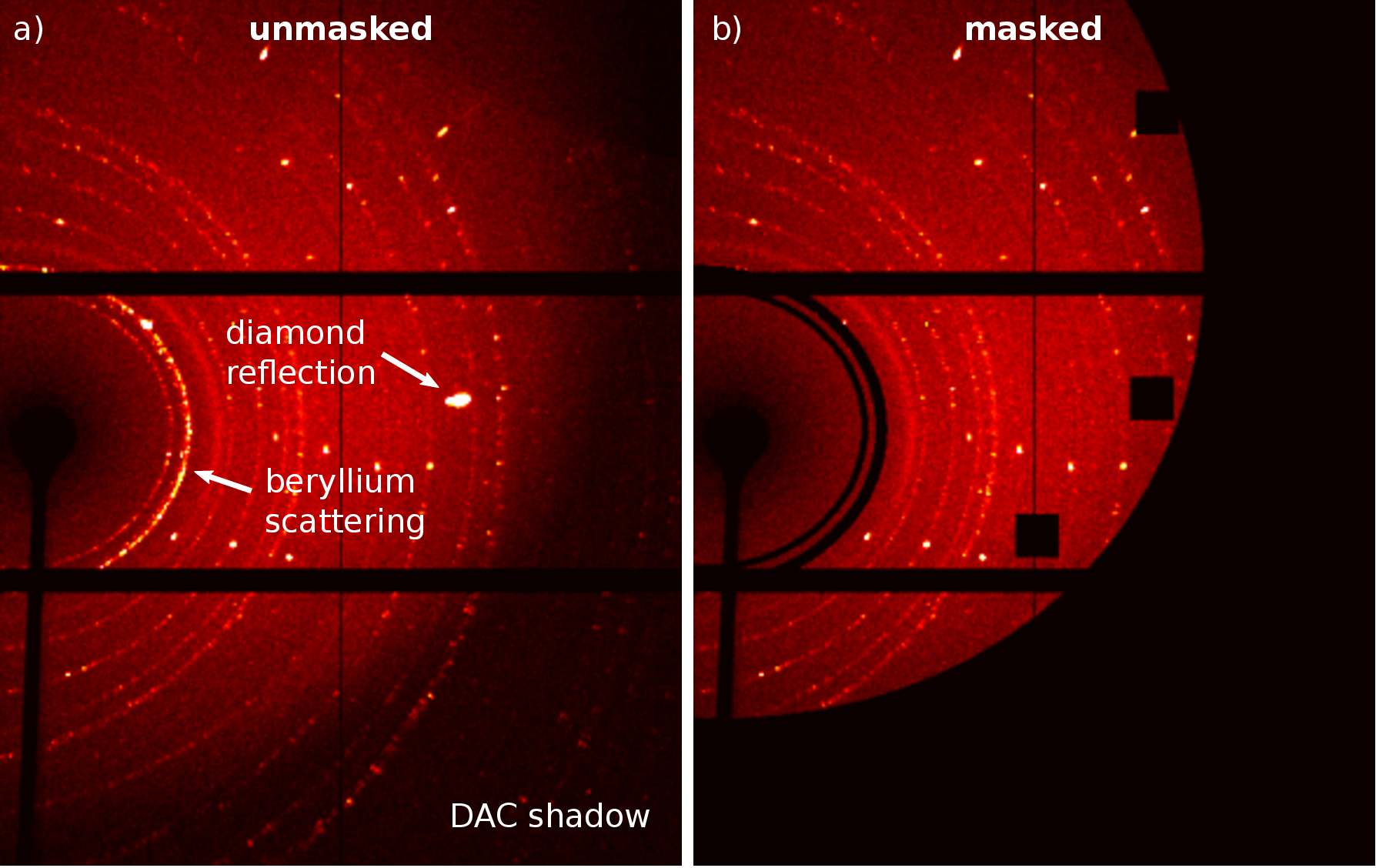}
  \caption{X-ray diffraction image from a high-pressure
    low-temperature experiment (a)~before and (b)~after application of
    masks to cover regions shadowed by the DAC body, diamond
    reflections and prominent Debye rings from the beryllium heat
    and radiation shield.}
  \label{fig:frame-mask}
\end{figure}

In case of ambient-pressure low-temperature studies without pressure
cell parasitic scattering from the beryllium heat and radiation
shields was determined and subtracted explicitly (procedure
illustrated in Fig.~\ref{fig:bg-subtraction}). For this purpose, the
investigated sample was translated reproducibly out of and into the
x-ray beam using a linear nanopositioner (ATTOCUBE). In case of
high-pressure studies masks were applied to regions of the diffraction
images shadowed by the DAC body or dominated by strong Bragg
reflections from the DAC diamonds (Fig.~\ref{fig:frame-mask}).
Additionally the two strongest innermost Debye rings of beryllium were
masked (compare Fig.~\ref{fig:frame-mask}a and
Fig.~\ref{fig:frame-mask}b) in case of high-pressure low-temperature
measurements.

Reflection intensities were evaluated employing the EVAL14 suite of
programs\cite{Duisenberg92,Duisenberg03} and subjected to scaling and
absorption correction employing SADABS/TWINABS.\cite{Krause15}
Structural models were refined with the program
JANA2006\cite{Petricek14} using the HKLF4 reflection format for
untwinned and the HKLF5 reflection format for systematically twinned
samples below the HT$\rightarrow$LT phase transition.\vspace{0.5cm}

\FloatBarrier
\hspace{1cm}

\newpage

\subsection{Overview of experiments}

\begin{table}[h]
  \centering
  \begin{tabular}{lccc}
    \hline
    & \textbf{1} & \textbf{2} & \textbf{3}\\
    \hline
    sample dim. [$\mu$m$^3$] & 40$\times$51$\times$290 & 55$\times$131$\times$141 & 68$\times$116$\times$126\\
    sample photo & Fig.~\ref{fig:crystal-ambient-pressure} &
                                                             Fig.~\ref{fig:crystals-tdac}a & Fig.~\ref{fig:crystals-tdac}b\\[0.7cm]
    pressure [GPa] & ambient & ambient / 0.6, 1.9, 4.0, 5.5 & ambient / 4\\
    pressure cell\footnote{sample / ruby spheres were wetted with perfluorinated polyether and placed in the pressure chamber.} & -- & Tozer DAC & Tozer DAC\\
    culet diam. [$\mu$m] & -- & 600 & 600\\
    press. chamber & \multirow{2}{*}{--} & \multirow{2}{*}{285} & \multirow{2}{*}{270}\\[-0.3cm]
    diam. [$\mu$m]\\
    press. chamber & \multirow{2}{*}{--} & \multirow{2}{*}{90} & \multirow{2}{*}{88}\\[-0.3cm]
    height. [$\mu$m]\\
    press. determ.\footnote{performed at $T =$~294~K.} & -- & -- / ruby fluoresc. & -- / ruby fluoresc.\\
    press. transm. & \multirow{2}{*}{--} & \multirow{2}{*}{-- / Daphne~7575} & \multirow{2}{*}{-- / Daphne~7575}\\[-0.3cm]
    medium\\
    hydrostat. lim. [GPa] & \multirow{2}{*}{--} & \multirow{2}{*}{-- / 3.9 - 4 [\!\!\citenum{murata_development_2016}]} & \multirow{2}{*}{-- / 3.9 - 4 [\!\!\citenum{murata_development_2016}]}\\[-0.3cm]
    at $T =$~RT\\[0.7cm]
    
    temperature [K] & 11, 70, 100 & 22 - 300 & 36, 106 / 37, 107\\
    sample cryostat & closed cycle He & closed cycle He & closed cycle He\\
    vacuum chamber & beryllium domes & steel cubes & beryllium dome\\
    parasitic scatt./ & \multirow{2}{*}{subtracted} & \multirow{2}{*}{--} & \multirow{2}{*}{masked}\\[-0.3cm]
    shadows &&&\\
    \hline
  \end{tabular}
  \caption{Overview of single-crystal x-ray diffraction experiments.}
  \label{tab:single-crystal-xrd-experiments}
\end{table}

\begin{table}[h]
  \centering
  \begin{tabular}{lccc}
    \hline
    & \textbf{4} & \textbf{5} & \textbf{6}\\
    \hline
    sample dim. [$\mu$m$^3$] & 48$\times$150$\times$152 & 73$\times$120$\times$186 & 70$\times$98$\times$98\\
    sample photo & Fig.~\ref{fig:crystals-tdac}c &
                            Fig.~\ref{fig:crystals-bdac}a & Fig.~\ref{fig:crystals-bdac}b\\[0.7cm]
    pressure [GPa] & ambient / 4.5 & 0.2, 4.2 & 0.1, 5.4, 6.7, 8.1, 10.1\\
    pressure cell\footnote{sample / ruby spheres were wetted with perfluorinated polyether and placed in the pressure chamber.} & -- / Tozer DAC & Boehler DAC & Boehler DAC\\
    culet diam. [$\mu$m] & -- / 600 & 600 & 600\\
    press. chamber & \multirow{2}{*}{-- / 270}  & \multirow{2}{*}{240} & \multirow{2}{*}{255}\\[-0.3cm]
    diam. [$\mu$m]\\
    press. chamber & \multirow{2}{*}{-- / 85}  & \multirow{2}{*}{94} & \multirow{2}{*}{92}\\[-0.3cm]
    height. [$\mu$m]\\
    press. determ.\footnote{performed at $T =$~294~K.} & -- / ruby fluoresc. & ruby fluoresc. & ruby fluoresc.\\
    press. transm. & \multirow{2}{*}{-- / Daphne~7575} & \multirow{2}{*}{4:1 MeOH:EtOH} & \multirow{2}{*}{4:1 MeOH:EtOH}\\[-0.3cm]
    medium\\
    hydrostat. lim. [GPa] & \multirow{2}{*}{-- / 3.9 - 4 [\!\!\citenum{murata_development_2016}]} & \multirow{2}{*}{$\approx$10 [\!\!\citenum{piermarini_hydrostatic_1973}]} & \multirow{2}{*}{$\approx$10 [\!\!\citenum{piermarini_hydrostatic_1973}]}\\[-0.3cm]
    at $T =$~RT &&&\\[0.7cm]
    temperature [K] & 13, RT / 27, RT & RT & RT\\
    sample cryostat & closed cycle He & -- & --\\
    vacuum chamber & steel cubes & -- & --\\
    parasitic scatt./ & \multirow{2}{*}{--} & \multirow{2}{*}{masked} & \multirow{2}{*}{masked}\\[-0.3cm]
    shadows &&&\\
    \hline
  \end{tabular}
  \caption{Overview of single-crystal x-ray diffraction experiments (continued).}
  \label{tab:single-crystal-xrd-experiments1}
\end{table}
\FloatBarrier
\hspace{1cm}

\newpage

\subsection{Ambient-pressure low-temperature measurements without
  pressure cell}

\enlargethispage{1cm}

\begin{table}[h]
  \centering
  \begin{tabular}{p{0.3\textwidth}>{\centering}p{0.23\textwidth}
    >{\centering}p{0.23\textwidth}>{\centering\arraybackslash}p{0.23\textwidth}}
    \hline
    \centering $T$ [K] & 11 & 70 & 100\\
    \hline
    \multirow{5}{*}{unit cell dimensions} & $a =$~5.53630(10)~\AA & $a =$~5.53600(10)~\AA & $a =$~5.53720(10)~\AA\\
            & $b =$~12.0210(2)~\AA & $b =$~12.0167(2)~\AA & $b =$~12.00370(10)~\AA\\
            & $c =$~5.53640(10)~\AA & $c =$~5.53590(10)~\AA & $c =$~5.53710(10)~\AA\\
            & $\beta =$~104.8070(10)\degr & $\beta =$~104.7720(10)\degr & $\beta =$~104.4620(10)\degr\\
            & $V =$~356.222(11)~\AA$^3$ & $V =$~356.101(11)~\AA$^3$ & $V =$~356.372(10)~\AA$^3$\\
    calculated density & 4.5095~g$\cdot$cm$^{-3}$ & 4.511~g$\cdot$cm$^{-3}$ & 4.5076~g$\cdot$cm$^{-3}$\\
    crystal size &  & 40$\times$51$\times$290~$\mu$m$^3$ & \\
    wave length &  & 0.56087~\AA & \\
    transm. ratio (max/min) & 0.747 / 0.686 & 0.747 / 0.633 & 0.747 / 0.646\\
    absorption coefficient & 5.016~mm$^{-1}$ & 5.017~mm$^{-1}$ & 5.014~mm$^{-1}$\\
    $F(000)$ &  & 456 & \\
    $\theta$ range & 3\degr\ to 36\degr & 3\degr\ to 37\degr & 3\degr\ to 37\degr\\
    range in $hkl$ &  & $\pm$11, $\pm$25, $\pm$11 & \\
    total no. reflections & 8720 & 8390 & 8509\\
    independent reflections & 2142 ($R_\mathrm{int} =$~0.0123) & 2112 ($R_\mathrm{int} =$~0.0158) & 2153 ($R_\mathrm{int} =$~0.0148)\\
    reflections with $I \geq 2\sigma(I)$ & 2007 & 1986 & 1947\\
    data / parameters & 2142 / 43  & 2112 / 43 & 2153 / 43\\
    goodness-of-fit on $F^2$ & 1.27 & 1.79 & 1.44\\
    \multirow{2}{*}{final $R$ indices [$I \geq 2\sigma(I)$]} & $R =$~0.0220 & $R =$~0.0381 & $R =$~0.0302\\
    & $wR =$~0.0414 & $wR =$~0.0665 & $wR =$~0.0535\\
    \multirow{2}{*}{$R$ indices (all data)} & $R =$~0.0271 & $R =$~0.0446 & $R =$~0.0389\\
    & $wR =$~0.0424 & $wR =$~0.0673 & $wR =$~0.0549\\
    extinction coefficient & 0.0461(14) & 0.043(2) & 0.0231(15)\\
    largest diff. peak and hole & 1.97 / -2.18~e$\cdot$\AA$^{-3}$ & 0.65 / -0.86~e$\cdot$\AA$^{-3}$ & 1.94 / -2.20~e$\cdot$\AA$^{-3}$\\
    \hline
  \end{tabular}
  \caption{Crystal data and structure refinements for ambient-pressure
    single-crystal x-ray diffraction experiments without pressure cell
    and at temperatures of 11~K, 70~K and 100~K.}
  \label{tab:comparison-ambient-low-T-crysdata-ref}
\end{table}

\begin{table}[htb]
  \centering
  \begin{tabular}{>{\centering}p{0.1\textwidth}>{\centering}p{0.1\textwidth}
    >{\centering}p{0.15\textwidth}>{\centering}p{0.15\textwidth}
    >{\centering}p{0.15\textwidth}>{\centering\arraybackslash}p{0.15\textwidth}}
    & \textbf{\textit{T}} & \multicolumn{3}{c}{\textbf{fractional atomic coordinates}} &
    \textbf{\textit{U}\textsubscript{eq}}\\
    \hline
    atom & [K] & $x$ & $y$ & $z$ & [\AA$^2$]\\
    \hline
    \multirow{3}{*}{Co} & 11 & 0.26595(2) & 0 & 0.26673(2) & 0.00204(3)\\
    & 70 & 0.26559(3) & 0 & 0.26635(3) & 0.00240(5)\\
    & 100 & 0.25987(2) & 0 & 0.26046(2) & 0.00243(5)\\[0.3cm]
    \multirow{3}{*}{Sc1} & 11 & 0.75582(3) & 0 & 0.24273(3) & 0.00207(6)\\
    & 70 & 0.75570(4) & 0 & 0.24290(4) & 0.00241(9)\\
    & 100 & 0.75383(3) & 0 & 0.24498(3) & 0.00239(10)\\[0.3cm]
    \multirow{3}{*}{Sc2} & 11 & 0 & 0.187417(10) & 0 & 0.00210(9)\\
    & 70 & 0 & 0.187442(16) & 0  & 0.00244(11)\\
    & 100 & 0 & 0.187747(13) & 0  & 0.00233(10)\\[0.3cm]
    \multirow{3}{*}{Sc3} & 11 & 0 & 0.311540(10) & 0.5 & 0.00210(9)\\
    & 70 & 0 & 0.311550(16) & 0.5 & 0.00244(11)\\
    & 100 & 0 & 0.311642(13) & 0.5 & 0.00235(10)\\[0.3cm]
    \multirow{3}{*}{C1} & 11 & 0.4110(3) & 0.12557(5) & 0.0766(2) & 0.0031(2)\\
    & 70 & 0.4103(4) & 0.12568(7) & 0.0763(4) & 0.0035(3)\\
    & 100 & 0.4109(3) & 0.12514(5) & 0.0773(3) & 0.0033(3)\\[0.3cm]
    \multirow{3}{*}{C2} & 11 & 0.0889(3) & 0.12487(5) & 0.4233(2) & 0.0030(2)\\
    & 70 & 0.0896(4) & 0.12490(7) & 0.4238(4) & 0.0034(3)\\
    & 100 & 0.0890(3) & 0.12471(5) & 0.4228(3) & 0.0033(3)\\
    \hline
  \end{tabular}
  \caption{Refined fractional atomic coordinates and mean-square
    atomic displacement parameters obtained from ambient-pressure
    single-crystal x-ray diffraction experiments without pressure cell
    and at temperatures of 11~K, 70~K and 100~K.}
  \label{tab:comparison-ambient-low-T}
\end{table}

\begin{table}[htb]
  \centering
  \begin{tabular}{>{\centering}p{0.07\textwidth}>{\centering}p{0.07\textwidth}
    >{\centering}p{0.12\textwidth}>{\centering}p{0.12\textwidth}
    >{\centering}p{0.12\textwidth}>{\centering}p{0.12\textwidth}
    >{\centering}p{0.12\textwidth}>{\centering\arraybackslash}p{0.12\textwidth}}
    & \textbf{\textit{T}} & \multicolumn{6}{c}{\textbf{mean-square atomic displacement
                            parameters} [\AA$^2$]}\\
    \hline
    atom & [K] & $U_{11}$ & $U_{22}$ & $U_{33}$ & $U_{12}$ & $U_{13}$ & $U_{23}$\\
    \hline
    \multirow{3}{*}{Co} & 11 & 0.00229(6) & 0.00176(4) & 0.00219(6) & $\ast$ & 0.00082(3) & $\ast$\\
    & 70 & 0.00239(9) & 0.00275(7) & 0.00238(9) & $\ast$ & 0.00120(6) & $\ast$\\
    & 100 & 0.00295(10) & 0.00192(5) & 0.00276(10) & $\ast$ & 0.00136(4) & $\ast$\\[0.3cm]
    \multirow{3}{*}{Sc1} & 11 & 0.00223(12) & 0.00202(5) & 0.00195(12) & $\ast$ & 0.00053(5) & $\ast$\\
    & 70 & 0.00228(17) & 0.00291(9) & 0.00216(16) & $\ast$ & 0.00079(9) & $\ast$\\
    & 100 & 0.00280(19) & 0.00224(6) & 0.00220(19) & $\ast$ & 0.00078(5) & $\ast$\\[0.3cm]
    \multirow{3}{*}{Sc2} & 11 & 0.00295(18) & 0.00198(5) & 0.00144(17) & $\ast$ & 0.00068(4) & $\ast$\\
    & 70 & 0.0036(2) & 0.00283(9) & 0.00114(19) & $\ast$ & 0.00100(9) & $\ast$\\
    & 100 & 0.0034(2) & 0.00221(7) & 0.00152(19) & $\ast$ & 0.00090(5) & $\ast$\\[0.3cm]
    \multirow{3}{*}{Sc3} & 11 & 0.00295(18) & 0.00196(5) & 0.00147(17) & $\ast$ & 0.00071(4) & $\ast$\\
    & 70 & 0.0035(2) & 0.00281(9) & 0.00119(19) & $\ast$ & 0.00094(9) & $\ast$\\
    & 100 & 0.0035(2) & 0.00218(7) & 0.00153(19) & $\ast$ & 0.00087(5) & $\ast$\\[0.3cm]
    \multirow{3}{*}{C1} & 11 & 0.0017(4) & 0.00353(17) & 0.0037(4) & 0.0005(3) & 0.00026(16) & 0.0002(3)\\
    & 70 & 0.0015(5) & 0.0049(3) & 0.0041(6) & 0.0002(6) & 0.0010(3) & 0.0001(6)\\
    & 100 & 0.0027(5) & 0.0035(2) & 0.0035(5) & 0.0002(4) & 0.0008(2) & -0.0001(4)\\[0.3cm]
    \multirow{3}{*}{C2} & 11 & 0.0019(4) & 0.00337(17) & 0.0037(4) & -0.0002(3) & 0.00039(16) & 0.0000(3)\\
    & 70 & 0.0015(5) & 0.0050(3) & 0.0038(6) & -0.0001(6) & 0.0005(3) & 0.0002(6)\\
    & 100 & 0.0025(5) & 0.0036(2) & 0.0036(5) & 0.0000(4) & 0.0004(2) & 0.0003(4)\\
    \hline
  \end{tabular}
  \caption{Refined mean-square atomic displacement parameters obtained
    from ambient-pressure single-crystal x-ray diffraction experiments
    without pressure cell and at temperatures of 11~K, 70~K and
    100~K. Parameters marked by an asterisk are forbidden by symmetry.}
  \label{tab:comparison-ambient-low-T-ADPs}
\end{table}
\FloatBarrier

\newpage

\subsection{High-pressure low-temperature measurements}

\enlargethispage{1cm}

\begin{table}[h]
  \centering
  \begin{tabular}{p{0.3\textwidth}>{\centering}p{0.23\textwidth}
    >{\centering\arraybackslash}p{0.23\textwidth}}
    \hline
    \centering $T$ [K] & 37 & 107\\
    \hline
    \multirow{5}{*}{unit cell dimensions} & $a =$~5.5046(3)~\AA & $a =$~5.5082(2)~\AA\\
            & $b =$~11.9698(5)~\AA & $b =$~11.9729(6)~\AA\\
            & $c =$~5.4723(7)~\AA & $c =$~5.4832(8)~\AA\\
            & $\beta =$~105.038(3)\degr & $\beta =$~104.949(4)\degr\\
            & $V =$~348.22(5)~\AA$^3$ & $V =$~349.37(6)~\AA$^3$\\
    calculated density & 4.6131~g$\cdot$cm$^{-3}$ & 4.5979~g$\cdot$cm$^{-3}$\\
    crystal size & \multicolumn{2}{c}{68$\times$116$\times$126~$\mu$m$^3$}\\
    wave length & \multicolumn{2}{c}{0.56087~\AA}\\
    transm. ratio (max/min) & 0.746 / 0.687 & 0.746 / 0.670\\
    absorption coefficient & 5.131~mm$^{-1}$ & 5.114~mm$^{-1}$\\
    $F(000)$ & \multicolumn{2}{c}{456}\\
    $\theta$ range & \multicolumn{2}{c}{3\degr\ to 33\degr}\\
    range in $hkl$ & -6/9, -13/19, -5/4 & -6/9, -13/20, -5/4\\
    total no. reflections & 716 & 713\\
    independent reflections & 212 ($R_\mathrm{int} =$~0.0048) & 209 ($R_\mathrm{int} =$~0.0077)\\
    reflections with $I \geq 2\sigma(I)$ & 190 & 178\\
    data / parameters & 190 / 18 & 178 / 18\\
    goodness-of-fit on $F^2$ & 3.66 & 3.81\\
    \multirow{2}{*}{final $R$ indices [$I \geq 2\sigma(I)$]} & $R =$~0.0425 & $R =$~0.0436\\
    & $wR =$~0.1002 & $wR =$~0.1053\\
    \multirow{2}{*}{$R$ indices (all data)} & $R =$~0.0425 & $R =$~0.0436\\
    & $wR =$~0.1002 & $wR =$~0.1053\\
    extinction coefficient & -- & --\\
    largest diff. peak and hole & 0.69 / -0.76~e$\cdot$\AA$^{-3}$ & 0.90 / -1.17~e$\cdot$\AA$^{-3}$\\
    \hline
  \end{tabular}
  \caption{Crystal data and structure refinements for single-crystal
    x-ray diffraction experiments at a pressure of 4~GPa and
    temperatures of 37~K and 107~K.}
  \label{tab:comparison-HP-low-T-crysdata-ref}
\end{table}

\begin{table}[htb]
  \centering
  \begin{tabular}{>{\centering}p{0.1\textwidth}>{\centering}p{0.1\textwidth}
    >{\centering}p{0.15\textwidth}>{\centering}p{0.15\textwidth}
    >{\centering}p{0.15\textwidth}>{\centering\arraybackslash}p{0.15\textwidth}}
    & \textbf{\textit{T}} & \multicolumn{3}{c}{\textbf{fractional atomic coordinates}} &
    \textbf{\textit{U}\textsubscript{eq}}\\
    \hline
    atom & [K] & $x$ & $y$ & $z$ & [\AA$^2$]\\
    \hline
    \multirow{2}{*}{Co} & 37 & 0.26648(19) & 0 & 0.2672(3) & 0.0027(3)\\
    & 107 & 0.2663(2) & 0 & 0.2666(3) & 0.0035(4)\\[0.3cm]
    \multirow{2}{*}{Sc1} & 37 & 0.7565(2) & 0 & 0.2417(3) & 0.0025(4)\\
    & 107 & 0.7564(3) & 0 & 0.2418(4) & 0.0033(4)\\[0.3cm]
    \multirow{2}{*}{Sc2} & 37 & 0 & 0.18764(9) & 0 & 0.0025(4)\\
    & 107 & 0 & 0.18771(10) & 0 & 0.0032(4)\\[0.3cm]
    \multirow{2}{*}{Sc3} & 37 & 0 & 0.31118(9) & 0.5 & 0.0028(4)\\
    & 107 & 0 & 0.31122(10) & 0.5 & 0.0037(4)\\[0.3cm]
    \multirow{2}{*}{C1} & 37 & 0.4147(12) & 0.1261(3) & 0.0837(17) & 0.0042(8)\\
    & 107 & 0.4137(14) & 0.1262(4) & 0.0840(19) & 0.0057(9)\\[0.3cm]
    \multirow{2}{*}{C2} & 37 & 0.0854(13) & 0.1250(3) & 0.4160(17) & 0.0042(8)\\
    & 107 & 0.0867(14) & 0.1248(4) & 0.4169(19) & 0.0057(9)\\
    \hline
  \end{tabular}
  \caption{Refined fractional atomic coordinates and mean-square
    atomic displacement parameters obtained from single-crystal
    x-ray diffraction experiments at a pressure of 4~GPa and
    temperatures of 37~K and 107~K.}
  \label{tab:comparison-HP-low-T}
\end{table}

\FloatBarrier
\hspace{1cm}

\newpage

\subsection{Superstructure reflections at room
  temperature}
\label{sec:superstruc-RT}

\begin{figure}[htb]
  \centering
  \includegraphics[width=1.0\textwidth]{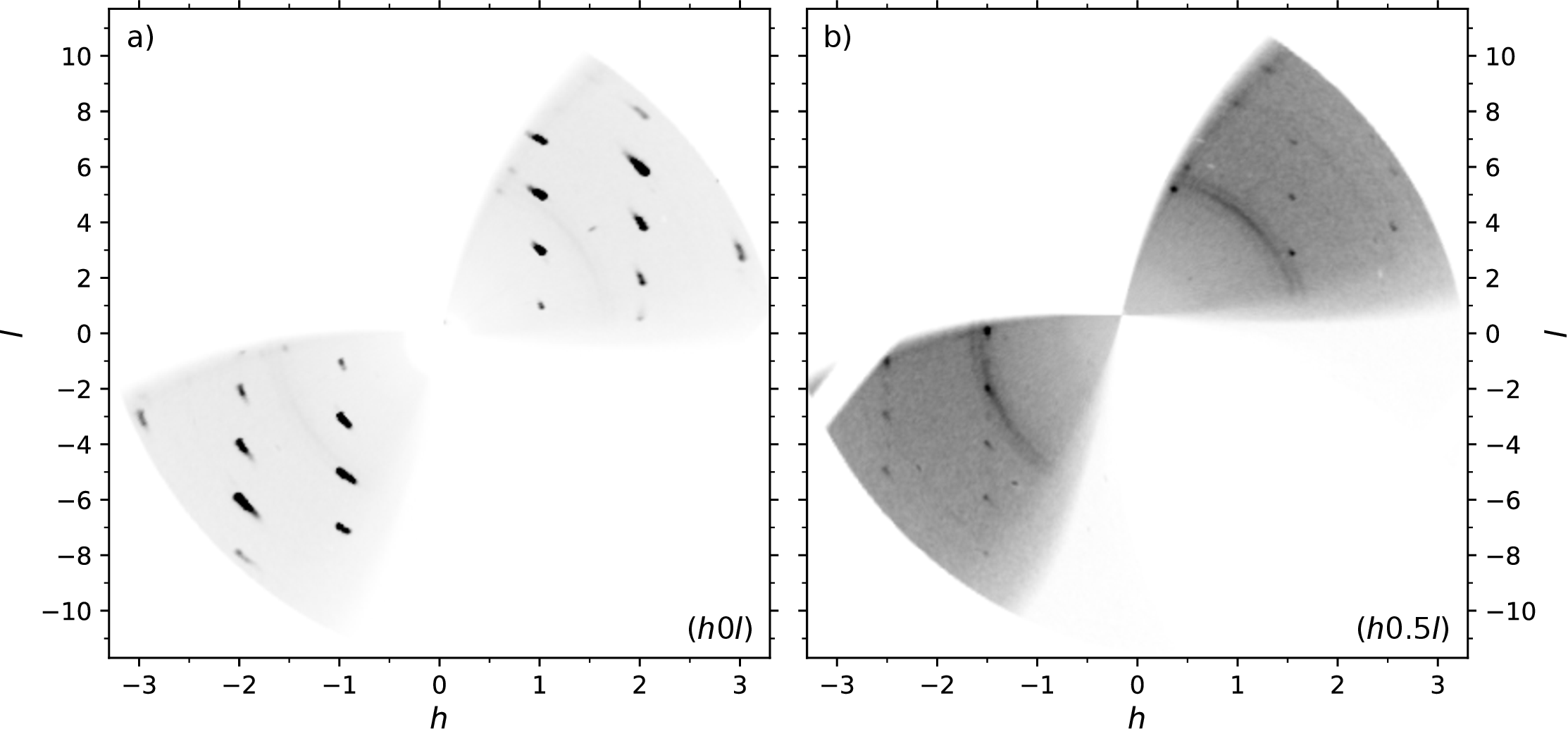}
  \caption{Reconstructions of reciprocal-space planes (a)~$(h0l)$ and
    (b)~$(h0.5l)$ from room-temperature x-ray diffraction data
    collected after applying a pressure of 5.5~GPa, cooling to 22~K
    and heating to room temperature again.}
  \label{fig:rec-planes-5p4gpa-rt}
\end{figure}

Superstructure reflections indicative of the low-temperature phase of
\Co\ could be preserved up to room temperature at a pressure of
5.5~GPa, \textit{i.e.} above the hydrostatic limit of the employed
pressure medium Daphne~7575\cite{murata_development_2016}
(experiment~2 in Tab.~\ref{tab:single-crystal-xrd-experiments}; see
reconstruction of the $(h0.5l)$ reciprocal-space plane in
Fig.~\ref{fig:rec-planes-5p4gpa-rt}b). But this was only possible
after cooling the sample to 22~K and heating it back to room
temperature and at the cost of a deterioration of the sample
crystallinity (see broadened reflections in the reconstructed $(h0l)$
plane in Fig.~\ref{fig:rec-planes-5p4gpa-rt}a).

\begin{figure}[htbp]
  \centering
  \includegraphics[width=1.0\textwidth]{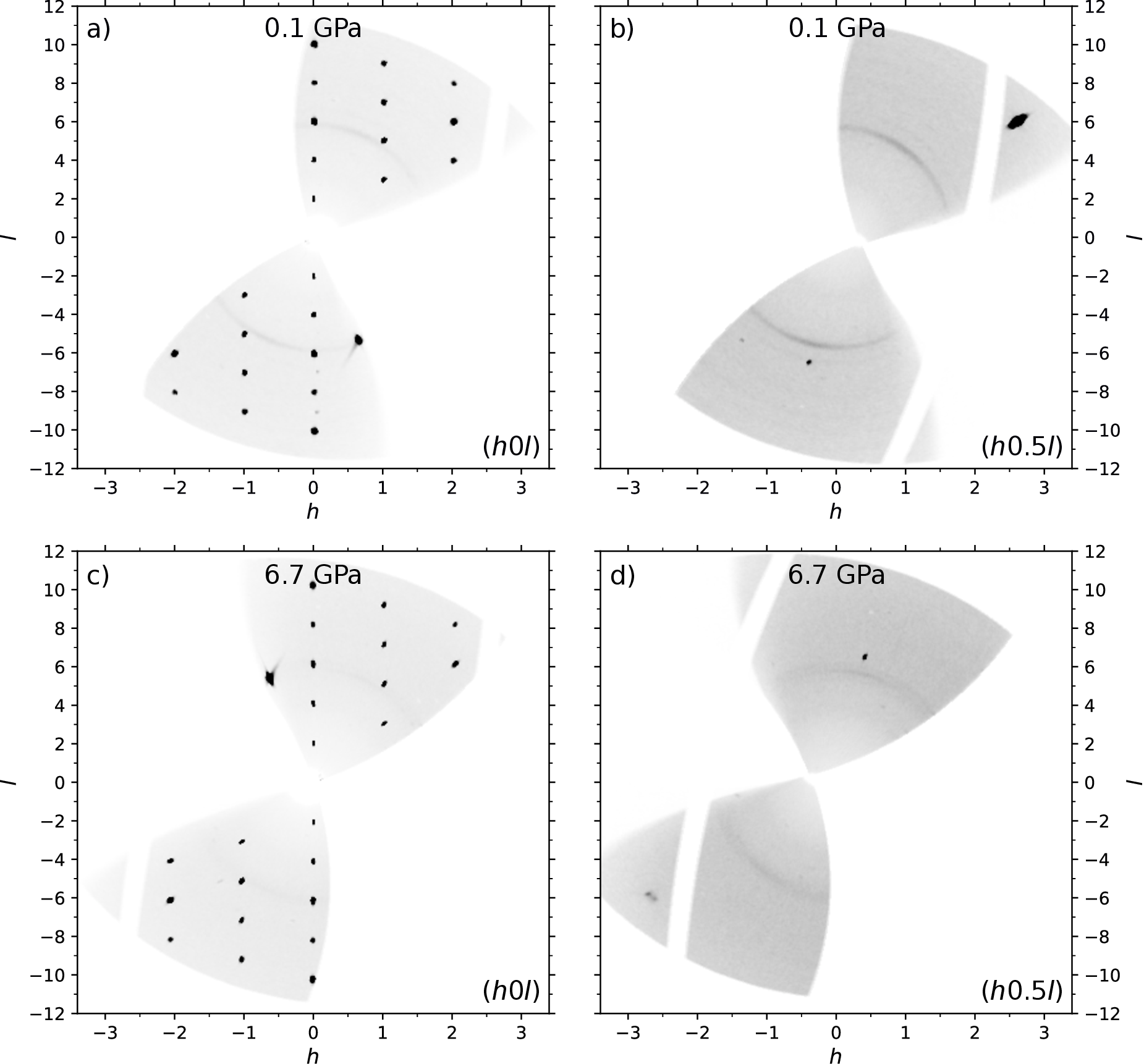}
  \caption{Reconstructions of reciprocal-space planes $(h0l)$ and
    $(h0.5l)$ from x-ray diffraction data collected at
    room-temperature and at pressures of 0.1~GPa (a, b) and 6.7~GPa
    (c, d). Parasitic scattering from gasket and diamonds leads to
    ring-shaped features and broad intense reflections, respectively.}
  \label{fig:rec-space-bdac}
\end{figure}

\begin{figure}[h]
  \centering
  \includegraphics[width=1.0\textwidth]{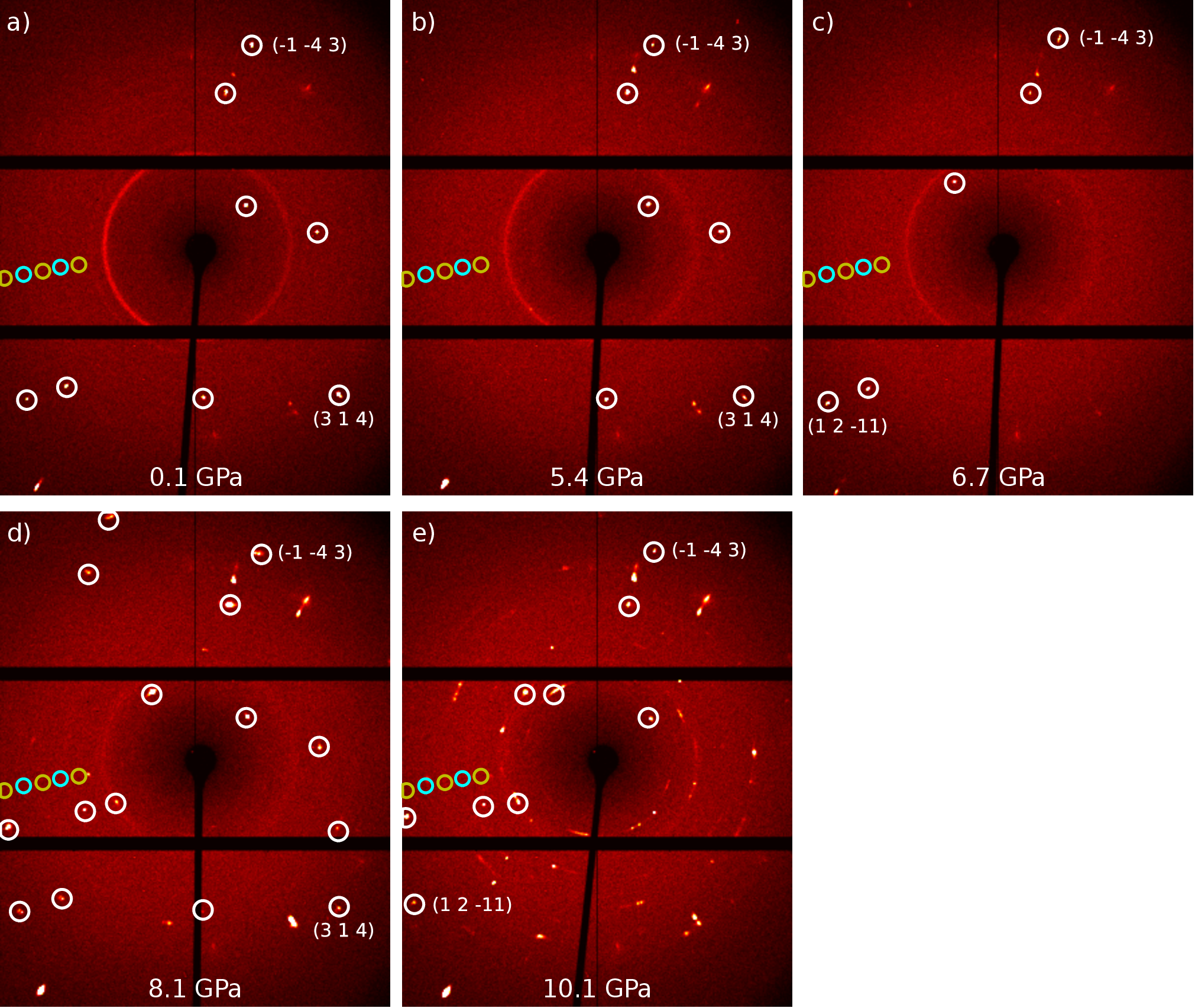}
  \caption{Selected x-ray diffraction images collected at room
    temperature and up to a maximum pressure of 10.1~GPa with nearly
    unchanged orientation of the pressure cell. Main reflections
    expected for the high-temperature phase of \Co\ are indicated by
    white circles. The positions of superstructure reflections
    indicative of a transition into the low-temperature phase are
    marked by yellow (twin domain~1) and cyan circles (twin
    domain~2).}
  \label{fig:frames-bdac}
\end{figure}

We did not succeed in inducing the phase transition between
high-temperature and low-temperature phase by pressure application at
room temperature alone (experiment~6 in
Tab.~\ref{tab:single-crystal-xrd-experiments1}). As can be recognized
from the reconstructed $(h0.5l)$ planes in
Fig.~\ref{fig:rec-space-bdac} and the x-ray diffraction images in
Fig.~\ref{fig:frames-bdac}, no superstructure reflections could be
observed in a room-temperature high-pressure experiment up to a
maximum pressure of 10.1~GPa.

\vspace{0.5cm}

\FloatBarrier

\newpage

\subsection{Test for sample degradation in
  high-pressure/low-temperature studies}

\begin{figure}[p]
  \centering
  \includegraphics[width=0.85\textwidth]{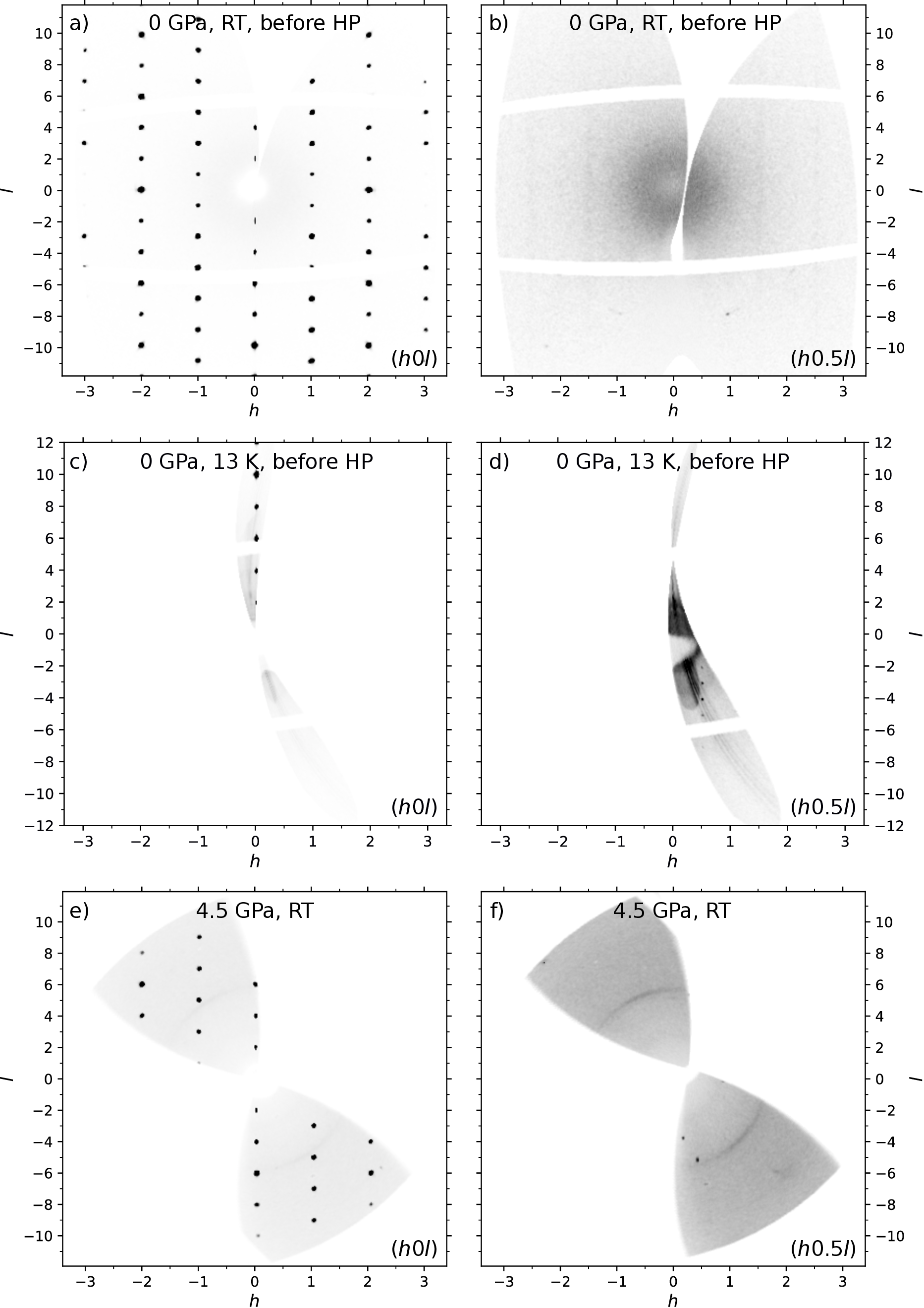}
  \caption{Reconstructions of reciprocal-space planes $(h0l)$ and
    $(h0.5l)$ from x-ray diffraction data collected under various
    conditions, \textit{i.e.} at room-temperature (a, b) and 13~K (c, d)
    before performing a high-pressure (HP) experiment, and again at
    room-temperature but with an applied pressure of 4.5~GPa (e, f).}
  \label{fig:rec-planes-before-after-hp}
\end{figure}

To exclude irreversible changes or a degradation of the sample quality
in our combined high-pressure/low-temperature studies we reenacted a
typical experimental procedure and collected x-ray diffraction data
after each step (experiment~4 in
Tab.~\ref{tab:single-crystal-xrd-experiments1}). Reconstructions of
reciprocal-space planes $(h0l)$ and $(h0.5l)$ obtained for a \Co\
single crystal at ambient conditions (see
Fig.~\ref{fig:rec-planes-before-after-hp}a and
Fig.~\ref{fig:rec-planes-before-after-hp}b) and at 13~K
(Fig.~\ref{fig:rec-planes-before-after-hp}c and
Fig.~\ref{fig:rec-planes-before-after-hp}d) demonstrate the quality of
the investigated sample. Notably, a column of superstructure
reflections along the $c^\ast$ axis with equal contributions from two
distinct twin domains can be recognized in
Fig.~\ref{fig:rec-planes-before-after-hp}d.

\begin{figure}[p]
  \centering
  \includegraphics[width=0.85\textwidth]{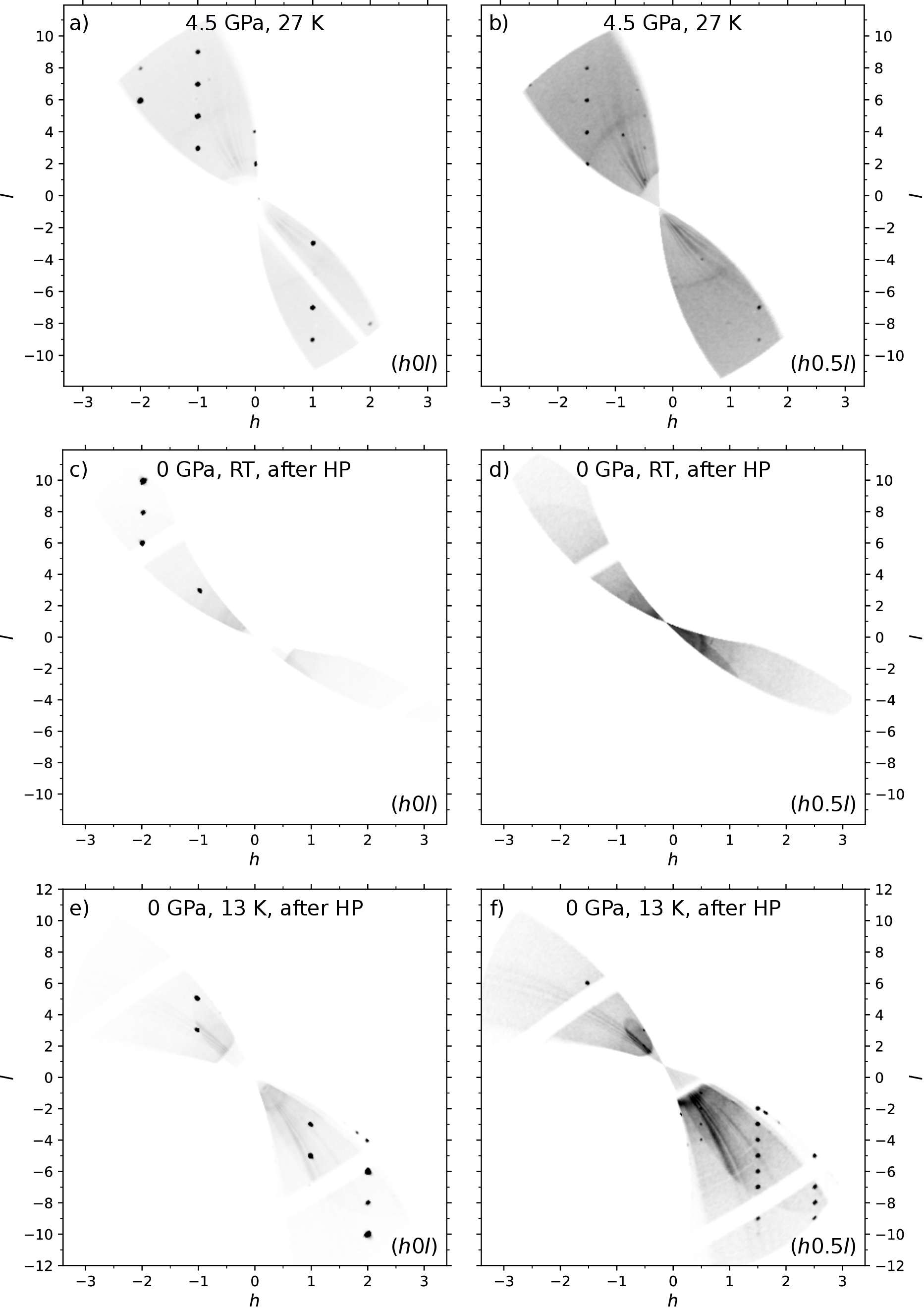}
  \caption{Reconstructions of reciprocal-space planes $(h0l)$ and
    $(h0.5l)$ from x-ray diffraction data collected under various
    conditions, \textit{i.e.} at 27~K with an applied pressure of
    4.5~GPa (a, b), at room temperature after removing the crystal from
    the high-pressure (HP) cell (c, d) and after cooling to 13~K again
    (e, f).}
  \label{fig:rec-planes-before-after-hp1}
\end{figure}

Inserting the single crystal into the pressure chamber of a Tozer-type
diamond anvil cell (DAC) and applying a pressure of 4.5~GPa leaves the
sample crystallinity unchanged
(Fig.~\ref{fig:rec-planes-before-after-hp}e and
Fig.~\ref{fig:rec-planes-before-after-hp}f; ring-shaped features and
additional reflections are due to parasitic x-ray scattering from the
gasket and the DAC diamonds, respectively). This does not change by
cooling the pressure cell to 27~K
(Fig.~\ref{fig:rec-planes-before-after-hp1}a and
Fig.~\ref{fig:rec-planes-before-after-hp1}b). As pointed out in the
main paper, every second reflection in the columns of superstructure
reflections along $c^\ast$ in
Fig.~\ref{fig:rec-planes-before-after-hp1}b is now absent due to a
pressure-induced detwinning process.

Still, all pressure-induced changes to the sample are fully
reversible. This is pointed out by sharp profiles of the Bragg
reflections in the $(h0l)$ plane
(Fig.~\ref{fig:rec-planes-before-after-hp1}c) and the absence of
scattered intensity in the $(h0.5l)$ plane
(Fig.~\ref{fig:rec-planes-before-after-hp1}d) after heating to room
temperature and removing the single crystal from the pressure chamber
of the DAC. Furthermore, the effect of cooling to 13~K again without
applied pressure is consistent with the measurements before the
high-pressure study (compare
Fig.~\ref{fig:rec-planes-before-after-hp1}e and
Fig.~\ref{fig:rec-planes-before-after-hp1}f with
Fig.~\ref{fig:rec-planes-before-after-hp}c and
Fig.~\ref{fig:rec-planes-before-after-hp}d). Namely, reflections in
the $(h0l)$ plane are preserved and complete columns of superstructure
reflections along $c^\ast$ featuring contributions from two twin
domains appear.

\FloatBarrier

\newpage

\subsection{High-pressure measurements at room temperature}

As pointed out in Sec.~\ref{sec:superstruc-RT}, we did not succeed in
forcing the transition between the high-temperature and
low-temperature phase of \Co\ in high-pressure experiments at room
temperature. Still, the HT phase structure provides some flexibility
to react to applied pressure, \textit{e.g.} the position of the carbon
atoms. Therefore, we investigated the effect of pressure on the HT
phase structure by performing x-ray diffraction experiments at 0.2~GPa
and 4.2~GPa (experiment~5 in
Tab.~\ref{tab:single-crystal-xrd-experiments1}, crystal and refinement
details are available in
Tab.~\ref{tab:comparison-0GPa-4GPa-RT-crysdata-ref}, fractional
coordinates and mean-square atomic displacement parameters in
Tab.~\ref{tab:comparison-0GPa-4GPa-RT} and
Tab.~\ref{tab:comparison-0GPa-4GPa-RT-ADPs}).

\begin{figure}[htb]
  \centering
  \includegraphics[width=0.6\textwidth]{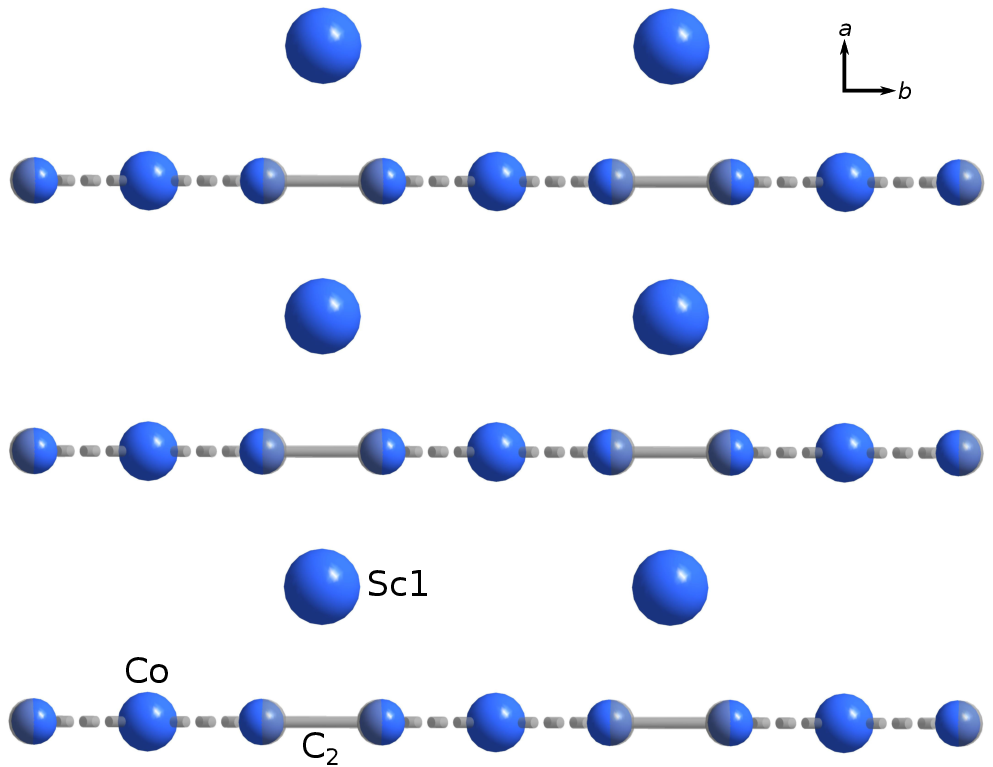}
  \caption{Overlay of the refined atomic positions within a layered
    building unit of \Co\ at room temperature and applied pressures of
    0.2~GPa (gray, semi-transparent) and 4.2~GPa (blue,
    non-transparent). All atom displacements are exaggerated
    seven-fold, Sc2 and Sc3 atoms have been omitted for clarity. The
    given coordinate system refers to the orthorhombic unit cell of
    the high-temperature phase.}
  \label{fig:overlay-0GPa-4GPa-RT}
\end{figure}

Upon pressure application the distance between adjacent
[Co(C$_2$)$_2$]$_\infty$ ribbons along the $a$ axis is reduced from
3.3998(9)~\AA\ to 3.3701(9)~\AA\ (all values are given with their
threefold standard deviation). Analogously, a compression from
6.0061(8)~\AA\ to 5.9670(5)~\AA\ is found for the distance between
adjacent quasi-2D Sc1-Co-C layers. Inspection of the overlaid atomic
positions at 0.2~GPa and 4.2~GPa in
Fig.~\ref{fig:overlay-0GPa-4GPa-RT} reveals no major pressure-induced
changes. Only the C--C bond distance expands insignificantly from
1.452(9)~\AA\ to 1.455(9)~\AA, while the Co--C bond distance is
compressed from 2.094(4)~\AA\ to 2.078(3)~\AA.

\vspace{0.5cm}

\begin{table}[h]
  \centering
  \begin{tabular}{p{0.3\textwidth}>{\centering}p{0.23\textwidth}
    >{\centering\arraybackslash}p{0.23\textwidth}}
    \hline
    \centering $p$ [GPa] & 0.2 & 4.2\\
    \hline
    \multirow{5}{*}{unit cell dimensions} & $a =$~3.3998(3)~\AA & $a =$~3.3701(3)~\AA\\
            & $b =$~4.3738(2)~\AA & $b =$~4.35220(10)~\AA\\
            & $c =$~12.0121(5)~\AA & $c =$~11.9340(3)~\AA\\
            & $V =$~178.620(19)~\AA$^3$ & $V =$~175.040(17)~\AA$^3$\\
    calculated density & 4.4966~g$\cdot$cm$^{-3}$ & 4.5886~g$\cdot$cm$^{-3}$\\
    crystal size & \multicolumn{2}{c}{73$\times$120$\times$186~$\mu$m$^3$}\\
    wave length & \multicolumn{2}{c}{0.56087~\AA}\\
    transm. ratio (max/min) & 0.747 / 0.677 & 0.747 / 0.659\\
    absorption coefficient & 5.001~mm$^{-1}$ & 5.104~mm$^{-1}$\\
    $F(000)$ & \multicolumn{2}{c}{228}\\
    $\theta$ range & \multicolumn{2}{c}{3\degr\ to 35\degr}\\
    range in $hkl$ &  \multicolumn{2}{c}{-2/2, -8/5, -14/24}\\
    total no. reflections &  880 & 871\\
    independent reflections & 178 ($R_\mathrm{int} =$~0.0086) & 172 ($R_\mathrm{int} =$~0.0087)\\
    reflections with $I \geq 2\sigma(I)$ & 169 & 165\\
    data / parameters & 178 / 14 & 172 / 14\\
    goodness-of-fit on $F^2$ & 1.99 & 2.03\\
    \multirow{2}{*}{final $R$ indices [$I \geq 2\sigma(I)$]} & $R =$~0.0172 & $R =$~0.0159\\
    & $wR =$~0.0473 & $wR =$~0.0475\\
    \multirow{2}{*}{$R$ indices (all data)} & $R =$~0.0180 & $R =$~0.0165\\
    & $wR =$~0.0474 & $wR =$~0.0476\\
    extinction coefficient & \multicolumn{2}{c}{--}\\
    largest diff. peak and hole & 0.55 / -0.60~e$\cdot$\AA$^{-3}$ & 0.39 / -0.46~e$\cdot$\AA$^{-3}$\\
    \hline
  \end{tabular}
  \caption{Crystal data and structure refinements for
    single-crystal x-ray diffraction experiments at room
    temperature and at pressures of 0.2~GPa and 4.2~GPa.}
  \label{tab:comparison-0GPa-4GPa-RT-crysdata-ref}
\end{table}

\begin{table}[htb]
  \centering
  \begin{tabular}{>{\centering}p{0.1\textwidth}>{\centering}p{0.1\textwidth}
    >{\centering}p{0.15\textwidth}>{\centering}p{0.15\textwidth}
    >{\centering}p{0.15\textwidth}>{\centering\arraybackslash}p{0.15\textwidth}}
    & \textbf{\textit{p}} & \multicolumn{3}{c}{\textbf{fractional atomic coordinates}} &
    \textbf{\textit{U}\textsubscript{iso}/\textit{U}\textsubscript{eq}}\\
    \hline
    atom & [GPa] & $x$ & $y$ & $z$ & [\AA$^2$]\\
    \hline
    \multirow{2}{*}{Co} & 0.2 & 0 & 0.5 & 0 & 0.0039(3)\\
    & 4.2 & 0 & 0.5 & 0 & 0.0035(3)\\
    \multirow{2}{*}{Sc1} & 0.2 & 0.5 & 0 & 0 & 0.0035(4)\\
    & 4.2 & 0.5 & 0 & 0 & 0.0038(4)\\
    \multirow{2}{*}{Sc2} & 0.2 & 0.5 & 0.5 & 0.18808(3) & 0.0034(3)\\
    & 4.2 & 0.5 & 0.5 & 0.18827(2) & 0.0032(4)\\
    \multirow{2}{*}{C1} & 0.2 & 0.5 & 0.6660(3) & 0.37515(10) & 0.0047(2)\\
    & 4.2 & 0.5 & 0.6671(3) & 0.37516(9) & 0.0039(2)\\
    \hline
  \end{tabular}
  \caption{Refined fractional atomic coordinates and mean-square
    atomic displacement parameters obtained from single-crystal x-ray
    diffraction experiments at room temperature and at pressures of
    0.2~GPa and 4.2~GPa. Note that the carbon atom was refined
    isotropically.}
  \label{tab:comparison-0GPa-4GPa-RT}
\end{table}

\begin{table}[htb]
  \vspace{0.5cm}
  \centering
  \begin{tabular}{>{\centering}p{0.07\textwidth}>{\centering}p{0.07\textwidth}
    >{\centering}p{0.12\textwidth}>{\centering}p{0.12\textwidth}
    >{\centering}p{0.12\textwidth}>{\centering}p{0.12\textwidth}
    >{\centering}p{0.12\textwidth}>{\centering\arraybackslash}p{0.12\textwidth}}
    & \textbf{\textit{p}} & \multicolumn{6}{c}{\textbf{mean-square atomic displacement
                            parameters} [\AA$^2$]}\\
    \hline
    atom & [GPa] & $U_{11}$ & $U_{22}$ & $U_{33}$ & $U_{12}$ & $U_{13}$ & $U_{23}$\\
    \hline
    \multirow{2}{*}{Co} & 0.2 & 0.0051(10) & 0.00355(18) & 0.00301(13) & $\ast$ & $\ast$ & $\ast$\\
    & 4.2 & 0.0042(9) & 0.00328(18) & 0.00294(12) & $\ast$ & $\ast$ & $\ast$\\
    \multirow{2}{*}{Sc1} & 0.2 & 0.0029(12) & 0.0040(2) & 0.00372(16) & $\ast$ & $\ast$ & $\ast$\\
    & 4.2 & 0.0036(13) & 0.0037(2) & 0.00390(16) & $\ast$ & $\ast$ & $\ast$\\
    \multirow{2}{*}{Sc2} & 0.2 & 0.0031(10) & 0.00355(18) & 0.00348(12) & $\ast$ & $\ast$ & $\ast$\\
    & 4.2 & 0.0031(10) & 0.00316(18) & 0.00337(13) & $\ast$ & $\ast$ & $\ast$\\
    \hline
  \end{tabular}
  \caption{Refined mean-square atomic displacement parameters obtained
    from single-crystal x-ray diffraction experiments at room
    temperature and at pressures of 0.2~GPa and 4.2~GPa. Parameters
    marked by an asterisk are forbidden by symmetry. The carbon atom
    was refined isotropically.}
  \label{tab:comparison-0GPa-4GPa-RT-ADPs}
\end{table}

\FloatBarrier

\hspace{1cm}

\newpage

\section{Phonon dispersion relations under uniaxial
  strain} \label{sec:phonon-disp}

\begin{figure}[htb]
  \centering
  \includegraphics[width=0.97\textwidth]{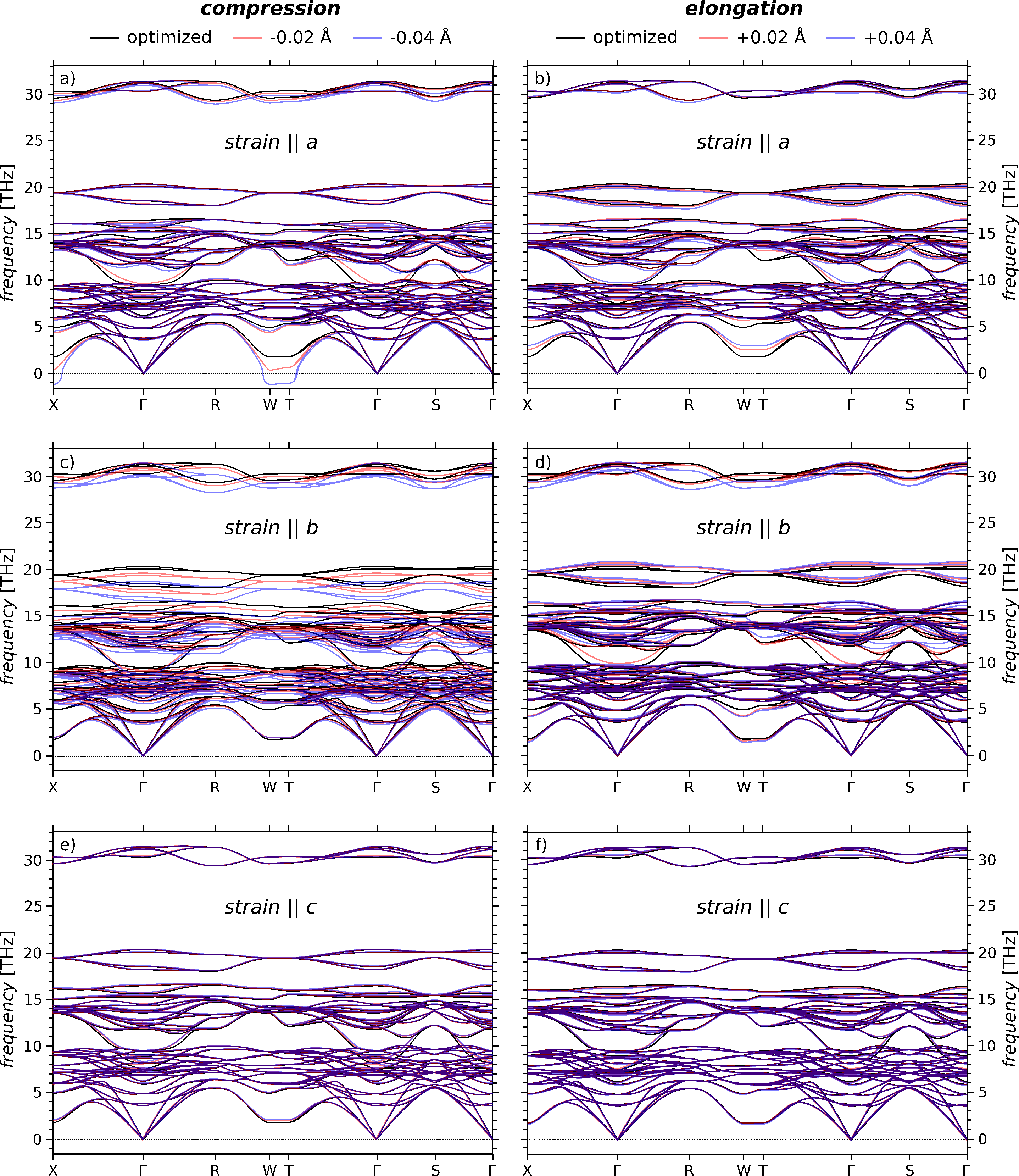}
  \caption{Response of the phonon dispersion of HT \Co\ (DFT study) to
    a compression or elongation of the lattice parameters $a$
    (a, b), $b$ (c, d) and $c$ (e, f).}
  \label{fig:phonon-disp}
\end{figure}

\FloatBarrier
\newpage

\section{Analysis of experimental data}
\label{sec:analysis}
\subsection{Calculation of atom displacements}

\begin{figure}[htb]
  \centering
  \includegraphics[width=1.0\textwidth]{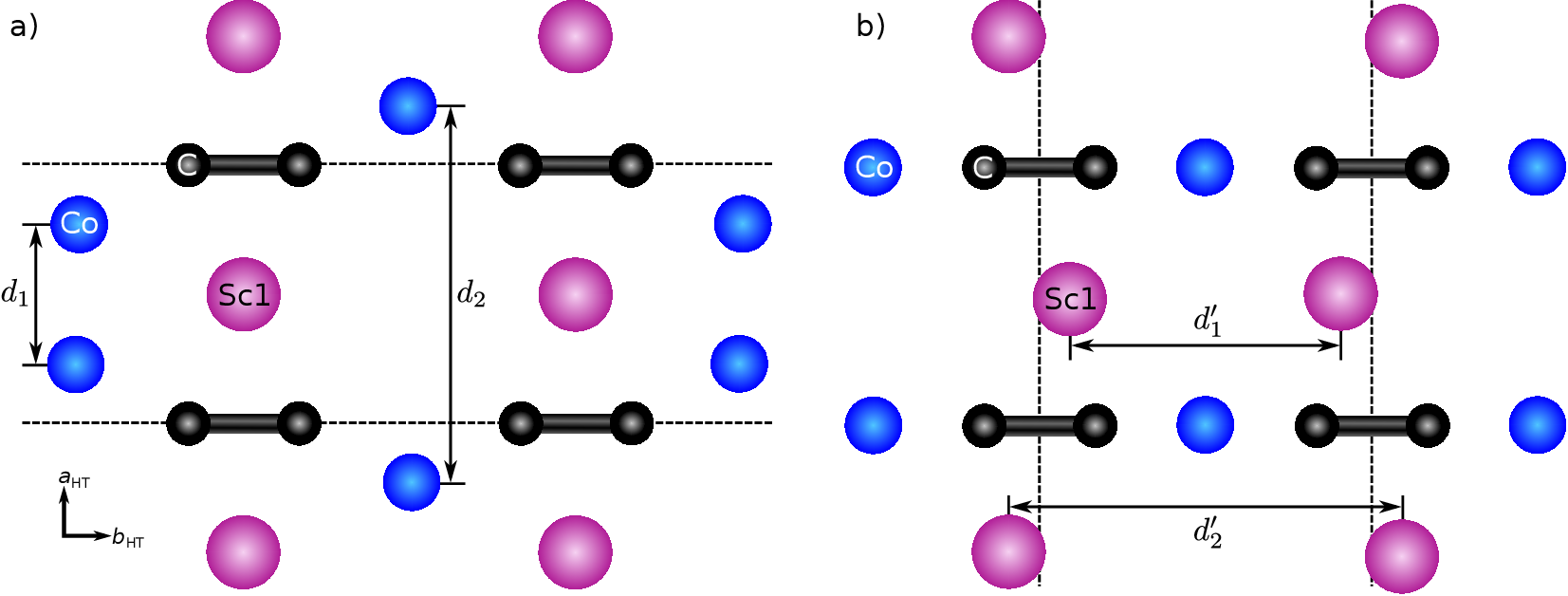}
  \caption{Location of short and long distances between (a)~cobalt and
    (b)~scandium (Sc1) atoms in the low-temperature (LT) phase
    structure of \Co. For clarity, only the displacements of the Co
    (a) or Sc1 atoms (b) from their high-temperature (HT) phase
    positions are depicted. The given coordinate system refers to the
    orthorhombic unit cell of the HT phase.}
  \label{fig:calc-co-sc1-displ}
\end{figure}

Displacements $\Delta r_{\mathrm{Co}}$ of the cobalt atoms from their
positions in the high-temperature phase of \Co\ were calculated from
the difference between the longer Co--Co distance $d_2$ and the
shorter Co--Co distance $d_1$ (see Fig.~\ref{fig:calc-co-sc1-displ}a):

\begin{align}
  \Delta r_{\mathrm{Co}} = \frac{1}{4} (d_2 - d_1)~.
\end{align}

An analogous procedure was applied for the calculation of the scandium
atom (Sc1) displacements $\Delta r_{\mathrm{Sc1}}$ (see
Fig.~\ref{fig:calc-co-sc1-displ}b):

\begin{align}
  \Delta r_{\mathrm{Sc1}} = \frac{1}{4} (d_2' - d_1')~.
\end{align}

\begin{figure}[htb]
  \centering
  \includegraphics[width=0.7\textwidth]{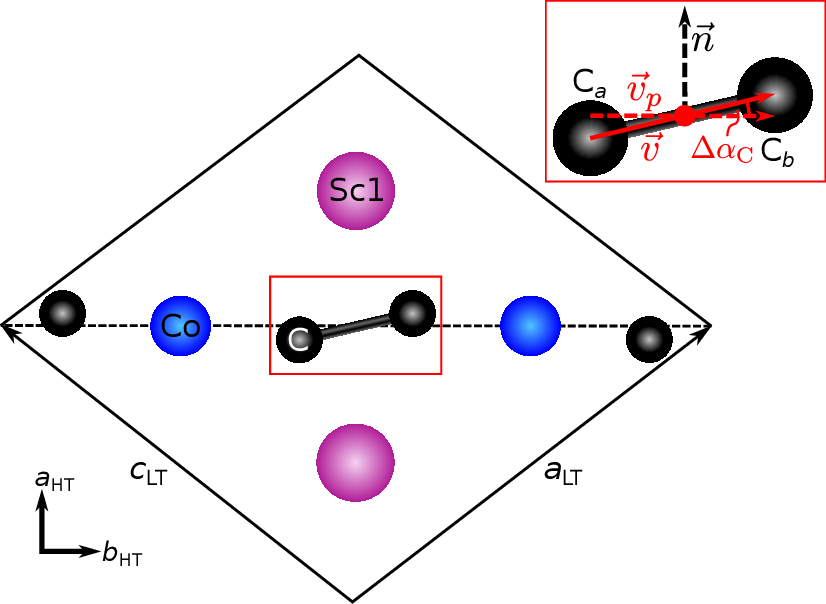}
  \caption{Projection of the low-temperature (LT) phase unit cell of
    \Co\ into the $a_\mathrm{LT}$/$c_\mathrm{LT}$ plane (coordinate
    system corresponding to the unit cell of the orthorhombic
    high-temperature (HT) phase in the lower left corner). For
    clarity, only the displacements of the carbon atoms from their HT
    phase positions are depicted. The location of vectors $\vec{v}$,
    $\vec{v}_p$ and $\vec{n}$ required for the calculation of the
    C$_2$ unit rotation angle $\Delta \alpha_\mathrm{C}$ (see text) is
    indicated in the inset.}
  \label{fig:calc-c-rot}
\end{figure}

The rotation angle $\Delta \alpha_\mathrm{C}$ of the C$_2$ units is spanned
by the vector

\begin{align}
  \vec{v} = A^T (X_{\mathrm{C}_b} - X_{\mathrm{C}_a})
\end{align}

\noindent connecting the two constituent carbon atoms $\mathrm{C}_a$
and $\mathrm{C}_b$ (red solid arrow in Fig.~\ref{fig:calc-c-rot}) and
its projection

\begin{align}
  \vec{v_p} = \vec{v} - (\vec{v} \cdot \vec{n})~\vec{n}
\end{align}

\noindent into the plane of undisplaced carbon atoms (red dashed arrow
in Fig.~\ref{fig:calc-c-rot}). $X_{\mathrm{C}_a}$ and $X_{\mathrm{C}_b}$
denote fractional coordinate matrices, while $A$ denotes the basis
vector matrix of the low-temperature phase unit cell

\begin{align}
  A^T = \left(\vec{a}_\mathrm{LT}~\vec{b}_\mathrm{LT}~\vec{c}_\mathrm{LT}\right).
\end{align}

\noindent Furthermore, the normal vector $\vec{n}$ of the plane of
undisplaced carbon atoms (black dashed arrow in
Fig.~\ref{fig:calc-c-rot}) is calculated from the basis vectors of the
low-temperature phase unit cell as

\begin{align}
  \vec{n} = \frac{(\vec{a}_\mathrm{LT} - \vec{c}_\mathrm{LT}) \times
  \vec{b}_\mathrm{LT}}{|(\vec{a}_\mathrm{LT} - \vec{c}_\mathrm{LT}) \times \vec{b}_\mathrm{LT}|}~.
\end{align}

In a final step, the value of $\Delta \alpha_\mathrm{C}$ can
be obtained using the inner product of $\vec{v}$ and $\vec{v}_p$

\begin{align}
  \Delta \alpha_\mathrm{C} = \arccos{\left(\frac{\vec{v}\cdot\vec{v}_p}{|\vec{v}||\vec{v}_p|}\right)}.
\end{align}

\FloatBarrier
\hspace{1cm}

\newpage

\subsection{Sensitivity of superstructure reflection intensities to
  atom displacements}

Information about the displacements of Co, Sc1 and C atoms from their
positions in the high-temperature (HT) phase of \Co\ is encoded in the
intensities of main and superstructure reflections. To further
elucidate the nature and size of these changes we repeatedly
calculated reflection intensities for rigid structural models with
different positions of Co, Sc1 and C atoms using the software
JANA2006.\cite{Petricek14}

\begin{figure}[htb]
  \centering
  \includegraphics[width=1.0\textwidth]{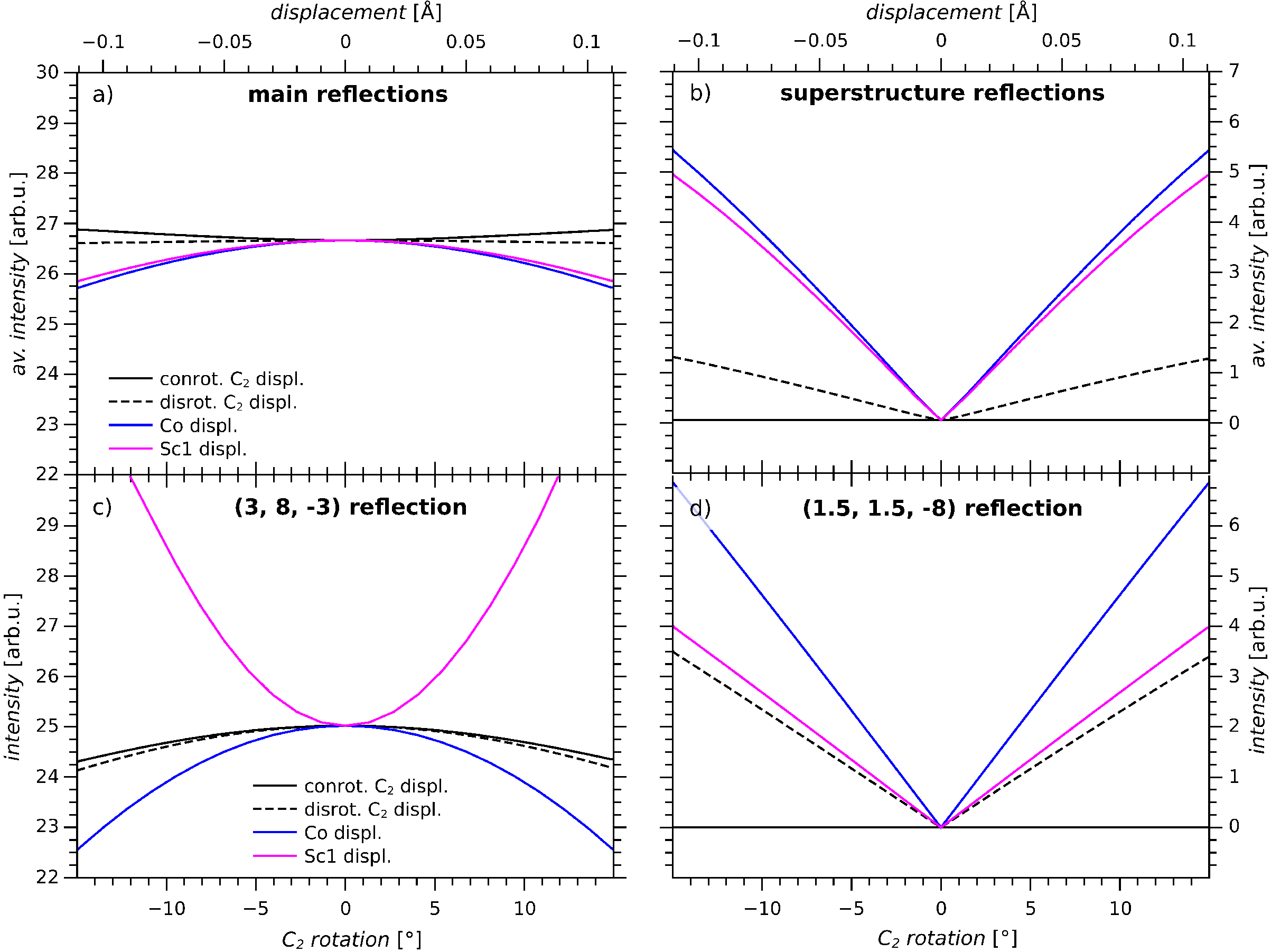}
  \caption{Impact of conrotatory and disrotatory C$_2$ displacements
    (solid and dashed black lines), cobalt (blue line) and scandium
    atom displacements (magenta line) from their HT phase positions on
    the averaged intensity of (a)~main and (b)~superstructure
    reflections.  The according effect on the intensity of an
    exemplary reflection from each of these categories is illustrated
    in (c)~for the main reflection $(3, 8, -3)$ and (d)~for the
    superstructure reflection $(1.5, 1.5, -8)$.  Carbon atom
    displacements corresponding to each C$_2$ rotation (lower
    abscissa) are specified on the upper abscissa.}
  \label{fig:impact-atom-displ-av-refl-intens}
\end{figure}

As a starting point the Co, Sc1 and C atoms in a structural model of
the ambient-pressure low-temperature (LT) phase at 11~K (see
Tab.~\ref{tab:comparison-ambient-low-T-crysdata-ref},
Tab.~\ref{tab:comparison-ambient-low-T} and
Tab.~\ref{tab:comparison-ambient-low-T-ADPs}) were reset to their
high-symmetry positions in the HT phase (see
Tab.~\ref{tab:refls-ints-starting-model}). Furthermore, anisotropic
atomic displacement parameters (ADPs) were replaced by isotropic ones
(Tab.~\ref{tab:refls-ints-starting-model}) and the twin ratio was
fixed to a value of 0.5. The atoms were then displaced individually
and stepwise into the direction of their LT phase positions,
\textit{i.e.} the Co and Sc1 atoms were moved linearly and the C$_2$
units were rotated in a conrotatory or disrotatory fashion. At each
step, only reflection intensities were calculated without refining the
model parameters. Ensuing changes in the averaged intensity of main
and superstructure reflections are displayed in
Fig.~\ref{fig:impact-atom-displ-av-refl-intens}a and
Fig.~\ref{fig:impact-atom-displ-av-refl-intens}b, while
Fig.~\ref{fig:impact-atom-displ-av-refl-intens}c and
Fig.~\ref{fig:impact-atom-displ-av-refl-intens}d show changes in the
intensity of a specific strongly reacting main and superstructure
reflection.

\begin{table}[htb]
  \centering
  \begin{tabular}{>{\centering}p{0.1\textwidth}
    >{\centering}p{0.15\textwidth}>{\centering}p{0.15\textwidth}
    >{\centering}p{0.15\textwidth}>{\centering\arraybackslash}p{0.15\textwidth}}
    & \multicolumn{3}{c}{\textbf{fractional atomic coordinates}} &
    \textbf{\textit{U}\textsubscript{iso}}\\
    \hline
    atom & $x$ & $y$ & $z$ & [\AA$^2$]\\
    \hline
    Co & 0.25 & 0 & 0.25 & 0.002035\\
    Sc1 & 0.75 & 0 & 0.25 & 0.002068\\
    Sc2 & 0 & 0.18801 & 0 & 0.002103\\
    Sc3 & 0 & 0.31199 & 0.5 & 0.002100\\
    C1 & 0.41693 & 0.12557 & 0.08307 & 0.003063\\
    C2 & 0.08307 & 0.12487 & 0.41693 & 0.003034\\
    \hline
  \end{tabular}
  \caption{Fractional atomic coordinates and mean-square
    atomic displacement parameters used in the initial
    structural model for the calculation of reflection
    intensities.}
  \label{tab:refls-ints-starting-model}
\end{table}

\FloatBarrier
\hspace{1cm}

\newpage

\subsection{Reliability of the refined atom displacements}

\begin{figure}[htb]
  \centering
  \includegraphics[width=1.0\textwidth]{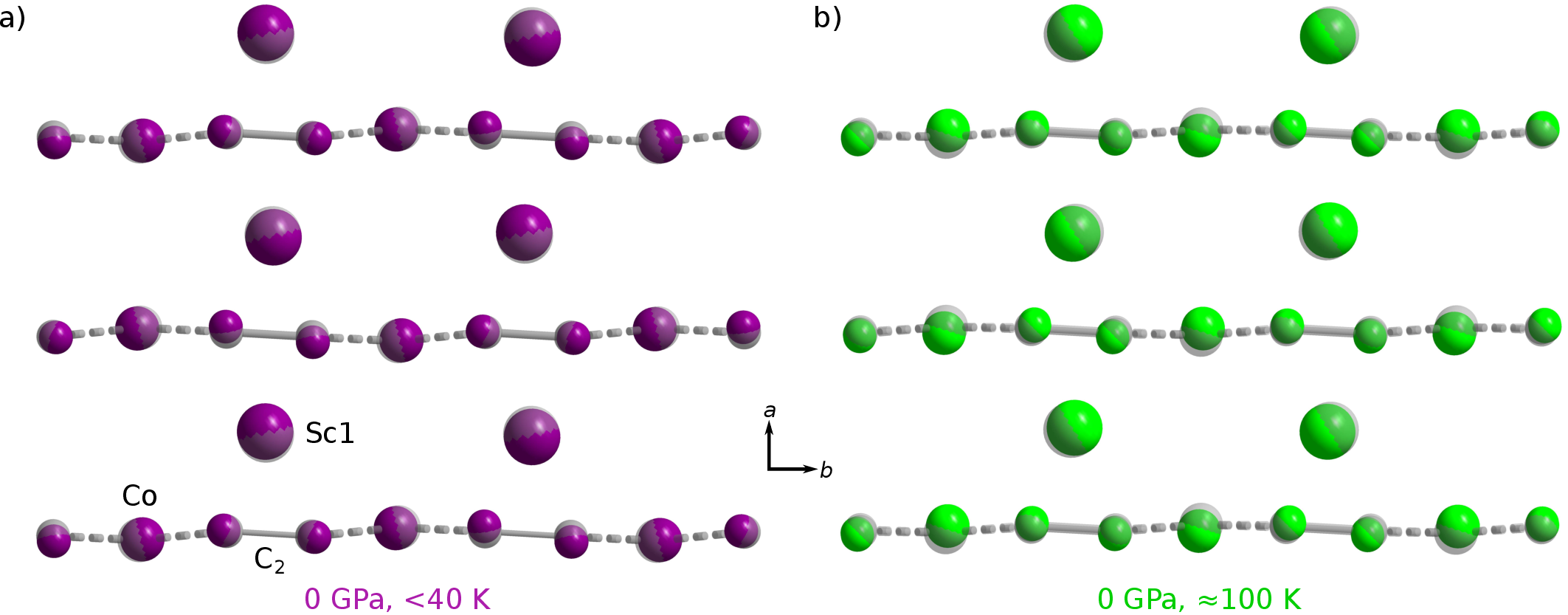}
  \caption{Overlays of the refined atom positions within a layered
    building unit of \Co\ at low temperatures and 0~GPa without (gray,
    semi-transparent) and with surrounding pressure cell (colored,
    non-transparent). In (a) the atom positions at below 40~K (without
    DAC: 11~K, with DAC: 36~K) are compared, and in (b) the atom
    positions at approx. 100~K (without DAC: 100~K, with DAC:
    106~K). Note that all atom displacements are exaggerated
    seven-fold and that Sc2 and Sc3 atoms have been omitted for
    clarity. The given coordinate system refers to the orthorhombic
    unit cell of the high-temperature phase.}
  \label{fig:overlay-0GPa-Base-DAC-Nadel}
\end{figure}

We examined the significance of the observed changes in the rotation
angles of the C$_2$ units between 0~GPa and 4~GPa (experiment~3 in
Tab.~\ref{tab:single-crystal-xrd-experiments}) in more detail. The use
of a closed-cycle sample cryostat and a diamond anvil cell (DAC) in
our low-temperature x-ray diffraction experiments reduces the
achievable data quality in various ways, \textit{e.g.} by a limitation
of the accessible reciprocal space and by parasitic scattering from
the beryllium vacuum shroud (see Sec.~\ref{sec:SCXRD} for more
details). To examine the effect of these factors, we collected
low-temperature x-ray diffraction data sets ($T <$~40~K and
$T \approx$~100~K) for a \Co\ single crystal inside a DAC before and
after filling the cell with a pressure transmitting medium and
application of 4~GPa. In agreement with our results for a
single-crystalline \Co\ needle without surrounding pressure cell
($T =$~11~K and 100~K, experiment~1 in
Tab.~\ref{tab:single-crystal-xrd-experiments}) structural refinements
on the ambient-pressure DAC data ($T =$~36~K and 106~K) point out
conrotatory displacements of neighboring C$_2$ units along the
[Co(C$_2$)$_2$]$_\infty$ ribbons with clearly non-zero rotation angles
between 6(2)$^\circ$ and 7(2)$^\circ$ (all angles are specified with
their threefold standard deviation). Overlays of the structural
models of the ambient-pressure low-temperature phase obtained without
and with surrounding DAC are given in
Fig.~\ref{fig:overlay-0GPa-Base-DAC-Nadel}a ($T <$~40~K) and
Fig.~\ref{fig:overlay-0GPa-Base-DAC-Nadel}b ($T \approx$~100~K).
Crystal and refinement details, fractional coordinates and mean-square
atomic displacement parameters are compared in
Tab.~\ref{tab:comparison-DAC-non-DAC-40K-crysdata-ref} and
Tab.~\ref{tab:comparison-dac-non-dac-40K} ($T <$~40~K), and
Tab.~\ref{tab:comparison-DAC-non-DAC-100K-crysdata-ref} and
Tab.~\ref{tab:comparison-dac-non-dac-100K} ($T \approx$~100~K).

\begin{figure}[h]
  \centering
  \includegraphics[width=1.0\textwidth]{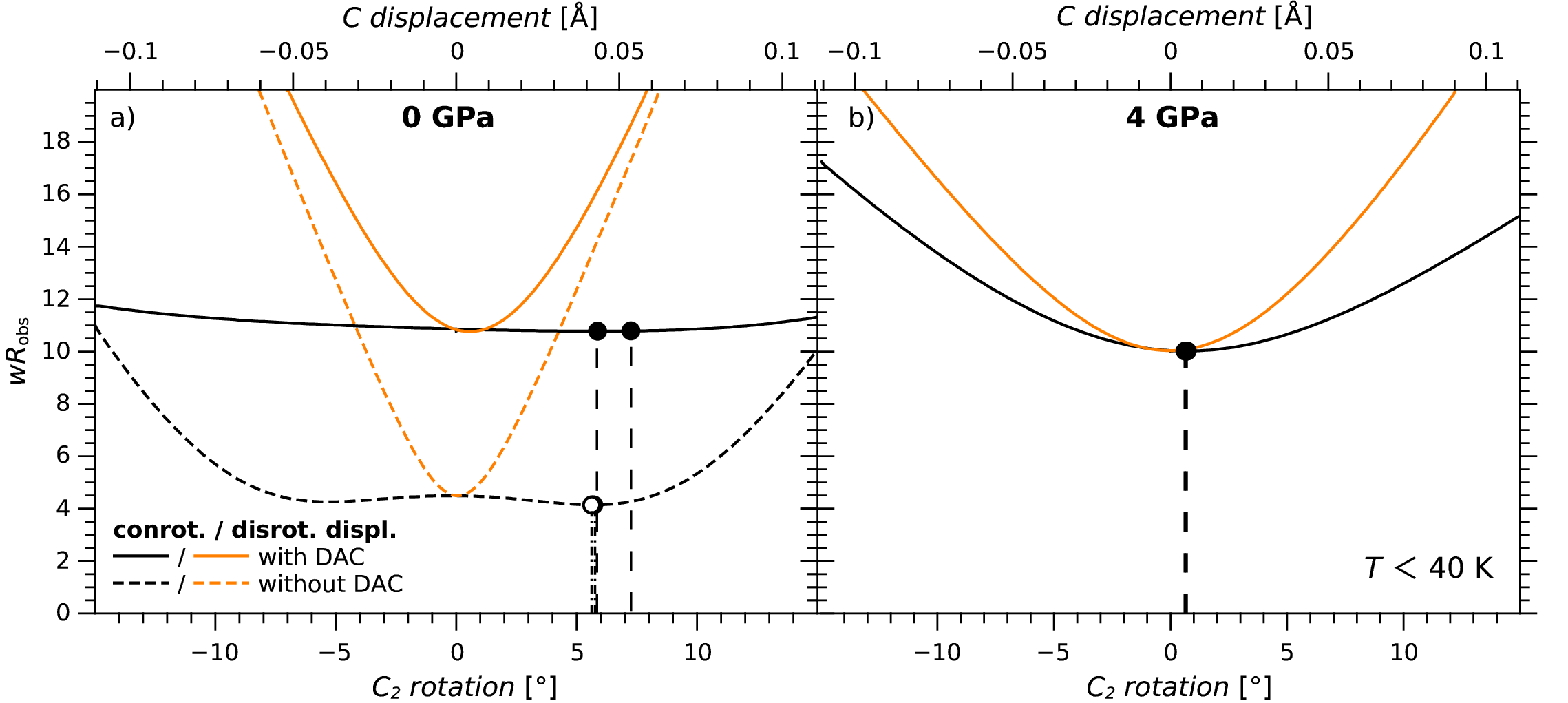}
  \caption{Variation of the weighted $R$ value $wR_\mathrm{obs}$ with the
    rotation angle of the C$_2$ units (lower abscissa) or the
    corresponding carbon atom displacement (upper abscissa) in rigid
    structural models for different low-temperature x-ray diffraction
    data sets (see text for more detailed explanation). In (a) the
    behavior of $wR_\mathrm{obs}$ for ambient-pressure data sets collected
    without ($T =$ 11~K, dashed lines) and with ($T =$ 36~K, solid
    lines) pressure cell is given, while (b) shows the behavior for a
    4~GPa data set ($T =$ 37~K, solid lines).  In each case, the
    rotation angles of neighboring symmetry-independent C$_2$ units
    were constrained to follow conrotatory (black lines) or
    disrotatory displacement patterns (orange lines). Rotation angles
    obtained from unconstrained structural refinements are indicated
    by open and filled black circles.}
  \label{fig:c2rot-rvalues}
\end{figure}

\begin{figure}[h]
  \centering
  \includegraphics[width=1.0\textwidth]{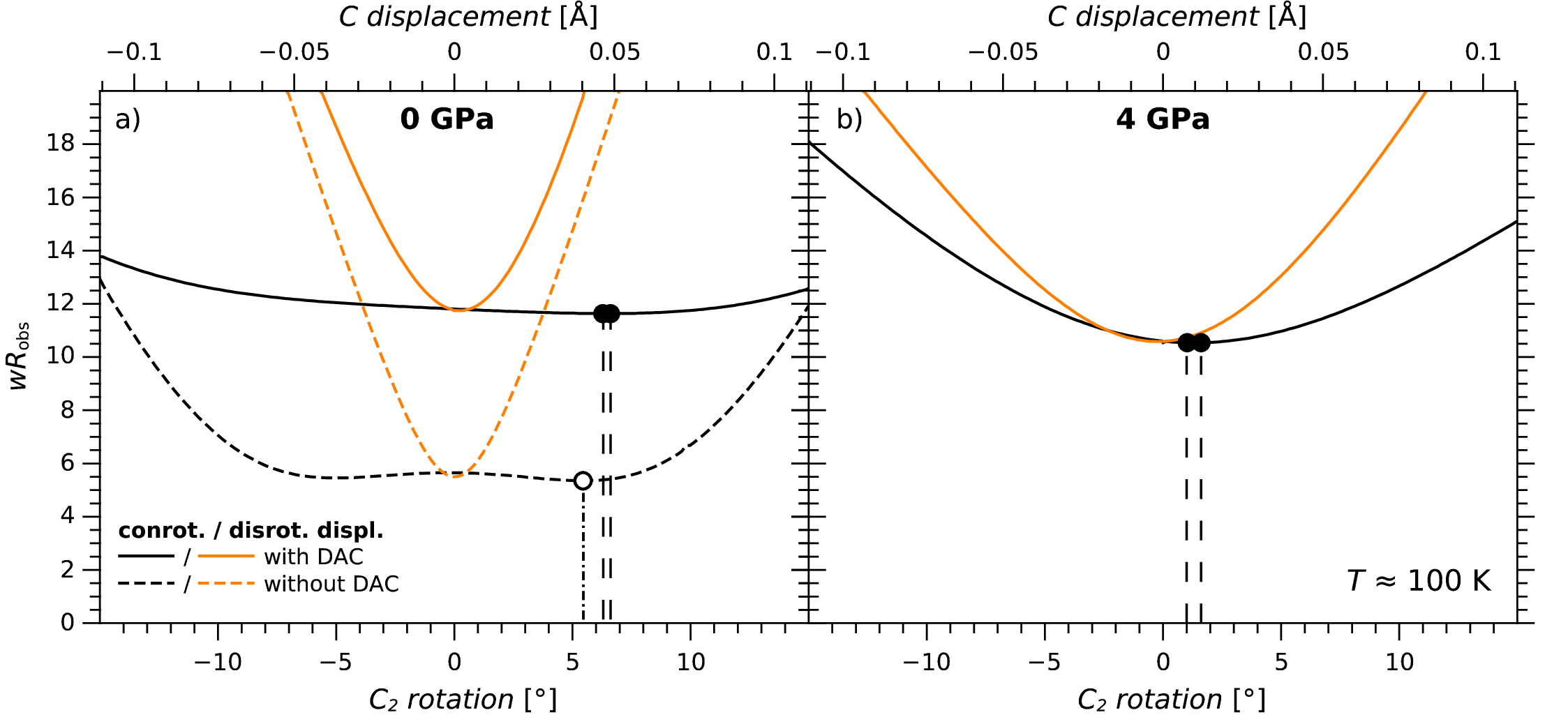}
  \caption{Variation of the weighted $R$ value $wR_\mathrm{obs}$ with the
    rotation angle of the C$_2$ units (lower abscissa) or the
    corresponding carbon atom displacement (upper abscissa) in rigid
    structural models for different low-temperature x-ray diffraction
    data sets. In (a) the behavior of $wR_\mathrm{obs}$ for ambient-pressure
    data sets collected without ($T =$ 100~K, dashed lines) and with
    ($T =$ 106~K, solid lines) pressure cell is given, while (b) shows
    the behavior for a 4~GPa data set ($T =$ 107~K, solid lines).  In
    each case, the rotation angles of neighboring symmetry-independent
    C$_2$ units were constrained to follow conrotatory (black
    lines) or disrotatory displacement patterns (orange
    lines). Rotation angles obtained from unconstrained structural
    refinements are indicated by open and filled black circles.}
  \label{fig:c2rot-rvalues-100K}
\end{figure}

Furthermore, we checked the sensitivity of the fit quality indicator
$wR_\mathrm{obs}$ (weighted $R$ value) in structural refinements of the x-ray
diffraction data collected with and without DAC to changes in the
C$_2$ rotation angles (see Fig.~\ref{fig:c2rot-rvalues} for $T <$~40~K
and Fig.~\ref{fig:c2rot-rvalues-100K} for $T \approx$~100~K). To this
end, the rotation angles of the two symmetry-independent C$_2$ units
in the LT phase structure of \Co\ were constrained to follow
conrotatory or disrotatory displacement patterns. After an initial
relaxation the structural model was kept rigid and the value of
$wR_\mathrm{obs}$ was recorded while incrementing the C$_2$ rotation angle in
0.1\degr\ steps between -15\degr\ and 15\degr.

The course of $wR_\mathrm{obs}$ with varying rotation angle at ambient
pressure without DAC is indicated by dashed lines in
Fig.~\ref{fig:c2rot-rvalues}a and Fig.~\ref{fig:c2rot-rvalues-100K}a.
Conrotatory displacements of neighboring C$_2$ units lead to a curve
with two minima (black lines), whereas disrotatory displacements lead
to a curve with a single minimum (orange lines).  Consistent with the
results of unconstrained refinements (marked by black open circles)
the minima in $wR_\mathrm{obs}$ for disrotatory displacements at 0\degr\ are
located at slightly higher $wR_\mathrm{obs}$ values than the minima at
non-zero angles for conrotatory displacements ($T =$~11~K:
$\Delta wR_\mathrm{obs} = 0.35$; $T =$~100~K: $\Delta wR_\mathrm{obs} = 0.13$). This
$wR_\mathrm{obs}$ difference between the minima of the curves for conrotatory
and disrotatory C$_2$ displacements degrades for the low-temperature
DAC measurements at 0~GPa ($T =$~36~K: $\Delta wR_\mathrm{obs} = -0.02$;
$T =$~106~K: $\Delta wR_\mathrm{obs} = 0.11$), although an unconstrained
refinement still reliably converges to a structural model with
conrotatory displacements and non-zero rotation angles (filled
circles). Still, the flat course of $wR_\mathrm{obs}$ with varying
conrotatory C$_2$ displacements directly relates to the large
(threefold) standard deviation of the obtained rotation angles in the
range of 2\degr. Applying a pressure of 4~GPa makes the situation much
more clear-cut (see Fig.~\ref{fig:c2rot-rvalues}b for $T =$~37~K and
Fig.~\ref{fig:c2rot-rvalues-100K}b for $T =$~107~K): Both, the curves
for conrotatory (solid black lines) and disrotatory C$_2$ unit
displacements (solid orange lines) have a substantial curvature and a
single minimum close to 0\degr\ in confirmation of the results of
unconstrained structural refinements (filled black circles).

\begin{table}[htb]
  \centering
  \begin{tabular}{p{0.3\textwidth}>{\centering}p{0.23\textwidth}
    >{\centering\arraybackslash}p{0.23\textwidth}}
    \hline
    \centering DAC & no & yes\\
    \hline
    \multirow{5}{*}{unit cell dimensions} & $a =$~5.53630(10)~\AA & $a =$~5.53940(10)~\AA\\
            & $b =$~12.0210(2)~\AA & $b =$~12.0309(2)~\AA\\
            & $c =$~5.53640(10)~\AA & $c =$~5.53850(10)~\AA\\
            & $\beta =$~104.8070(10)\degr & $\beta =$~104.8280(10)\degr\\
            & $V =$~356.222(11)~\AA$^3$ & $V =$~356.816(11)~\AA$^3$\\
    calculated density & 4.5095~g$\cdot$cm$^{-3}$ & 4.502~g$\cdot$cm$^{-3}$\\
    crystal size & 40$\times$51$\times$290~$\mu$m$^3$ & 68$\times$116$\times$126~$\mu$m$^3$\\
    wave length & \multicolumn{2}{c}{0.56087~\AA}\\
    transm. ratio (max/min) & 0.747 / 0.686 & 0.747 / 0.565\\
    absorption coefficient & 5.016~mm$^{-1}$ & 5.007~mm$^{-1}$\\
    $F(000)$ & \multicolumn{2}{c}{456}\\
    $\theta$ range & 3\degr\ to 36\degr & 3\degr\ to 33\degr\\
    range in $hkl$ & -11/11, -25/25, -11/11 & -6/10, -20/21, -6/10\\
    total no. reflections & 8720 & 1291\\
    independent reflections & 2142 ($R_\mathrm{int} =$~0.0123) & 390 ($R_\mathrm{int} =$~0.0101)\\
    reflections with $I \geq 2\sigma(I)$ & 2007 & 318\\
    data / parameters & 2142 / 43  & 318 / 19\\
    goodness-of-fit on $F^2$ & 1.27 & 3.50\\
    \multirow{2}{*}{final $R$ indices [$I \geq 2\sigma(I)$]} & $R =$~0.0220 & $R =$~0.0342\\
    & $wR =$~0.0414 & $wR =$~0.1064\\
    \multirow{2}{*}{$R$ indices (all data)} & $R =$~0.0271 & $R =$~0.0342\\
    & $wR =$~0.0424 & $wR =$~0.1064\\
    extinction coefficient & 0.0461(14) & --\\
    largest diff. peak and hole & 1.97 / -2.18~e$\cdot$\AA$^{-3}$ & 0.41 / -0.43~e$\cdot$\AA$^{-3}$\\
    \hline
  \end{tabular}
  \caption{Crystal data and structure refinements for ambient-pressure
    single-crystal x-ray diffraction experiments without ($T =$~11~K)
    and with surrounding unpressurized Tozer-type diamond anvil cell
    ($T =$~36~K).}
  \label{tab:comparison-DAC-non-DAC-40K-crysdata-ref}
\end{table}

\begin{table}[htb]
  \centering
  \begin{tabular}{>{\centering}p{0.1\textwidth}>{\centering}p{0.1\textwidth}
    >{\centering}p{0.15\textwidth}>{\centering}p{0.15\textwidth}
    >{\centering}p{0.15\textwidth}>{\centering\arraybackslash}p{0.15\textwidth}}
    && \multicolumn{3}{c}{\textbf{fractional atomic coordinates}} &
    \textbf{\textit{U}\textsubscript{iso}/\textit{U}\textsubscript{eq}}\\
    \hline
    atom & DAC & $x$ & $y$ & $z$ & [\AA$^2$]\\
    \hline
    \multirow{2}{*}{Co} & no & 0.26595(2) & 0 & 0.26673(2) & 0.00204(3)\\
    & yes & 0.26649(12) & 0 & 0.26564(12) & 0.0026(2)\\
    \multirow{2}{*}{Sc1} & no & 0.75582(3) & 0 & 0.24273(3) & 0.00207(6)\\
    & yes & 0.75700(14) & 0 & 0.24355(13) & 0.0029(3)\\
    \multirow{2}{*}{Sc2} & no & 0 & 0.187417(10) & 0 & 0.00210(9)\\
    & yes & 0 & 0.18746(6) & 0  & 0.0029(3)\\
    \multirow{2}{*}{Sc3} & no & 0 & 0.311540(10) & 0.5 & 0.00210(9)\\
    & yes & 0 & 0.31145(6) & 0.5 & 0.0030(3)\\
    \multirow{2}{*}{C1} & no & 0.4110(3) & 0.12557(5) & 0.0766(2) & 0.0031(2)\\
    & yes & 0.410(3) & 0.1257(2) & 0.077(3) & 0.0049(5)\\
    \multirow{2}{*}{C2} & no & 0.0889(3) & 0.12487(5) & 0.4233(2) & 0.0030(2)\\
    & yes & 0.091(3) & 0.1251(2) & 0.425(3) & 0.0049(5)\\
    \hline
  \end{tabular}
  \caption{Refined fractional atomic coordinates and mean-square
    atomic displacement parameters obtained from low-temperature
    single-crystal x-ray diffraction experiments without ($T =$~11~K)
    and with surrounding unpressurized Tozer-type diamond anvil cell
    (DAC; $T =$~36~K). Note that the fractional coordinates refined
    from DAC data have been transformed in order to correspond to the
    same twin individual as the coordinates from non-DAC data.}
  \label{tab:comparison-dac-non-dac-40K}
\end{table}

\begin{table}[h]
  \centering
  \begin{tabular}{p{0.3\textwidth}>{\centering}p{0.23\textwidth}
    >{\centering\arraybackslash}p{0.23\textwidth}}
    \hline
    \centering DAC & no & yes\\
    \hline
    \multirow{5}{*}{unit cell dimensions} & $a =$~5.53720(10)~\AA & $a =$~5.5386(2)~\AA\\
            & $b =$~12.00370(10)~\AA & $b =$~12.0071(3)~\AA\\
            & $c =$~5.53710(10)~\AA & $c =$~5.5365(2)~\AA\\
            & $\beta =$~104.4620(10)\degr & $\beta =$~104.386(2)\degr\\
            & $V =$~356.372(10)~\AA$^3$ & $V =$~356.65(2)~\AA$^3$\\
    calculated density & 4.5076~g$\cdot$cm$^{-3}$ & 4.5041~g$\cdot$cm$^{-3}$\\
    crystal size & 40$\times$51$\times$290~$\mu$m$^3$ & 68$\times$116$\times$126~$\mu$m$^3$\\
    wave length & \multicolumn{2}{c}{0.56087~\AA}\\
    transm. ratio (max/min) & 0.747 / 0.646 & 0.746 / 0.583\\
    absorption coefficient & 5.014~mm$^{-1}$ & 5.01~mm$^{-1}$\\
    $F(000)$ & \multicolumn{2}{c}{456}\\
    $\theta$ range & 3\degr\ to 37\degr & 3\degr\ to 32\degr\\
    range in $hkl$ & -11/11, -25/25, -11/11 & -6/10, -21/21, -6/10\\
    total no. reflections & 8509 & 1234\\
    independent reflections & 2153 ($R_\mathrm{int} =$~0.0148) & 394 ($R_\mathrm{int} =$~0.0143)\\
    reflections with $I \geq 2\sigma(I)$ & 1947 & 291\\
    data / parameters & 2153 / 43  & 291 / 19\\
    goodness-of-fit on $F^2$ & 1.44 & 3.25\\
    \multirow{2}{*}{final $R$ indices [$I \geq 2\sigma(I)$]} & $R =$~0.0302 & $R =$~0.0437\\
    & $wR =$~0.0535 & $wR =$~0.1153\\
    \multirow{2}{*}{$R$ indices (all data)} & $R =$~0.0389 & $R =$~0.0437\\
    & $wR =$~0.0549 & $wR =$~0.1153\\
    extinction coefficient & 0.0231(15) & --\\
    largest diff. peak and hole & 1.94 / -2.20~e$\cdot$\AA$^{-3}$ & 0.42 / -0.47~e$\cdot$\AA$^{-3}$\\
    \hline
  \end{tabular}
  \caption{Crystal data and structure refinements for ambient-pressure
    single-crystal x-ray diffraction experiments without ($T =$~100~K)
    and with surrounding unpressurized Tozer-type diamond anvil cell
    ($T =$~106~K).}
  \label{tab:comparison-DAC-non-DAC-100K-crysdata-ref}
\end{table}

\begin{table}[htb]
  \centering
  \begin{tabular}{>{\centering}p{0.1\textwidth}>{\centering}p{0.1\textwidth}
    >{\centering}p{0.15\textwidth}>{\centering}p{0.15\textwidth}
    >{\centering}p{0.15\textwidth}>{\centering\arraybackslash}p{0.15\textwidth}}
    && \multicolumn{3}{c}{\textbf{fractional atomic coordinates}} &
    \textbf{\textit{U}\textsubscript{iso}/\textit{U}\textsubscript{eq}}\\
    \hline
    atom & DAC & $x$ & $y$ & $z$ & [\AA$^2$]\\
    \hline
    \multirow{2}{*}{Co} & no & 0.25987(2) & 0 & 0.26046(2) & 0.00243(5)\\
    & yes & 0.25814(11) & 0 & 0.25757(11) & 0.0033(3)\\
    \multirow{2}{*}{Sc1} & no & 0.75383(3) & 0 & 0.24498(3) & 0.00239(10)\\
    & yes & 0.75366(14) & 0 & 0.24668(13) & 0.0034(3)\\
    \multirow{2}{*}{Sc2} & no & 0 & 0.187747(13) & 0 & 0.00233(10)\\
    & yes & 0 & 0.18785(6) & 0  & 0.0036(3)\\
    \multirow{2}{*}{Sc3} & no & 0 & 0.311642(13) & 0.5 & 0.00235(10)\\
    & yes & 0 & 0.31168(6) & 0.5 & 0.0035(3)\\
    \multirow{2}{*}{C1} & no & 0.4109(3) & 0.12514(5) & 0.0773(3) & 0.0033(3)\\
    & yes & 0.4103(19) & 0.1247(2) & 0.076(2) & 0.0047(6)\\
    \multirow{2}{*}{C2} & no & 0.0890(3) & 0.12471(5) & 0.4228(3) & 0.0033(3)\\
    & yes & 0.0893(19) & 0.1247(2) & 0.425(2) & 0.0047(6)\\
    \hline
  \end{tabular}
  \caption{Refined fractional atomic coordinates and mean-square
    atomic displacement parameters obtained from low-temperature
    single-crystal x-ray diffraction experiments without ($T =$~100~K)
    and with surrounding unpressurized Tozer-type diamond anvil cell
    (DAC; $T =$~106~K). Note that the fractional coordinates refined
    from DAC data have been transformed in order to correspond to the
    same twin individual as the coordinates from non-DAC data.}
  \label{tab:comparison-dac-non-dac-100K}
\end{table}

\FloatBarrier

\newpage

\subsection{Differences between twin domains}

\begin{figure}[htb]
  \centering
  \includegraphics[width=1.0\textwidth]{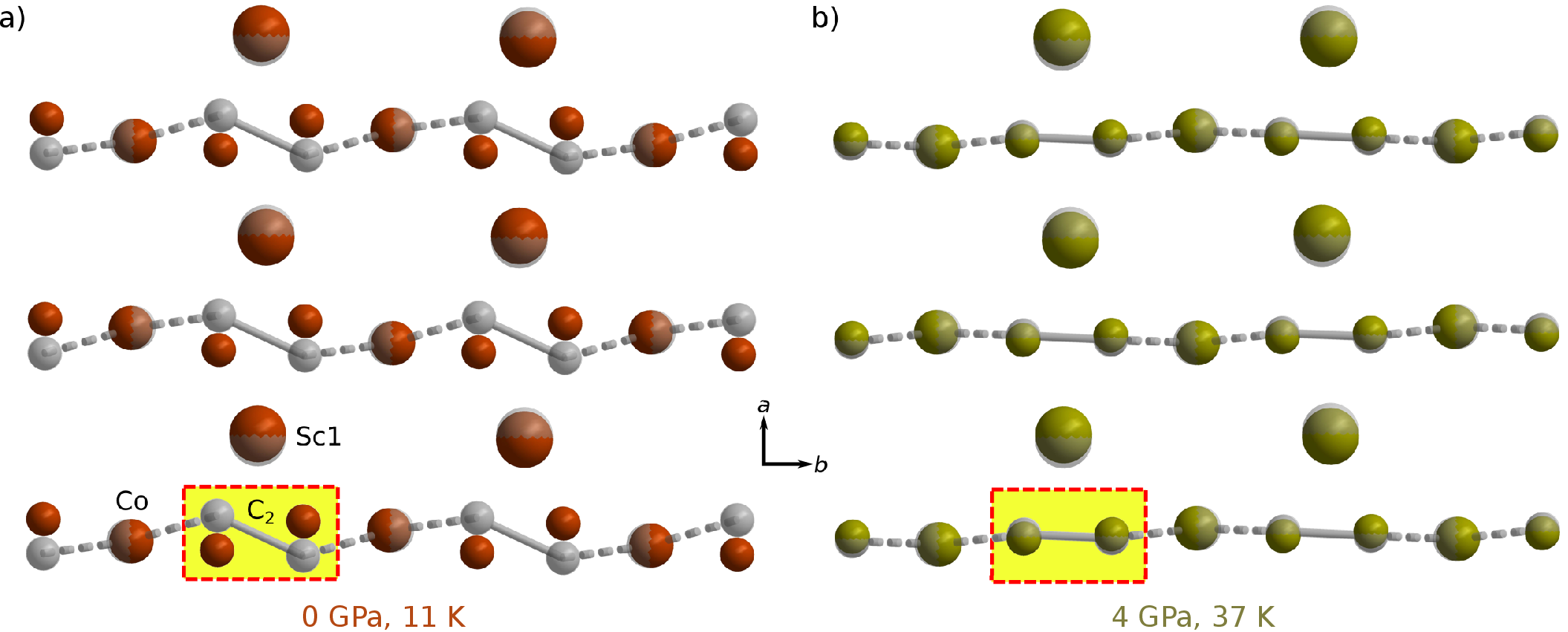}
  \caption{Overlay of the atomic positions within the possible twin
    domains~1 (colored, non-transparent spheres) and 2 (gray,
    semi-transparent spheres) of \Co\ (a)~in its ambient-pressure and
    (b)~in its high-pressure low-temperature phase. All atom
    displacements are exaggerated seven-fold. For clarity, only the
    atoms within a layered building unit are shown, and Sc2 and Sc3
    atoms have been omitted. The given coordinate system refers to the
    orthorhombic unit cell of the high-temperature phase.}
  \label{fig:diff-twin-domains}
\end{figure}

Single crystals of \Co\ are subjected to systematic twinning in the
phase transition from the high-temperature (HT) to the low-temperature
(LT) phase. This twinning process is due to a $t2$ step followed by an
$i2$ step in the symmetry reduction (\textit{translationengleiche} and
isomorphic groub-subgroup relationship, respectively) from the
orthorhombic space-group $Immm$ of the HT phase structure to the
monoclinic space-group $C2/m$ of the LT phase
structure.\cite{Eickerling13,Vogt09} An exemplary twinning operation
that transforms the atomic positions in one twin domain into the
atomic positions of the other is a $m$ plane perpendicular to the $a$
axis of the orthorhombic HT unit cell ([[-1~0~0], [0~1~0],
[0~0~1]]). Its effect is demonstrated for the structural models of the
LT phase at 0~GPa and 11~K (Fig.~\ref{fig:diff-twin-domains}a) and at
4~GPa and 37~K (Fig.~\ref{fig:diff-twin-domains}b). Whereas the twin
domains at 0~GPa can be differentiated easily by the sense of rotation
of the C$_2$ units, this is barely possible for the (hypothetical) twin
domains with nearly unrotated C$_2$ units at 4~GPa.

\vspace{0.5cm}

\FloatBarrier

%